\documentclass[11pt,a4paper]{article}

\usepackage[utf8]{inputenc}
\usepackage{amsmath,latexsym}
\usepackage{xcolor}
\usepackage{graphicx}
\usepackage{axodraw2}
\usepackage{slashed}
\usepackage{caption}
\usepackage{subcaption}
\usepackage{ytableau}
\usepackage{amsthm}
\usepackage{axodraw2}
\usepackage{xfp}
\usepackage{url}

\usepackage{tikz}
\usepackage[nomessages]{fp}
\usepackage{jheppub}

\usepackage{makecell}

\usepackage{bbm}
\usepackage{float}
\usepackage{empheq}
\usepackage{cleveref}
\usepackage{tensor}
\usepackage{booktabs}
\usepackage{multirow}
\usepackage{array}
\usepackage{tabularx}
\newcolumntype{Y}{>{\centering\arraybackslash}X}
\usepackage{makecell}
\usepackage{soul}
\usepackage{placeins}

\DeclareMathOperator{\tr}{tr}
\newcommand{\trId}{\tr{(\mathbbm{1})}}
\DeclareMathOperator{\sgn}{sgn}

\newcolumntype{?}{!{\vrule width 3\arrayrulewidth}}

\def\beq{\begin{equation}}   
\def\eeq{\end{equation}}
\def\bea{\begin{eqnarray}}  
\def\eea{\end{eqnarray}} 
\def\nn{\nonumber}

\def\f21{{}_2F_{1}}

\def\muSet{{\mu_1\dots\mu_N}}

\newcommand{\FORM}{\textsc{Form}}
\newcommand{\OPITER}{\textsc{OPITeR}}

\newcommand{\ed}{\end{document}}

\newcommand{\ice}[1]{\relax}

\renewcommand{\arraystretch}{1.5}

\newcommand{\ixset}[2]{{#1_1\dots #1_{#2}}}
\newcommand{\measure}[2]{d^D#1_1 \dots d^D#1_#2}

\newcommand\pinkline[4]{
\Line[color=RubineRed,width=1.5,dash,dsize=2](#1,#2)(#3,#4)
\Vertex(#1,#2){1.5}
\Vertex(#3,#4){1.5}
}

\newcommand\Pinkline[4]{
\Line[color=RubineRed,width=6,dash,dsize=6](#1,#2)(#3,#4)
\Vertex(#1,#2){6}
\Vertex(#3,#4){6}
}

\newcommand\Pinkbezier[8]{
\Bezier[color=RubineRed,width=6,dash,dsize=6](#1,#2)(#3,#4)(#5,#6)(#7,#8)
\Vertex(#1,#2){6}
\Vertex(#7,#8){6}
}

\newcommand\pinkspin[2]{\CCirc(#1,#2){10}{White}{White}
\DashArc[color=RubineRed,width=1](#1,#2)(10,0,360){3}}
\newcommand\Pinkspin[2]{\CCirc(#1,#2){10}{White}{White}
\DashArc[color=RubineRed,width=3,dsize=4](#1,#2)(10,0,360){3}}

\newcommand\Blackspin[2]{\CCirc(#1,#2){10}{White}{White}
\Arc[width=3](#1,#2)(10,0,360)}

\ytableausetup{smalltableaux}
\preprint{Nikhef 2024-015}
\title{Tensor Reduction for Feynman Integrals with Lorentz and Spinor Indices}

\author[a]{Jae Goode,}
\author[a]{Franz Herzog,}
\author[a]{Anthony Kennedy,}
\author[a]{Sam Teale,}
\author[b]{and Jos Vermaseren}

\affiliation[a]{Higgs Centre for Theoretical Physics, School of Physics and Astronomy, The University of Edinburgh, Edinburgh EH9 3FD, Scotland, UK
}
\affiliation[b]{Nikhef Theory Group, Science Park 105, 1098 XG Amsterdam, The Netherlands}

\emailAdd{jgoode2@ed.ac.uk}
\emailAdd{fherzog@ed.ac.uk}
\emailAdd{tony.kennedy@ed.ac.uk}
\emailAdd{sam.teale@ed.ac.uk}
\emailAdd{t68@nikhef.nl}

\abstract{
We present an efficient graphical approach to construct projectors for the tensor reduction of multi-loop Feynman integrals with both Lorentz and spinor indices in $D$ dimensions. An ansatz for the projectors is constructed making use of its symmetry properties via an orbit partition formula. The graphical approach allows to identify and enumerate the orbits in each case. For the case without spinor indices we find a 1 to 1 correspondence between orbits and integer partitions describing the cycle structure of certain bi-chord graphs. This leads to compact combinatorial formulae for the projector ansatz. With spinor indices the graph-structure becomes more involved, but the method is equally applicable. Our spinor reduction formulae are based on the antisymmetric basis of $\gamma$ matrices, and make use of their orthogonality property. We also provide a new compact formula to pass into the antisymmetric basis. We compute projectors for vacuum tensor Feynman integrals with up to 32 Lorentz indices and up to 4 spinor indices. We discuss how to employ the projectors in problems with external momenta.
}

\begin{document}

\keywords{}

\maketitle

\section{Introduction}
\label{sec:intro}
Calculations of multi-loop corrections in perturbative Quantum Field Theory have played an indispensable role in the validation of the Standard Model. The most fruitful method for such calculations is to reduce loop-integrals, appearing in Feynman-diagrammatic calculations, to a basis of so-called master integrals (MIs). These MIs are a set of independent (Lorentz) scalar integrals. The procedure to write a Feynman amplitude into this basis generally consists of two steps. The first is to reduce the tensor integrals into scalar integrals. This procedure -- known as tensor reduction -- is the main focus of this paper. After tensor reduction the scalar integrals are typically further reduced to MIs by solving a large system of linear equations, generated from integration-by-parts identities (IBPs) \cite{Tkachov:1981wb,Chetyrkin:1981qh,Laporta:2000dsw}.

A general method to deal with the problem of tensor reduction is known as Passarino-Veltmann reduction \cite{Passarino:1978jh}. Here the idea is to first write down an ansatz in terms of all possible Lorentz structures composed of metric tensors and momenta, on which the integral could depend. Subsequently, one contracts the integral and ansatz with all these possible structures to obtain a system of equations, the solution of which will yield expressions for the scalar coefficients in the ansatz in terms of scalar integrals one has to compute. The problem with this approach is that this dense system quickly becomes large and intractable.

In the meantime a host of new methods have been introduced to deal with the problem. In particular, at one-loop, elegant solutions to the tensor reduction problem exist, which build on a variety of different methodologies \cite{Ezawa:1990dh,Devaraj:1997es,Denner:2005nn,Binoth:2008uq,Diakonidis:2008ij,Diakonidis:2009fx, Fleischer:2010sq,Fleischer:2011bi,Fleischer:2011zz,Fleischer:2011zx,Fleischer:2011hc,Fleischer:2011nt} or circumvent the problem via unitarity-based methods \cite{Ossola:2006us,Forde:2007mi,Giele:2008ve,Berger:2008sj}. Another method developed in recent years, so far mostly at one-loop, is based on the contraction with an auxiliary vector \cite{Feng:2022hyg,Feng:2021enk,Hu:2021nia,Feng:2022uqp,Feng:2022rfz,Feng:2022iuc}. Also, at two loops,
a variety of methods have been proposed and successfully employed. These include the Passarino-Veltmann technique applied to on-shell amplitudes \cite{Binoth:2002xg}, unitarity-based approaches \cite{Ita:2015tya,Abreu:2017hqn,Badger:2017jhb}, projectors based on  differential operators \cite{Davydychev:1995nq}, as well as dimensional shift identities together with IBPs in the parametric representation \cite{Tarasov:1996br,Anastasiou:1999bn,ReFiorentin:2015kri}. Projectors have also been developed for multi-loop on-shell amplitudes in the 't Hooft-Veltmann scheme in refs.\ \cite{Peraro:2019cjj,Peraro:2020sfm} and were recently employed in \cite{ Gehrmann:2022vuk,Gehrmann:2023zpz}.

Particularly tough tensor integrals were encountered in an automated approach to compute the UV counterterms for individual Feynman diagrams via the $R^*$-method \cite{Chetyrkin:1982nn,Chetyrkin:1984xa,Smirnov:1986me,Chetyrkin:2017ppe} as implemented in the approach in ref.~\cite{Herzog:2017bjx}. For a UV divergent Feynman diagram of degree $\omega$ this method requires Taylor expanding the integral to order $\omega$ in the external momenta - to extract the dependence on the external momenta. In this way only vacuum tensor integrals require reduction. However, the price of this procedure is that the rank of the integrals produced can grow quickly. Indeed, in 5-loop calculations \cite{Herzog:2017bjx,Herzog:2017dtz,Herzog:2018kwj} such as the 5-loop beta function \cite{Herzog:2017ohr}, tensors with rank up to around 14 were encountered. The number of possible structures that could appear is large, of order $10^5$, and the application of standard methods is not practical.

To overcome this problem, a new idea was proposed in Appendix A of ref.\ \cite{Herzog:2017ohr} and summarised also in ref.\ \cite{Ruijl:2018poj}. The basic idea is to use the permutation symmetry of a given product of metric tensors to write down a simple ansatz for its dual/projector, in terms of an \textit{orbit partition formula}. This turns out to radically simplify the system of equations one has to solve. In fact, the system required to solve the general projector at rank 20 is reduced from $10^8$ to only $42$ equations.

This approach is in fact not limited to vacuum integrals, but can be employed also for general loop integrals that depend on an arbitrary number of external momenta \cite{Anastasiou:2023koq}. This can be achieved by performing a transverse decomposition, see e.g. refs. \cite{Mastrolia:2016czu,Abreu:2017xsl}, of the loop momenta into components parallel and transverse to the external momenta. Then only tensor integrals of transverse loop momenta remain, which can be reduced using the transverse versions of the vacuum projectors. This strategy was first employed in ref.\ \cite{Anastasiou:2023koq} and, furthermore, a single closed-form idempotent projector, which organises the combinatorics of the general tensor reduction in terms of Wick contractions, was derived. The usefulness of this idempotent projector was also demonstrated in a 3-loop $2\to 2$ amplitude calculation.

In comparison to some of the other tensor reduction approaches mentioned above, this approach, which we shall refer to as the orbit partition approach, can be very useful in generic situations when not much is known about tensors under consideration. Beyond its application to multi-loop $R^*$ calculations it should be particularly useful for  asymptotic expansions \cite{Smirnov:1990rz,Smn94,BnkSmn97,SmnRkmt99} in momentum space on a diagram-by-diagram basis. In particular there now exists a new subgraph approach for developing asymptotic expansions around Minkowskian limits in momentum space \cite{Gardi:2022khw,Ma:2023hrt,Herzog:2023sgb,Guan:2024hlf}. Given the increasing complexity of modern-day amplitude and cross-section calculations, asymptotic expansions are becoming a more and more important method for obtaining theoretical predictions, since they allow one to circumvent a number of the bottlenecks, e.g. complicated master integrals and and their IBP-reduction, present in these calculations; see, e.g., refs.\ \cite{Anastasiou:2013mca,Anastasiou:2015yha,Davies:2023npk,Davies:2018ood}. However, at higher orders the Taylor expansion operator will lead to high-rank tensor integrals for which our approach can be employed.

The purpose of this paper is threefold. First, we will elucidate the group theory, and its realisation in terms of graphs, that governs the orbit partition formula and use it to develop a method to construct projectors for vacuum tensors with up to $N$ Lorentz and up to 4 spinor indices in $D$ dimensions in conventional dimensional regularization. Second, we will apply this method and construct projectors with up to $32$ Lorentz indices. We also construct projectors with two spinor indices and up to 15 Lorentz indices, and for four spinor indices with up to 7 Lorentz indices. Finally, we discuss the algorithmic implementation of the projectors and how to optimise their use for particular problems. The implementation of these projectors in an efficient \FORM{} \cite{Ruijl:2017dtg,Tentyukov:2007mu,Vermaseren:2000nd} procedure will be the subject of a separate work \cite{opiter}.

%

We begin \cref{sec:noSpin} with a motivational example illustrating the basic idea of the method. We then discuss the general case of vacuum tensors with $N$ Lorentz indices. We use chord diagrams to represent products of metric tensors, from which we can straightforwardly determine the structure of the orbit partition formula. We find that the number of terms in the orbit partition formula equals the number of integer partitions of half the tensor rank. We discuss the solution of the projector and provide a more efficient representation in terms of symmetric tensors. We also discuss how to optimise the algorithm to take advantage of integrand symmetries.

We then move on to tensors with spinor indices as well as Lorentz indices. A convenient basis for both integrand and projectors is the antisymmetric basis, since it eliminates Clifford algebra relations and fulfills an orthogonality condition; see, e.g., ref.\ \cite{Kennedy:1981kp}. In \cref{sec:1fermion line} we introduce tensors with $N$ Lorentz indices and two spinor indices, which we refer to as \emph{one fermion line}. We find a simple extension to the graphical method which yields the relevant orbits and allows the one fermion line projectors to be constructed. We also provide a new highly efficient relation to rewrite products of gamma matrices in the antisymmetric basis. In \cref{sec:2fermion lines} we extend the method to two fermion lines. A complication, which arises for tensors with four or more spinor indices, is that the general basis is infinite-dimensional. However, thanks to orthogonality, we show that each integral only requires a finite basis in our approach. Nevertheless, the construction of two fermion line projectors is more involved than the other cases.

\Cref{sec:external} outlines the extension of the orbit partition method beyond the vacuum case, to include integrals with external momenta.
As an example we discuss one external momentum in detail. Finally, \cref{sec:testing} compares two sample implementations of the method and gives indicative computation times for the reduction of a variety of example integrals with spinor indices. We give our conclusions and discuss our outlook in \cref{sec:conclusions}.

\section{$N$ Lorentz indices}
\label{sec:noSpin}

\subsection{Motivating example}
\label{sec:motivating-example}

The perhaps most standard way of reducing a tensor Feynman integral to scalar Feynman integrals is due to Passarino and Veltmann in ref.\ \cite{Passarino:1978jh}. Let us consider the example of a general Lorentz invariant integral with 4 Lorentz indices $I^{\mu\nu\rho\sigma}$. For now we will only consider the vacuum case, where the integral does not depend on external momenta. We can express this integral as a linear combination of Lorenz invariant tensors composed only of metric tensors $g^{\mu\nu}$. The result of this reduction is the following:
\begin{align}\label{eq:Fourexample}
I^{\mu\nu\rho\sigma}= g^{\mu\nu}g^{\rho\sigma}\,A_1+g^{\mu\rho}g^{\nu\sigma}\,A_2+g^{\mu\sigma}g^{\nu\rho}\,A_3,
\end{align}
where the $A_i$ are scalar integrals. Using a more condensed notation, we write
\begin{equation}\label{eq:intro-ansatz}
  I= \sum_{i} t_i \,A_i
\end{equation}
where $t_i$ are the independent Lorentz invariant tensor structures and the indices have been dropped. In the standard approach, we find the $A_i$ by contracting $I^{\mu\nu\rho\sigma}$ with each of the structures in the expansion.
The result of this can be represented in matrix form,

The $A_i$ are obtained by inverting the matrix $M$:
\begin{align}
  \begin{bmatrix}
    A_1\\A_2\\A_3
  \end{bmatrix}& =
  \frac{1}{D(D + 2) (D - 1)}
  \begin{bmatrix}
    D+1 & -1 &-1\\
    -1 & D+1 & -1\\
    -1 & -1 & D+1
  \end{bmatrix}
  \begin{bmatrix}I\indices{^\mu_\mu^\rho_\rho}\\I\indices{^\mu_\rho_\mu^\rho}\\I\indices{^\mu_\rho^\rho_\mu}\end{bmatrix}.
\end{align}
As we will show in \cref{sec:nospingeneralcase}, the size of the basis grows factorially with the rank of the tensor. 
With 10 external Lorentz indices we are already faced with inverting a $945\times 945$ dense matrix, which is computationally rather challenging. In the following we will demonstrate a method which drastically reduces the size of the matrix we need to invert.

As an alternative approach we wish to define projectors $P_i$ that are orthogonal to the $t_i$ in the sum in \cref{eq:intro-ansatz}, such that they satisfy
\begin{equation}
P_i\cdot t_j = \delta_{ij}, \qquad P_i\cdot I = A_i\,,
\end{equation}
where the ``$\cdot$'' represents the contraction of all indices,
$A\cdot B=A^{\mu_1\mu_2\dots\mu_N}B_{\mu_1\mu_2\dots\mu_N}$. The projector is  expressible as a linear combination in the same basis:
\begin{equation}
  P_i = \sum_j t_j B_j.
\end{equation}
Let us focus on the projector $P_1$, for the $g^{\mu\nu}g^{\rho\sigma}$ term of the tensor in eq.\ \ref{eq:Fourexample}.
We can improve the ansatz for the projector by demanding that it should have the same symmetry properties under exchange of Lorentz indices as $g^{\mu\nu}g^{\rho\sigma}$. Enforcing this condition requires $B_2=B_3$ so we are left with
\begin{equation}\label{eq:intro-proj-ansatz}
P_1^{\mu\nu\rho\sigma}= g^{\mu\nu}g^{\rho\sigma}\,B_1+(g^{\mu\rho}g^{\nu\sigma}+g^{\mu\sigma}g^{\nu\rho})\,B_2.
\end{equation}

The above is a simple example of what we will refer to as an \emph{orbit partition formula}, which rewrites the projector ansatz as an explicit sum over groupings of tensors with the desired overall symmetry property. In this example it was easy to find the appropriate way to group the tensors, although the need will quickly arise for a general algorithmic approach.

To determine the unknown coefficients in \cref{eq:intro-proj-ansatz} we set up the following system of equations:
\begin{align}
  \begin{bmatrix}
    P_1\cdot t_1\\P_1\cdot t_2
  \end{bmatrix} =
  \begin{bmatrix}
    1\\0
  \end{bmatrix} =
  \underbrace{\begin{bmatrix}
    D^2 & 2D \\ D & D(D+1)
  \end{bmatrix}}_{=\widetilde{M}}
  \begin{bmatrix}B_1\\ B_2\end{bmatrix}\,.
\end{align}
Now, rather than inverting a $3\times 3$ matrix, the size of the matrix to be inverted is only $2\times 2$.
This may not seem like a great simplification, but the same method does lead to far more impressive simplifications when applied to problems involving more Lorentz indices.
To take an extreme example, in an integral with 20 external Lorentz indices, the relevant matrix is reduced in size from $\sim 10^9\times 10^9$ to just $42\times 42$.

We note that the other projectors $P_2$ and $P_3$ are just index permutations of $P_1$.
Therefore, we have effectively reduced the problem from inverting the $3\times3$ matrix $M$, to inverting the $2\times2$ matrix $\widetilde{M}$.
We express the unknown coefficients $A_i$ in terms of the projectors as follows:
\begin{align}
A_1&=P_1\cdot I=B_1 I\indices{^\mu_\mu^\rho_\rho}+B_2 (I\indices{^\mu_\rho_\mu^\rho} +I\indices{^\mu_\rho^\rho_\mu})\nn\\
A_2&=P_2\cdot I=B_1 I\indices{^\mu_\rho_\mu^\rho}+B_2 (I\indices{^\mu_\mu^\rho_\rho} +I\indices{^\mu_\rho^\rho_\mu})\nn\\
A_3&=P_3\cdot I=B_1 I\indices{^\mu_\rho^\rho_\mu}+B_2 (I\indices{^\mu_\rho_\mu^\rho} +I\indices{^\mu_\mu^\rho_\rho}).\nn
\end{align}
The repeated coefficients in $M^{-1}$ are then naturally identified in terms of the independent coefficients $B_i$ as follows:
\begin{align}
 A_i=M^{-1}_{ij}\,(t_j\cdot I)\,,\qquad  M^{-1} =
 \begin{bmatrix}
    B_1 & B_2 & B_2\\
    B_2 & B_1 & B_2\\
    B_2 & B_2 & B_1
  \end{bmatrix}.
\end{align}

\paragraph{Example.} We will now work through a full example in both methods to illustrate how the projectors may be applied in a simple integral. We will examine the 2-loop vacuum example
\begin{equation}
  I^{\mu\nu\rho\sigma} = \int d^Dp_1d^Dp_2\, \,\frac{p_1^{\mu}p_1^{\nu}p_2^{\rho}p_2^{\sigma}}{\mathcal{N}(p_1,p_2,\dots)}\,,
\end{equation}
where ${\mathcal{N}(p_1,p_2;\dots)}$ is a polynomial in the loop momenta and any further mass scales in the problem. The ansatz for the decomposition is then given by
\begin{equation}
  I^{\mu\nu\rho\sigma}= g^{\mu\nu}g^{\rho\sigma}\,A_1+g^{\mu\rho}g^{\nu\sigma}\,A_2+g^{\mu\sigma}g^{\nu\rho}\,A_3\,.
\end{equation}
In the standard approach we contact $ I^{\mu\nu\rho\sigma}$ with each of the tensor structures and set up the following set of equations
\begin{align}
  \renewcommand*{\arraystretch}{1.5}
  \begin{bmatrix}\int d^Dp_1d^Dp_2\,\frac{p_1^2 p_2^2}{\mathcal{N}(p_1,p_2,\dots)}\\\int d^Dp_1d^Dp_2\,\frac{(p_1\cdot p_2)^2 }{\mathcal{N}(p_1,p_2,\dots)}\\\int d^Dp_1d^Dp_2\,\frac{(p_1\cdot p_2)^2 }{\mathcal{N}(p_1,p_2,\dots)}
  \end{bmatrix}& =
  \begin{bmatrix}
    D^2 & D &D\\
    D & D^2 & D\\
    D & D & D^2
  \end{bmatrix}
  \begin{bmatrix}
    A_1\\A_2\\A_3
  \end{bmatrix}.
\end{align}
Solving this we have 
\begin{align}
 A_1&=  \int d^Dp_1d^Dp_2\,\frac{(D+1)p_1^2 p_2^2 -2(p_1\cdot p_2)^2}{D(D + 2) (D - 1)\,\mathcal{N}(p_1,p_2,\dots)}\,, \\
 A_2&=  \int d^Dp_1d^Dp_2\,\frac{-p_1^2 p_2^2 +D(p_1\cdot p_2)^2}{D(D + 2) (D - 1)\,\mathcal{N}(p_1,p_2,\dots)}\,, \\
 A_3&=  \int d^Dp_1d^Dp_2\,\frac{-p_1^2 p_2^2 +D(p_1\cdot p_2)^2}{D(D + 2) (D - 1)\,\mathcal{N}(p_1,p_2,\dots)} \,.
\end{align}
Alternatively we may use the projector defined in \cref{eq:intro-proj-ansatz} which has the explicit form 
\begin{equation}
  P_1^{\mu\nu\rho\sigma}= g^{\mu\nu}g^{\rho\sigma}\,\frac{D+1}{D(D + 2) (D - 1)}+(g^{\mu\rho}g^{\nu\sigma}+g^{\mu\sigma}g^{\nu\rho})\,\frac{-1}{D(D + 2) (D - 1)}\,.
\end{equation} 
Using the orthogonality of the projector with the different tensor structures we may rewrite the integral as
\begin{equation}
  I^{\mu\nu\rho\sigma}= g^{\mu\nu}g^{\rho\sigma}\,\underbrace{P_1^{\mu'\nu'\rho'\sigma'}I_{\mu'\nu'\rho'\sigma'}}_{=A_1}+g^{\mu\rho}g^{\nu\sigma}\,\underbrace{P_1^{\mu'\rho'\nu'\sigma'}I_{\mu'\nu'\rho'\sigma'}}_{=A_2}+g^{\mu\sigma}g^{\nu\rho}\,\underbrace{P_1^{\mu'\sigma'\nu'\rho'}I_{\mu'\nu'\rho'\sigma'}}_{=A_3}\,.
\end{equation}
The contractions match exactly the scalar integrals.
%

%


\subsection{General case}
\label{sec:nospingeneralcase}
Following on from the introductory example, the remainder of this section is devoted to the tensor reduction of a general $L$-loop vacuum Feynman integral $I^{\mu_1\dots\mu_N}$ containing $N$ (even) open Lorentz indices. The most general ansatz for its reduction can be written as follows:
\begin{equation}
  I^{\mu_1\dots\mu_N}=\sum_{\sigma\in S_{2}^{N}} \,g^{\mu_{\sigma(1)}\mu_{\sigma(2)}} \dots g^{\mu_{\sigma(N-1)}\mu_{\sigma(N)}} I_{\sigma(1)\dots\sigma(N)} \,,
\end{equation}
or in shorthand notation:
\begin{equation}\label{eq:nospin-reduction}
  I=\sum_{\sigma\in S_{2}^{N}} g(\sigma)I(\sigma)\,.
\end{equation}
Here we denote by $S_{2}^{N}$ a set of permutations which generates all independent products of metric tensors $g$.
We obtain this set by considering the stabiliser subgroup, $H\subset S_N$, of a generic basis element.
This is given by the product group $H=(S_2)^{N/2}\times S_{N/2}$, where the $S_2$s are due to $g$ being a symmetric tensor, while the $S_{N/2}$ is due to the invariance under interchanging pairs of indices between $g$s. We partition the full $S_N$ group into left cosets given by the quotient $S_N/H$, in which each coset corresponds to the set of all permutations that map to an identical product of $g$s. The set $S_{2}^{N}$ is thus obtained by selecting one element of each coset in $S_N/H$. While the set $S_{2}^{N}$ is not unique it can be specified via some canonical ordering.

The number of cosets in $S_N/H$ is thus also the number of elements in the set $S_{2}^{N}$, and is given by
\begin{equation}
 |S_{2}^{N}|=\frac{|S_N|}{|H|}=\frac{N!}{2^{N/2}(N/2)!}=(N-1)!!\,.
\end{equation}

\begin{table}[h]
\begin{center}
\begin{tabular}{ c c c c c c c c c c c }
\Xhline{3\arrayrulewidth}
 $N$ & 2  & 4 & 6 & 8 & 10 & 12 & 14 & 16 & 18 & 20 \\
 \hline
$|S_{2}^{N}|$ & 1 & 3 & 15 & 105 & 945 & 10,395 & 135,135 & 2,027,025 & 34,459,425 & 654,729,075  \\
\Xhline{3\arrayrulewidth}
\end{tabular}
\end{center}
\caption{The table gives the number, $|S_{2}^{N}|$, of independent tensor structures which can be obtained by permuting a product of $g^{\mu\nu}$s, of total tensor rank $N$, in all possible ways.}
\label{tab:S2nsize}
\end{table}
As can be seen from table \ref{tab:S2nsize}, the number of independent tensor structures increases very rapidly. For large $N$ this makes the standard approach, described in section \ref{sec:motivating-example}, intractable. It is therefore desirable to construct projectors for the coefficients $I_{\sigma(1)\dots \sigma(N)}$ in a general combinatorial way. We define these projectors
to be orthonormal to the products of metric tensors,
\begin{equation}
  \label{eq:orthonormality}
  P(\sigma)\cdot g(\sigma')
  =\delta_{\sigma\sigma'}\,,
\end{equation}
where the dot represents index contraction. A major advantage of this approach is also that the projectors are related by permutations, i.e.
\begin{equation}
 P(\sigma)=P^{\mu_{\sigma(1)}\dots\mu_{\sigma(N)}}\,,
\end{equation}
where $P^{\mu_{1}\dots\mu_{N}}$ is the projector for $g^{\mu_1\mu_2}\dots g^{\mu_{N-1}\mu_N}$.
This feature is discussed in more detail in appendix \ref{sec:method}.
The projectors therefore also define a dual basis to the products of metric tensors. It follows that:
\begin{equation}
  I(\sigma)=P(\sigma)\cdot I\,.
\end{equation}
The coefficients $I(\sigma)$ can thus be extracted simply from the knowledge of the $P(\sigma)$. In the following we systematically construct expressions for these projectors. Now, by general covariance arguments, we can write that
\begin{equation}
\label{eq:proj}
P(\sigma)=\sum_{\sigma'} \,c(\sigma,\sigma')\, g(\sigma'),
\end{equation}
where the $c(\sigma,\sigma')$-coefficients can depend only on the dimension of space-time $D$. However, we can constrain this expression further by demanding that $P(\sigma)$ is invariant under elements of the stabiliser group $H(\sigma)$, which leaves $g(\sigma)$ unchanged. That is to say, we construct the stabiliser of a particular element $g(\sigma)$,
\begin{equation}
\label{eq:stablizer}
  H(\sigma) = \{\tau \ \vert \ g(\tau\circ\sigma) = g(\sigma) \},
\end{equation}
and demand that the projector satisfy
\begin{equation}
\label{eq:Psymmetry}
P(h\circ\sigma) = P(\sigma) \quad \text{for any}\quad  h\in H(\sigma)\,.
\end{equation}
The elements of $H(\sigma)$ are not necessarily symmetries of the other basis tensors. However, the space of tensors must be closed under the action of $H(\sigma)$, naturally leading to the notion of orbits. Two tensors $g(\sigma_1)$ and $g(\sigma_2)$ are then defined to be in the same orbit if there exists some element $h\in H(\sigma)$ such that $g(\sigma_1)=g(h\circ \sigma_2)$, i.e.\ they are related by a symmetry of $g(\sigma)$. Labelling each orbit by a particular integer $k$ we then define $C_k(\sigma)$ as a set of permutations which generates each tensor in the orbit exactly once.

As a consequence of \cref{eq:Psymmetry} we see that any two tensors in $P(\sigma)$ which are in the same orbit must have the same coefficient. This allows us to simplify \cref{eq:proj} as follows:
\begin{equation}
\label{eq:orbitpartitionformula}
  P(\sigma) = \sum_k c_k \sum_{\sigma'\in C_k(\sigma)} g(\sigma')\,,
\end{equation}
where the sum on $k$ runs over the different orbits. We refer to eq.\ \eqref{eq:orbitpartitionformula} as the \emph{orbit partition formula}, given that the orbits partition the set of basis tensors. We note the appearance in the above expression of the quantity
\begin{equation}\label{eq:invariant-sum-nospin}
  T_k(\sigma)=\sum_{\sigma' \in C_{k}(\sigma)} g(\sigma')\,,
\end{equation}
which we refer to as the \emph{invariant sum} over the orbit $k$. It is easy to check that this quantity is invariant under the action of $H(\sigma)$ so, by building the projector from a linear combination  of these invariant sums, we ensure the correct overall symmetry property; this is demonstrated explicitly in \cref{sec:method}.

\subsection{Enumeration of partitions}
\label{sec:puremetricpartitions}
Let us fix $\sigma$ to be the identity permutation, $\sigma=e$, and investigate the structure of the orbits under the action of $H(e)$.
The base element, i.e.\ the tensor we are trying to project out, is
\begin{equation}
g(e)=g^{\mu_1\mu_2}\dots g^{\mu_{N-1}\mu_N}.
\end{equation}
To understand the structure of the orbits we map the tensors $g(\sigma)$ to graphs as follows:
\begin{itemize}
 \item each index is mapped to a vertex,
 \begin{equation*}
 \begin{axopicture}{(100,15)(-150,0)}
\Text(-120, 0)[rc]{$\mu_1,\mu_2,\dots \quad \to$}
\Vertex(-100,0){1.5}
\Text(-103, 10){1}
\Vertex(-70,0){1.5}
\Text(-67, 10){2}
\Text(-40, 0){$\dots$}
\end{axopicture}
 \end{equation*}

 \item the indices which are paired in metric tensors of the base element $g(e)$ are connected by dashed (pink) lines,
  \begin{equation*}
 \begin{axopicture}{(100,20)(-150,0)}
\Text(-120, 0)[rc]{base term  $=g(e)=g^{\mu_1\mu_2} \quad \to$}
\pinkline{-100}{0}{-70}{0}
\Text(-103, 10){1}
\Text(-67, 10){2}
\Text(-40, 0){$\dots$}
\end{axopicture}
 \end{equation*}
 \item the indices which are paired in metric tensors of $g(\sigma)$ are connected by solid black lines.
   \begin{equation*}
 \begin{axopicture}{(100,50)(-150,-25)}
\Text(-120, 0)[rc]{$g(\sigma) = g^{\mu_1\mu_3}g^{\mu_2\mu_4}\dots \quad \to$}
\pinkline{-100}{15}{-70}{15}
\pinkline{-100}{-15}{-70}{-15}
\Text(-103, 25){1}
\Text(-67, 25){2}
\Text(-103,-25){3}
\Text( -67,-25){4}
\Line[width=1](-100,15)(-100,-15)
\Line[width=1](-70,-15)(-70,15)
\Text(-20, 0){$\dots$}
\end{axopicture}
 \end{equation*}
\end{itemize}

\begin{table}[h]
\centering
\begin{tabular}{ c c c }
\Xhline{3\arrayrulewidth}
\quad  \begin{minipage}{0.2\textwidth}\centering$\lambda$\end{minipage} & \begin{minipage}{0.2\textwidth}\centering Tensor\end{minipage}  &  \begin{minipage}{0.3\textwidth}\centering Graph\end{minipage} \\
 \hline
  (1,1,1)  &   $g^{\mu_1\mu_2}g^{\mu_3\mu_4}g^{\mu_5\mu_6}$  &\multicolumn{1}{m{110pt}}{
\begin{axopicture}{(90,90)(-45,-45)}
\Arc[width=1](0,33.33)(16.66,193,347)
\Arc[width=1](-28.83,-16.650)(16.66,-42,110)
\Arc[width=1](28.83,-16.650)(16.66,70,222)
\pinkline{16.666}{28.867}{-16.666}{28.867}
\pinkline{-33.33}{0.0}{-16.666}{-28.867}
\pinkline{16.666}{-28.867}{33.33}{0.0}
\Text(21.666, 37.528){1}
\Text(-21.666, 37.527){2}
\Text(-43.333, -0.0){3}
\Text(-21.670, -37.526){4}
\Text(21.671, -37.525){5}
\Text(43.333, 0.){6}
\end{axopicture}}
\\
 \hline
  (1,2)  &   $g^{\mu_1\mu_2}g^{\mu_3\mu_6}g^{\mu_4\mu_5}$  &
\multicolumn{1}{m{110pt}}{
\begin{axopicture}{(90,90)(-45,-45)}
\pinkline{16.666}{28.867}{-16.666}{28.867}
\pinkline{-33.33}{0.0}{-16.666}{-28.867}
\pinkline{16.666}{-28.867}{33.33}{0.0}
\Arc[width=1](0,33.33)(16.66,193,347)
\Line[width=1](-16.666,-28.867)(16.666,-28.867)
\Line[width=1](33.33,0.0)(-33.33,0.0)
\Text(21.666, 37.528){1}
\Text(-21.666, 37.527){2}
\Text(-43.333, -0.0){3}
\Text(-21.670, -37.526){4}
\Text(21.671, -37.525){5}
\Text(43.333, 0.){6}
\end{axopicture}}
\\
 \hline
  (3)  &   $g^{\mu_1\mu_6}g^{\mu_2\mu_3}g^{\mu_4\mu_5}$  & 
\multicolumn{1}{m{110pt}}{
\begin{axopicture}{(90,90)(-45,-45)}
\pinkline{16.666}{28.867}{-16.666}{28.867}
\pinkline{-33.33}{0.0}{-16.666}{-28.867}
\pinkline{16.666}{-28.867}{33.33}{0.0}
\Line[width=1](16.666,28.867)(33.33,0.0)
\Line[width=1](-16.666,28.867)(-33.33,0.0)
\Line[width=1](-16.666,-28.867)(16.666,-28.867)
\Text(21.666, 37.528){1}
\Text(-21.666, 37.527){2}
\Text(-43.333, -0.0){3}
\Text(-21.670, -37.526){4}
\Text(21.671, -37.525){5}
\Text(43.333, 0.){6}
\end{axopicture}}
\\
\Xhline{3\arrayrulewidth}
 \end{tabular}
\caption{The table lists examples of the three orbits for $N=6$. The left column lists integer partitions, $\lambda\vdash 3$, corresponding to the cycle structure of the graph.}
\label{tab:Pgraph}
\end{table}

Some graphs for the particular case $N=6$ are shown in table \ref{tab:Pgraph}. Although the other tensors in this example will generate more graphs, they will always be isomorphic to one of these three, in the sense that they differ only by a relabelling of the vertices. The result is a graphical representation of the orbits of \cref{eq:orbitpartitionformula}.
This follows immediately from the way the graphs were constructed, since the permutations belonging to $H(e)$ can only swap indices connected by dashed lines or pairs of dashed-line-connected indices.
Neither of these transformations change the overall structure (or \emph{topology} in physicist's jargon) of the graph.
We conclude that in the present example the tensors are partitioned into three orbits, represented by the graphs in \cref{tab:Pgraph}.
Note that the map from tensors to graphs is bijective and thus any conclusions drawn from the graphs apply also to the corresponding tensors.

For a general value of $N$ the graphs will consist of one or more disjoint  cycles, i.e. closed loops in the graph,   formed by joining up the dashed and solid lines in all possible ways. It is fairly easy to see that the structure of the graph is entirely specified by its cycles which we label by $i=1\dots l$ and their corresponding number of solid lines $\lambda_i$. Since the number of solid lines (or dashed lines) that can be drawn in the diagram is fixed to $N/2$ we arrive at the conclusion that the graph structure can be written as an integer partition $\lambda$ of $N/2$, which we write as $\lambda\vdash N/2$, as follows:
\begin{equation}
\lambda=(\lambda_1,\lambda_2,\dots,\lambda_{l}):\quad \sum_{i=1}^l \lambda_i =\frac{N}{2}\,.
\end{equation}
In \cref{tab:Pgraph} these integer partitions are listed in the left column.
It then follows that the number of distinct graphs that can be drawn, and consequently the number of independent coefficients entering the projector in eq.\ \eqref{eq:orbitpartitionformula}, is $p(N/2)$, the number of integer partitions of $N/2$. We plot this number in table \ref{table:nospin-system-size}, along with the number of basis tensors. This makes it clear that exploiting the symmetry drastically reduces the number of unknown coefficients in $P(\sigma)$.

\begin{table}[h]
\begin{tabular}{ c c c c c c c c c c c }
\Xhline{3\arrayrulewidth}
 $N$ & 2  & 4 & 6 & 8 & 10 & 12 & 14 & 16 & 18 & 20 \\
 \hline
$|S_{2}^{N}|$ & 1 & 3 & 15 & 105 & 945 & 10,395 & 135,135 & 2,027,025 & 34,459,425 & 654,729,075  \\
$p\big(\frac{N}{2}\big)$ & 1 & 2 & 3  & 5   & 7   & 11 & 15 & 22 & 30 & 42 \\
\Xhline{3\arrayrulewidth}
\end{tabular}
\caption{Table comparing the number of independent tensors, $|S_{2}^{N}|$, with the number of independent coefficients in the projector, $p(N/2)$, for $N$ Lorentz indices.}
\label{table:nospin-system-size}
\end{table}

\subsection{General structure of the projector}
\label{subsec:tensorstructure}
The general form of the projector can then be written as a sum over integer partitions, $\lambda\vdash N/2$, as follows
\begin{equation}
\label{eq:orbitpartition}  P(\sigma) = \sum_{\lambda\vdash N/2} c_\lambda T_\lambda(\sigma)\,,\qquad \sigma\in S^{N}_2\,.
\end{equation}
We will now give a concise expression for the orbit invariant sum $T_\lambda(\sigma)$. To obtain it we simply need to sum over all permutations corresponding to graphs with the particular cycle structure $\lambda$.

\begin{figure}
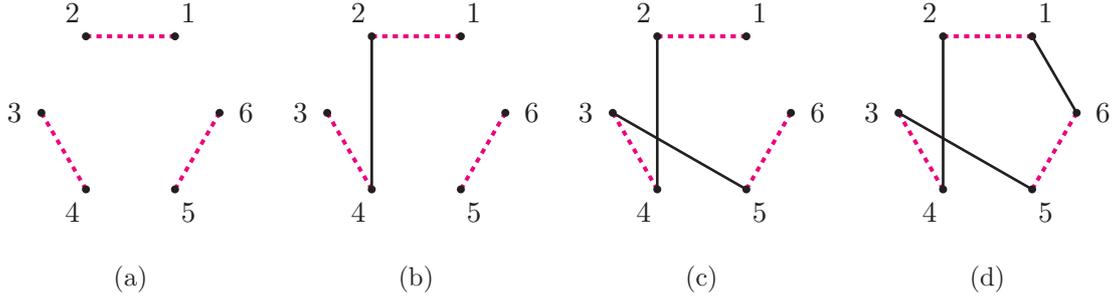

 \begin{subfigure}[t]{.24\textwidth}
  \centering
\begin{axopicture}{(100,100)(-50,-50)}
\pinkline{16.666}{28.867}{-16.666}{28.867}
\pinkline{-33.33}{0.0}{-16.666}{-28.867}
\pinkline{16.666}{-28.867}{33.33}{0.0}
\Text(21.666, 37.528){1}
\Text(-21.666, 37.527){2}
\Text(-43.333, -0.0){3}
\Text(-21.670, -37.526){4}
\Text(21.671, -37.525){5}
\Text(43.333, 0.){6}
\end{axopicture}
\caption{}
 \end{subfigure}
 \begin{subfigure}[t]{.24\textwidth}
\centering
\begin{axopicture}{(100,100)(-50,-50)}
\pinkline{16.666}{28.867}{-16.666}{28.867}
\pinkline{-33.33}{0.0}{-16.666}{-28.867}
\pinkline{16.666}{-28.867}{33.33}{0.0}
\Text(21.666, 37.528){1}
\Text(-21.666, 37.527){2}
\Text(-43.333, -0.0){3}
\Text(-21.670, -37.526){4}
\Text(21.671, -37.525){5}
\Text(43.333, 0.){6}
\Line[width=1](-16.666,28.867)(-16.666,-28.867)
\end{axopicture}
\caption{}
 \end{subfigure}
 \begin{subfigure}[t]{.24\textwidth}
  \centering
  \begin{axopicture}{(100,100)(-50,-50)}
\pinkline{16.666}{28.867}{-16.666}{28.867}
\pinkline{-33.33}{0.0}{-16.666}{-28.867}
\pinkline{16.666}{-28.867}{33.33}{0.0}
\Text(21.666, 37.528){1}
\Text(-21.666, 37.527){2}
\Text(-43.333, -0.0){3}
\Text(-21.670, -37.526){4}
\Text(21.671, -37.525){5}
\Text(43.333, 0.){6}
\Line[width=1](-16.666,28.867)(-16.666,-28.867)
\Line[width=1](-33.33,0.0)(16.666,-28.867)
\end{axopicture}
\caption{}
 \end{subfigure}
 \begin{subfigure}[t]{.24\textwidth}
\centering
\begin{axopicture}{(100,100)(-50,-50)}
\pinkline{16.666}{28.867}{-16.666}{28.867}
\pinkline{-33.33}{0.0}{-16.666}{-28.867}
\pinkline{16.666}{-28.867}{33.33}{0.0}
\Text(21.666, 37.528){1}
\Text(-21.666, 37.527){2}
\Text(-43.333, -0.0){3}
\Text(-21.670, -37.526){4}
\Text(21.671, -37.525){5}
\Text(43.333, 0.){6}
\Line[width=1](-16.666,28.867)(-16.666,-28.867)
\Line[width=1](-33.33,0.0)(16.666,-28.867)
\Line[width=1](33.33,0.0)(16.666,28.867)
\end{axopicture}
\caption{}
 \end{subfigure}
\caption{Graphical construction of a tensor contributing to orbit $C_{(3)}$ for $N=6$. Dashed lines connect indices which can \emph{not} occur in the same metric tensor. Solid lines represent metric tensors.}
\label{fig:evenfactorial}
\end{figure}

For the partition $\lambda=(N/2)$ the invariant sum is given by the set of permutations $C_{(N/2)}(\sigma)$ whose corresponding graphs form a single cycle. This leads to the following criterion. A permutation $\tau$ is in $C_{(N/2)}$ as long as it has no proper subset of $g$s of $g(\tau)$ with indices $\{\tau(i_1)\dots \tau(i_K)\}$ which correspond to the indices of any proper subset of $g$s in $g(\sigma)$. For example, we have for $N=4$:
\begin{equation}
\label{eq:T2}
T_{(2)}^{\mu_1\mu_2\mu_3\mu_4}(e)= g^{\mu_{1}\mu_{3}}g^{\mu_{2}\mu_{4}}+g^{\mu_{1}\mu_{4}}g^{\mu_{2}\mu_{3}}\ ,
\end{equation}
or for $N=6$:
\bea
T_{(3)}^{\mu_1\mu_2\mu_3\mu_4\mu_5\mu_6}(e) &&=
 g^{\mu_{1}\mu_{3}}g^{\mu_{2}\mu_{5}}g^{\mu_4\mu_6}
+g^{\mu_{1}\mu_{3}}g^{\mu_{2}\mu_{6}}g^{\mu_4\mu_5}
+g^{\mu_{1}\mu_{4}}g^{\mu_{2}\mu_{5}}g^{\mu_4\mu_6}\nn\\
&&
+g^{\mu_{1}\mu_{4}}g^{\mu_{2}\mu_{6}}g^{\mu_4\mu_5}
+g^{\mu_{1}\mu_{5}}g^{\mu_{2}\mu_{3}}g^{\mu_4\mu_6}
+g^{\mu_{1}\mu_{5}}g^{\mu_{2}\mu_{4}}g^{\mu_3\mu_6}\nn\\
&&
\label{eq:T3}
+g^{\mu_{1}\mu_{6}}g^{\mu_{2}\mu_{3}}g^{\mu_4\mu_5}
+g^{\mu_{1}\mu_{6}}g^{\mu_{2}\mu_{4}}g^{\mu_3\mu_5}\ .
\eea
It turns out that the number of elements of $C_{(N/2)}$ is given by the even double factorial:
\begin{equation}
\label{eq:CNm2}
|C_{(N/2)}|=(N-2)!!\,,
\end{equation}
which indeed agrees with the number of tensors produced in eqs. (\ref{eq:T2}) and (\ref{eq:T3}), i.e.
$|C_{(2)}|=2$ for $N=4$ and $|C_{(3)}|=4\cdot2=8$ for $N=6$. Eq. (\ref{eq:CNm2}) is best proven graphically. One starts by drawing $N$ vertices in a plane to represent the Lorentz indices. Next we draw dashed lines between all the pairs of indices ${12,34,56,..}$ as shown in figure \ref{fig:evenfactorial} (a), these are the pairs which can not occur in $C_{(N/2)}$. The set $C_{(N/2)}$ now corresponds to all graphs which connect all vertices such that it makes a single cycle. To count the number of such graphs, we first select a random vertex say $2$ and connect it to another vertex which is not its \textit{dashed} partner vertex 1. This gives $(N-2)$ possibilities; see \ref{fig:evenfactorial} (b). Next we take the \textit{dashed partner} of the new vertex and connect to another vertex which is also not vertex $1$, this gives another $(N-2-2)$ possibilities; see the figure \ref{fig:evenfactorial} (c). We then continue this procedure until only vertex $1$ remains, at which point we close the cycle. In this way the total number of graphs is clearly $(N-2)(N-2-2)...(N-(N-2))=(N-2)!!$.

Having defined the invariant sum of the largest orbit, $\lambda=(N/2)$, we now write down a recursive expression for any other partition $\lambda$ in terms of smaller invariant sums. For $\lambda=(\lambda_1,\lambda_2,\dots,\lambda_l)$, a partition of $N/2$, we have in particular that
\begin{equation}
\label{eq:pairshuffle}
T_{(\lambda_1,\dots,\lambda_l)}(\sigma)=\sum_{\sigma_1,\dots,\sigma_l\in \Sigma_\lambda(\sigma)} T_{(\lambda_1)} (\sigma_{1}) T_{(\lambda_2)} (\sigma_2)\cdots T_{(\lambda_l)}(\sigma_l)\,,
\end{equation}
where the sum over the set $\Sigma_{\lambda}(\sigma)$ goes over all inequivalent (in the sense that the resulting tensors would be the same) ways of distributing the pairs of indices in the $g$s of $g(\sigma)$ over the invariant sums $ T_{(\lambda_k)}(\sigma_k)$. For example at $N=4$, writing $ T_{(1,1)}(e)=T_{(1,1)}(1,2,3,4)$, consider
\begin{equation}
 T_{(1,1)}(1,2,3,4)=T_{(1)}(1,2)T_{(1)}(3,4) = g^{\mu_1\mu_2}g^{\mu_3\mu_4}\,,
\end{equation}
where we adopted the notation that the permutations in the arguments
are written as $T_\lambda(\sigma(1),\sigma(2),...\sigma(k))$.
At $N=6$ we have
\begin{equation}
 T_{(1,1,1)}(1,2,3,4,5,6)=T_{(1)}(1,2)T_{(1)}(3,4)T_{(1)}(5,6) = g^{\mu_1\mu_2}g^{\mu_3\mu_4}g^{\mu_5\mu_6}\,,
\end{equation}
and
\begin{align}
 T_{(2,1)}(1,\dots,6)&=T_{(2)}(1,2,3,4)T_{(1)}(5,6)+T_{(2)}(1,2,5,6)T_{(1)}(3,4)+T_{(2)}(3,4,5,6)T_{(1)}(1,2)\nonumber\\
 &=
 (g^{\mu_1\mu_3}g^{\mu_2\mu_4} +g^{\mu_1\mu_4}g^{\mu_2\mu_3})g^{\mu_5\mu_6}
+(g^{\mu_1\mu_5}g^{\mu_2\mu_6} +g^{\mu_1\mu_6}g^{\mu_2\mu_5})g^{\mu_3\mu_4}\nonumber\\
&+(g^{\mu_3\mu_5}g^{\mu_4\mu_6} +g^{\mu_3\mu_6}g^{\mu_4\mu_5})g^{\mu_1\mu_2}\,.
\end{align}
The derivation of \cref{eq:pairshuffle} follows more or less straightforwardly from the graph representation. One simply sums up all possible graphs whose cycle structure corresponds to a given partition $\lambda\vdash N/2$. The total number of tensors, or inequivalent permutations in an orbit, can then be counted
for a general partition
\begin{equation}
\qquad \lambda=(\underbrace{\ell_1,\dots,\ell_1}_{n_1},\underbrace{\ell_2,\dots,\ell_2}_{n_2},\dots,\underbrace{\ell_k,\dots,\ell_k}_{n_k}),
\end{equation}
where $n_i$ is the multiplicity of $\ell_i$ in $\lambda$. From \cref{eq:pairshuffle} we then obtain:
\begin{equation}
|C_{\lambda}|=\frac{(N/2)!}{\lambda_1!\dots\lambda_l!}\frac{(2\lambda_1-2)!!\dots(2\lambda_l-2)!!}{n_1!\dots n_k!},
\end{equation}
which follows by considering that the set of independent terms in $T_{\lambda}$ are generated by permutations in $S_{N/2}/S_{\lambda_1}/\dots/S_{\lambda_l}/S_{n_1}/\dots/S_{n_k}$, where $S_{N/2}$ is the number of exchanges of pairs of indices in $g$s of $g(\sigma)$ among the tensors $T_{(\lambda_i)}$ which each contain $(2\lambda_i-2)!!$ terms by \cref{eq:CNm2}, while $S_{\lambda_i}$ divides out the corresponding pair-exchange symmetry. Finally, the $S_{n_j}$ account for over-counting of identical $T_{(\ell_j)}$s.

\subsection{Solution for the projector} \label{sec:nospinsol}

As discussed above the projector $P(\sigma)$, which projects out a certain $g(\sigma)$, can be written as a sum over invariant sums corresponding to the orbits generated by the stabiliser subgroup $H(\sigma)$. The individual invariant sums are built by summing over all tensors of a particular orbit.
Therefore, the overall form of the projector is the following:
\begin{equation}\label{eq:nospin-projector-ansatz}
  P(\sigma)=\sum_\lambda c_\lambda T_\lambda(\sigma)
\end{equation}
Let us now determine the coefficients $c_\lambda$, which do not depend on $\sigma$. We can achieve this by setting up a system of equations:
\begin{equation}\label{eq:nospin-projector-definingproperty}
  g(\tau)\cdot P(\sigma)=\delta_{\sigma\tau}\,.
\end{equation}
In fact, it suffices to pick one representative from each orbit $C_\lambda(\sigma)$.
In the following we label these orbits by the integer $k\in \{1,\dots,p\}$ with $p=p(\frac{N}{2})$ for brevity.
As a result, the number of equations is equal to the number of unknowns. A corresponding representative of each orbit is written as $g(\sigma_k)$. This leads to the following linear system of equations:
\begin{equation}
  \begin{bmatrix}
    g(\sigma_1)\cdot T_{1}(\sigma) \phantom{A} &  g(\sigma_1)\cdot T_{2}(\sigma) &\cdots &  g(\sigma_1)\cdot T_{p}(\sigma) \\
    g(\sigma_2)\cdot T_{1}(\sigma) \phantom{A}  &  g(\sigma_2)\cdot T_{2}(\sigma) &\cdots &  g(\sigma_2)\cdot T_{p}(\sigma) \\
    \vdots & \vdots & \ddots & \vdots \\
    g(\sigma_p)\cdot T_{1}(\sigma) \phantom{A}  &  g(\sigma_p)\cdot T_{2}(\sigma) &\cdots & g(\sigma_p)\cdot T_{p}(\sigma)
  \end{bmatrix}
  \begin{bmatrix}
    c_{1}\\
    c_{2}\\
    \vdots\\
    c_{p}
  \end{bmatrix}
  =
  \begin{bmatrix}
    1\\
    0\\
    \vdots\\
    0
  \end{bmatrix}
  \,.
\end{equation}
Solutions for the coefficients up to $N=8$ are presented in \cref{tab:coeffs}.

\begin{table}[h]
  \renewcommand*{\arraystretch}{2.5}
  \centering
  \begin{tabular}{ccc}
    \Xhline{3\arrayrulewidth}\vspace{-10ex}\\
    $N$ & Graph & $c_\lambda$ \\\hline
    \fbox{$N=2$} & 
    {\scalebox{0.3}{
      \begin{axopicture}(40,20)
        \Pinkline{0}{0}{40}{0}
                  \Bezier[width=4](0,0)(5,20)(35,20)(40,0) 
      \end{axopicture}}}
    &
      $ c_{1} =\frac{1}{D}$
 \\
    \hline
    \fbox{$N=4$} & {\scalebox{0.3}{
      \begin{axopicture}(40,20)
        \Pinkline{0}{-10}{40}{-10}
        \Pinkline{0}{30}{40}{30}  
        \Line[width=4](0,-10)(0,30)
        \Line[width=4](40,-10)(40,30)        
      \end{axopicture}}}
    & $ c_{2}=\frac{1}{D(D+2)(D-1)},$\\

        &
        {\scalebox{0.3}{
      \begin{axopicture}(110,20)
        \Pinkline{0}{0}{40}{0}
        \Bezier[width=4](0,0)(5,20)(35,20)(40,0) 
        \Pinkline{70}{0}{110}{0}
        \Bezier[width=4](70,0)(75,20)(105,20)(110,0) 
      \end{axopicture}}}&$ c_{11}=\frac{D+1}{D(D+2)(D-1)}$\vspace{2ex}
  \\
      \hline
   \fbox{$N=6$} & 
   {\scalebox{0.3}{
    \begin{axopicture}(66,60)(-30,-20)
      \Pinkline{16.666}{28.867}{-16.666}{28.867}
      \Pinkline{-33.33}{0.0}{-16.666}{-28.867}
      \Pinkline{16.666}{-28.867}{33.33}{0.0}
      \Line[width=4](16.666,28.867)(33.33,0.0)
      \Line[width=4](-16.666,28.867)(-33.33,0.0)
      \Line[width=4](-16.666,-28.867)(16.666,-28.867)
    \end{axopicture}}}
    &   $ c_{3} =  \frac{2}{D(D+2)(D-1)(D+4)(D-2)},$\\
    &
    {\scalebox{0.3}{
      \begin{axopicture}(110,20)
        \Pinkline{0}{-10}{40}{-10}
        \Pinkline{0}{30}{40}{30}  
        \Line[width=4](0,-10)(0,30)
        \Line[width=4](40,-10)(40,30) 
        \Pinkline{70}{0}{110}{0}
        \Bezier[width=4](70,0)(75,20)(105,20)(110,0)        
      \end{axopicture}}}&  $ c_{21} =  -\frac{1}{D(D-1)(D-2)(D+4)},$ \\
    &{\scalebox{0.3}{
      \begin{axopicture}(180,20)
        \Pinkline{0}{0}{40}{0}
        \Bezier[width=4](0,0)(5,20)(35,20)(40,0) 
        \Pinkline{70}{0}{110}{0}
        \Bezier[width=4](70,0)(75,20)(105,20)(110,0) 
        \Pinkline{140}{0}{180}{0}
        \Bezier[width=4](140,0)(145,20)(175,20)(180,0) 
      \end{axopicture}}}
    &  $ c_{111} =  \frac{D^2+3D-2}{D(D+2)(D-1)(D+4)(D-2)}$
    \\
    \hline
    \fbox{$N=8$}
    &{\scalebox{0.3}{
      \begin{axopicture}(96.568,96.568)(0,-10)
        \Pinkline{0}{20}{0}{-20}
        \Pinkline{28.284}{-48.284}{68.284}{-48.284}
        \Pinkline{28.284}{48.284}{68.284}{48.284}
        \Pinkline{96.568}{20}{96.568}{-20}
        \Line[width=4](0,20)(28.284,48.284)
        \Line[width=4](0,-20)(28.284,-48.284)
        \Line[width=4](68.284,48.284)(96.568,20)
        \Line[width=4](68.284,-48.284)(96.568,-20)
      \end{axopicture}}}
    &  $ c_{4}  =  \frac{-(5D+6)}{D(D-1)(D-2)(D-3)(D+6)(D+4)(D+2)(D+1)},$\\
    &
    {\scalebox{0.3}{
    \begin{axopicture}(100,60)(-15,-10)
      \Pinkline{16.666}{28.867}{-16.666}{28.867}
      \Pinkline{-33.33}{0.0}{-16.666}{-28.867}
      \Pinkline{16.666}{-28.867}{33.33}{0.0}
      \Line[width=4](16.666,28.867)(33.33,0.0)
      \Line[width=4](-16.666,28.867)(-33.33,0.0)
      \Line[width=4](-16.666,-28.867)(16.666,-28.867)
      \Pinkline{60}{0}{100}{0}
        \Bezier[width=4](60,0)(65,20)(95,20)(100,0) 
    \end{axopicture}}}&  $ c_{31} =  \frac{2}{(D-1)(D-2)(D+2)(D-3)(D+6)(D+1)},$ \\
    &
    {\scalebox{0.3}{
      \begin{axopicture}(110,20)
        \Pinkline{0}{-10}{40}{-10}
        \Pinkline{0}{30}{40}{30}  
        \Line[width=4](0,-10)(0,30)
        \Line[width=4](40,-10)(40,30)    
        
        \Pinkline{70}{-10}{110}{-10}
        \Pinkline{70}{30}{110}{30}  
        \Line[width=4](70,-10)(70,30)
        \Line[width=4](110,-10)(110,30)  
      \end{axopicture}}}&  $ c_{22} =  \frac{D^2+5D+18}{D(D-1)(D-2)(D-3)(D+6)(D+4)(D+2)(D+1)}$\\
    &
    {\scalebox{0.3}{
      \begin{axopicture}(180,20)
        \Pinkline{0}{-10}{40}{-10}
        \Pinkline{0}{30}{40}{30}  
        \Line[width=4](0,-10)(0,30)
        \Line[width=4](40,-10)(40,30)

        \Pinkline{70}{0}{110}{0}
        \Bezier[width=4](70,0)(75,20)(105,20)(110,0) 
        \Pinkline{140}{0}{180}{0}
        \Bezier[width=4](140,0)(145,20)(175,20)(180,0) 
      \end{axopicture}}}&  $ c_{211} =  \frac{-(D^3+6D^2+3D-6)}{D(D-1)(D-2)(D-3)(D+6)(D+4)(D+2)(D+1)},$ \\
    &
    {\scalebox{0.3}{
      \begin{axopicture}(250,20)
        \Pinkline{0}{0}{40}{0}
        \Bezier[width=4](0,0)(5,20)(35,20)(40,0) 
        \Pinkline{70}{0}{110}{0}
        \Bezier[width=4](70,0)(75,20)(105,20)(110,0) 
        \Pinkline{140}{0}{180}{0}
        \Bezier[width=4](140,0)(145,20)(175,20)(180,0) 
        \Pinkline{210}{0}{250}{0}
        \Bezier[width=4](210,0)(215,20)(245,20)(250,0) 
      \end{axopicture}}}&  $ c_{1111} =  \frac{(D+3)(D^2+6D+1)}{D(D+4)(D+2)(D-1)(D-3)(D+6)(D+1)}$ 
    \\
    \Xhline{3\arrayrulewidth}
  \end{tabular}
  \caption{The table shows values for the unknown coefficients of \cref{eq:nospin-projector-ansatz} for the projectors for several values of $N$. The value for each $c_\lambda$ is shown next to the graph for the corresponding orbit. Each $\lambda$ is a partition of $N/2$ that we use to label the different orbits.}
  \label{tab:coeffs}
\end{table}

%
%
%
%

\subsection{Symmetric tensor representation}\label{sec:nospin-symmetric}
Let us define the fully symmetric tensor:
\begin{equation}
  \delta^{\mu_1\dots \mu_N}=\sum_{\sigma\in S^N_2} g^{\mu_{\sigma(1)}\mu_{\sigma(2)}}\dots\,. g^{\mu_{\sigma(N-1)}\mu_{\sigma(N)}}\,
\end{equation}
In our shorthand notation this can also be written as
\begin{equation}
  \delta=\sum_{\sigma\in S^N_2} g(\sigma)\,.
\end{equation}
Another way to write $\delta$ is as the sum over all invariant sums,
\begin{equation}
\label{eq:deltaTrel}
  \delta=\sum_{\lambda \vdash N/2} T_\lambda(\sigma)\,,
\end{equation}
since every tensor $g(\sigma)$ with $\sigma \in S^N_2$ will occur once in one of the orbits. E.g. at $N=2,4$:
\begin{align}
\delta^{\mu_1\mu_2}=T_{(1)}^{\mu_1\mu_2}\,, \qquad \delta^{\mu_1\mu_2\mu_3\mu_4}=T_{(2)}^{\mu_1\mu_2\mu_3\mu_4}+T_{(1,1)}^{\mu_1\mu_2\mu_3\mu_4}\,.\nonumber
\end{align}
For a given $N$ \cref{eq:deltaTrel} provides a 1-to-1 relation between $\delta$ and the largest invariant sum $T_{(N/2)}(\sigma)$. Indeed, since these largest invariant sums provide a basis
for all other $T_{\lambda}$ via \cref{eq:pairshuffle} we can invert \cref{eq:deltaTrel} to write all $T_{\lambda}$ in terms of $\delta$s. The explicit inversion formula is not important, however the implication is that we can equally well write the general solution to the projector in terms of symmetric tensors:
\begin{equation}
\label{eq:symProjector}
  P(\sigma)=\sum_{\lambda\vdash N/2} \hat{c}_{\lambda}\, \delta_{\lambda}(\sigma)\,,
\end{equation}
where we also introduced the $\delta$-analog of the invariant sum $T_\lambda$
 \begin{equation}
 \label{eq:deltapairshuffle}
 \delta_{(\lambda_1,\dots,\lambda_l)}(\sigma)=\sum_{\sigma_1,\dots,\sigma_l\in \Sigma_\lambda(\sigma)} \delta(\sigma_{1}) \delta(\sigma_2)\cdots \delta(\sigma_l)\,.
 \end{equation}
The sum over $\Sigma_\lambda$ was defined below \cref{eq:pairshuffle}. For instance, we have at $N=6$:
\begin{equation}
\begin{split}
P^{\mu_1\dots\mu_6}&=\hat{c}_{(3)}\,\delta^{\mu_1\dots\mu_6}+ \hat c_{(1,1,1)}\,\delta^{\mu_1\mu_2}\delta^{\mu_3\mu_4}\delta^{\mu_5\mu_6}\\
&+ \hat{c}_{(2,1)}\,(\delta^{\mu_1\mu_2\mu_3\mu_4}\delta^{\mu_5\mu_6}+
 \delta^{\mu_3\mu_4\mu_5\mu_6}\delta^{\mu_1\mu_2}+
 \delta^{\mu_1\mu_2\mu_5\mu_6}\delta^{\mu_3\mu_4})
 \,.
\end{split}
\end{equation}
While the symmetric tensor representation in \cref{eq:symProjector} leads to a larger number of terms when written out in terms of metric tensors, it is more suitable for manipulations in \FORM{} since \FORM{} has highly efficient built-in support for the symmetric tensors in the \verb|dd_| function. In particular contractions with several identical momenta can be performed faster in this representation. Furthermore, it turns out that the coefficients $c_{k_1\dots k_{N}}$ take a somewhat more compact form in the symmetric basis, then they take in the minimal basis. We include the coefficients in the symmetric basis up to rank 32 in an ancillary file. We also include a form routine to construct the corresponding projector.  

\paragraph{Checks.} We have checked the valididty of the projectors by explicitely checking orthonormality with randomly chosen representatives from each orbit.
For projectors up to rank $\sim 14$ this procedure has furthermore been validated through applications in calculations of physically meaningful quantities in the context of the $R^*$-method \cite{Herzog:2017bjx,Herzog:2017dtz,Herzog:2018kwj,Herzog:2017ohr}. For one-loop results the correctness of the projectors have been checked up to rank 20 via \cref{eq:onelooptensorN} to be discussed in the next subsection.

\subsection{Integrand symmetries}\label{sec:nospin_integrandsymm}

Enormous simplifications of the structure of the tensor integral are due to integrand symmetries.
A known result \cite{Anastasiou:2015yha} is that an even rank $N$ vacuum one-loop tensor integral is proportional to the fully symmetric tensor $\delta$,
\begin{equation}
\label{eq:onelooptensorN}
I_1^{\mu_1\dots \mu_N}=\int d^Dk \,k^{\mu_1}\cdots k^{\mu_N} (\dots) = \frac{\delta^{\mu_1\dots\mu_N}}{C(N)}  \int d^Dk (k^2)^{N/2} (\dots)\,,
\end{equation}
where the $(\dots)$ indicate the scalar part of the integrand and
\begin{equation}
C(N)=2^{N/2}\,\frac{\Gamma((D+N)/2)}{\Gamma(D/2)}\,,
\end{equation}
with $\Gamma$ being Euler's $\Gamma$-function. Eq.\ \eqref{eq:onelooptensorN} is very useful, providing a compact analytic expression for arbitrary $D$ and $N$. With the general projector framework developed in this section we can in principle obtain the corresponding expression at arbitrary loops $L$. For $L=N$ we generally expect that no symmetry is present, and that every product of $g$s would have a different coefficient after the reduction. In practice, however, $L$ might be significantly smaller than $N$, leading to a considerable residual symmetry in the integrand. Consider for example the 2-loop integral
\begin{equation}
\label{eq:twolooptensorN}
I_2^{\mu_1\dots \mu_{N_1};\nu_1\dots \nu_{N_2}}=\int d^Dk_1\,d^Dk_2 \,k_1^{\mu_1}\cdots k_1^{\mu_{N_1}} k_2^{\nu_1}\cdots k_2^{\nu_{N_2}}(\dots)\,, 
\end{equation}
with $N=N_1+N_2$. The answer will no longer be proportional to $\delta^{\mu_1\dots\mu_N}$ since the integrand is not fully symmetric under $S_{N_1+N_2}$. Instead, the symmetry group is reduced to $S_{N_1}\times S_{N_2}$, and the tensor structures will reflect this reduced invariance. Classifying a general basis of such tensors is beyond the scope of this work. However, there is a neat trick which can be used to obtain a compact expression with the algorithm we proposed, not just for this particular example, but also for more complicated ones at higher loops. To find the different invariant tensors one can simply contract the symmetric tensor $\delta^{\mu_1\dots\mu_N}$ with the integrand. This can be done efficiently in \FORM. For example, for $N_1=2$ and $N_2=4$, we find
\begin{equation}
\label{eq:k1k1k2k2k2k2}
 \delta^{k_1k_1k_2k_2k_2k_2}= \underbrace{3(k_1.k_1)(k_2.k_2)^2}_{\delta_1^{k_1k_1k_2k_2k_2k_2}}+ \underbrace{12(k_1.k_2)^2 (k_2.k_2)}_{\delta_2^{k_1k_1k_2k_2k_2k_2}}\,,
\end{equation}
where we introduced the \emph{Schoonschip notation} for a  vector contracted into a tensor: $k_\mu T^{\dots \mu \dots }= T^{\dots k \dots }$. This shows that the 15 tensors which make up $\delta$ at rank $N=6$ split into two invariant (under the internal symmetry)  tensors, we refer to them as $\delta_{1}$ and $\delta_2$, containing 3 and 12 terms each. We can actually reconstruct these tensors from \cref{eq:k1k1k2k2k2k2} by replacing the dot products with metric tensors, where the $k_1$ become an index $\mu_i$ and the $k_2$ become an index $\nu_i$, and symmetrising over the internal symmetry group $S_{2}\times S_{4}$. The first invariant is then
\begin{eqnarray}
 \delta_1^{\mu_1\mu_2\nu_1\nu_2\nu_3\nu_4}=3
 \frac{1}{2!}\frac{1}{4!} \sum_{\sigma\in S_2} \sum_{\tau\in S_4} g^{\mu_{\sigma(1)}\mu_{\sigma(2)}} g^{\nu_{\tau(1)}\nu_{\tau(2)}}g^{\nu_{\tau(3)}\nu_{\tau(4)}}=\delta^{\mu_1\mu_2}\delta^{\nu_1\nu_2\nu_3\nu_4}\,.
\end{eqnarray}
which factorises neatly. Somewhat less trivial is the second which is
\begin{eqnarray}
 \delta_2^{\mu_1\mu_2\nu_1\nu_2\nu_3\nu_4}=12
 \frac{1}{2!}\frac{1}{4!} \sum_{\sigma\in S_2} \sum_{\tau\in S_4} g^{\mu_{\sigma(1)}\nu_{\tau(1)}} g^{\mu_{\sigma(2)}\nu_{\tau(2)}}g^{\nu_{\tau(3)}\nu_{\tau(4)}}\,.
\end{eqnarray}
Note that the prefactors always cancel exactly, since the denominator factorises into the number of terms, present in the numerator, times an overcounting factor generated by the sum. Alternatively we can also generate $\delta_1$ and $\delta_2$ by applying Taylor differentiation operators to the contracted invariants in \cref{eq:k1k1k2k2k2k2}, e.g.
\begin{equation}
 \delta_1^{\mu_1\mu_2\nu_1\nu_2\nu_3\nu_4}=\frac{1}{2!}\frac{\partial}{\partial k_1^{\mu_1}} \frac{\partial}{\partial k_1^{\mu_1}} 
 \frac{1}{4!}\frac{\partial}{\partial k_2^{\nu_1}} \cdots \frac{\partial}{\partial k_2^{\nu_4}}  \delta_1^{k_1k_1k_2k_2k_2k_2}\,.
\end{equation}
We refrain from writing down here the most general formulae to describe the decomposition into invariant tensors, instead we hope that it is understood that there is no problem in reconstructing the entire tensor structures from the contracted symmetric tensor polynomial, like the one in \cref{eq:k1k1k2k2k2k2}. The important thing to note is that the projectors for a given term have the same symmetry properties as the term, and the independent contracted projectors are therefore in 1:1 correspondence to the invariant tensors. For the example at hand we thus obtain the following reduction
\begin{eqnarray}
\label{eq:twolooptensorN}
I_2^{\mu_1\mu_{2};\nu_1\cdots \nu_{4}}&=& \delta^{\mu_1\mu_2}\delta^{\nu_1\nu_2\nu_3\nu_4} \int d^Dk_1\,d^Dk_2  \, P^{k_1k_1k_2k_2k_2k_2}\,(\dots) \\
&&+ \delta_2^{\mu_1\mu_2\nu_1\nu_2\nu_3\nu_4} \int d^Dk_1\,d^Dk_2  \,P^{k_1k_2k_1k_2k_2k_2}\,(\dots) \nonumber
\end{eqnarray}
The implementation of this algorithm to arbitrary loops is presented in ref.~\cite{opiter}. Its main advantage is making the evaluation of these kinds of tensor integrals particularly fast, since it avoids the cumbersome writing out of all possible tensor structures, which represents the main bottleneck for large $N$. Also, only a few independent contractions of the projectors $P$ have to be evaluated. In fact one can easily tabulate the different possible contractions up to a given loop number in \FORM's table structure, meaning that the contractions of each kind only have to be computed once. The reduction of tensors with moderate loop numbers (up to 4 or 5) but with high rank (up to 20) can then be made very fast.

\section{$N$ Lorentz Indices and one Fermion Line}
\label{sec:1fermion line}
In Feynman diagram calculations involving fermions tensor-reduction becomes more complex, as the possible structures will now include gamma matrices. To deal with these structures it is convenient to use antisymmetric products of gamma matrices since they form a basis of the Clifford algebra, $\{\gamma^\mu,\gamma^\nu\}=2g^{\mu\nu}$, in arbitrary integer dimensions. As such we define
\begin{equation}\label{eq:antisymmetric-gamma}
    \left(\Gamma^{\mu_1\dots\mu_p}\right)_{i_1 i_2}=\left(\gamma^{[\mu_1}\dots\gamma^{\mu_p]}\right)_{i_1 i_2}= \frac{1}{p!}\,\delta^{\mu_1\dots\mu_p}_{\nu_1\dots\nu_p}\left(\gamma^{\nu_1}\dots\gamma^{\nu_p}\right)_{i_1 i_2},
\end{equation}
with 
\begin{equation}
\delta^{\mu_1\dots\mu_p}_{\nu_1\dots\nu_p} = p!\, \delta^{[\mu_1}_{\nu_1}\delta^{\mu_2}_{\nu_2}\,\dots\, \delta^{\mu_p]}_{\nu_p},
\end{equation}
and the brackets $[\mu_1\dots \mu_n]$ denoting the usual antisymmetrisation of the indices $\mu_1\dots \mu_n$. 
For instance,
\begin{equation}
M^{[\mu\nu]}=\frac{1}{2!}(M^{\mu\nu}-M^{\nu\mu})\,.
\end{equation}
Note that for the space-time dimension $D$ one can have up to $D$ indices in the $\Gamma$. To keep $D$ general, in line with conventional dimensional regularization, the maximum number of indices in the $\Gamma$ should thus be taken as arbitrary. This is of course in stark contrast to four dimensions where this maximum number is four. These antisymmetric gammas are orthogonal under traces  \cite{Kennedy:1981kp} with the relation given by
\begin{equation}\label{eq:gammaOrthog}
\tr \left(\Gamma_{\mu_{1} \dots \mu_{a}}\Gamma^{\nu_{1} \dots \nu_{b}}\right)=\delta_{a b}\, \trId \,\delta_{\mu_{1} \dots \mu_{a}}^{\nu_{1} \dots \nu_{b}}(-1)^{(a(a-1)/2)} \phantom{i}. 
\end{equation}
It turns out that a compact expression exists to express arbitrary strings of gamma matrices in terms of these antisymmetric gamma matrices and metric tensors:
\begin{equation}
\label{eq:gammatoGamma}
\gamma^{\mu_1}\dots\gamma^{\mu_n}=\sum_{k=0}^{n}\sum_{\pi\in \Sigma_n^k}\sgn(\pi)\Gamma^{\mu_{\pi{(1)}}\dots\mu_{\pi(k)}}\tr(\gamma^{\mu_{\pi(k+1)}}\dots\gamma^{\mu_{\pi(n)}})\,,
\end{equation}
where the sum over $\Sigma_{n}^k$ shuffles the first $k$ indices with the remaining $n-k$ indices over the two tensors, i.e. $\Sigma_{n}^k$ is the set of permutations which distributes the indices $\mu_1,\dots,\mu_n$ into two sets of size $n$ and $n-k$ which each respecting the original order, such that $\pi(1)<\pi(2)<...<\pi(k)$ and
$\pi(k+1)<\pi(k+2)<...<\pi(n)$. Further, $\sgn\pi$ denotes the usual sign of the permutation $\pi$. To the best of our knowledge \cref{eq:gammatoGamma} is new. A proof is given in appendix \ref{sec:gammaproof}.

\subsection{Construction of the projectors}

For a single fermion line the full set of basis tensors can be written in terms of a product of a single $\Gamma$ times multiple metric tensors with all possible permutations of the Lorentz indices. A particular basis element can therefore be written
\begin{equation}
t_{n_{\gamma}}^{\mu_{1} \dots \mu_{N}}(\sigma)=g^{\mu_{\sigma(1)} \mu_{\sigma(2)}} \dots g^{\mu_{\sigma(2 n_{g}-1)} \mu_{\sigma(2 n_{g})}} \Gamma^{\mu_{\sigma(2 n_{g}+1)} \dots \mu_{\sigma(N)}}\,,
\end{equation}
where $n_g$ and $n_\gamma$ denote the number of metric tensors and the number of indices on the antisymmetric gamma respectively, such that
\begin{equation}
    N = 2n_g + n_\gamma.
\end{equation}
We can decompose, via \cref{eq:gammatoGamma}, a Feynman integral $I^{\mu_{1} \dots \mu_{N}}$ with one fermion line as follows:
\begin{equation}\label{eq:1spin-integral}
I^{\mu_{1} \dots \mu_{N}}=\sum_{\substack{2 n_{g}+n_{\gamma}\\=N}} I^{\mu_{1} \dots \mu_{N}}_{n_\gamma}\,,
\end{equation}
into summands $ I^{\mu_{1} \dots \mu_{N}}_{n_\gamma}$ which each contain $\Gamma$s of only rank $n_\gamma$. Analogously to \cref{eq:nospin-reduction} we can write the reduction of each summand as follows:
\begin{equation}\label{eq:1spin-integral}
I^{\mu_{1} \dots \mu_{N}}_{n_\gamma}=\sum_{\substack{\sigma \in \\ S_{N;n_{\gamma}}}} t_{n_{\gamma}}^{\mu_{1} \dots \mu_{N}}(\sigma) I_{n_{\gamma}}(\sigma).
\end{equation}
Here $S_{N;n_{\gamma}}$ is the set of permutations which generate all non-identical basis tensors, up to a sign, with $n_\gamma$ indices in the antisymmetric part.
As before, this set is obtained by taking the quotient of the permutation group $S_N$ with the symmetry group (up to a sign) of a generic basis element, which we denote by $H$. In the present case, $H$ is the product group $H=(S_2)^{n_g}\times S_{n_g}\times S_{n_{\gamma}}$. The factor $S_{n_{\gamma}}$ describes the antisymmetry of the $\Gamma$, while the $(S_2)^{n_g}\times S_{n_g}$ refer to the symmetry group of a product of metric tensors, discussed in \cref{sec:noSpin}.

From the orthogonality relation, \cref{eq:gammaOrthog}, it is clear that $n_\gamma$ is a useful grading for the space of basis tensors. We can therefore decompose the space, $V^{(N)}$, of such vacuum tensors with $N$ indices into a direct sum of orthogonal subspaces as follows:
\begin{equation}
    \label{eq:1spin-grading}
    V^{(N)} = \bigoplus_{n_\gamma} V^{(N)}_{n_\gamma}\,,\vspace{-2.5ex}
\end{equation}
where $V^{(N)}_{n_\gamma}$ is the subspace spanned by vacuum tensors with $n_\gamma$ indices in the antisymmetrised gamma. In the following we will write a subscript $n_\gamma$ on several quantities to indicate that they are constructed from the particular subspace $V^{(N)}_{n_\gamma}$. The size of the subspace is given by
\begin{equation}
    \dim(V^{(N)}_{n_\gamma}) = 
    |S_{N;n_{\gamma}}|= \frac{N!}{n_g!\,2^{n_g}\,n_\gamma!} = \binom{N}{n_\gamma}(2n_g-1)!!\,.
\end{equation}
Therefore, the total number of basis elements of rank $N$ is given by
\begin{equation}
    \dim (V^{(N)}) = 
    \sum_{2n_g+n_\gamma=N}\binom{N}{n_\gamma}(2n_g-1)!!\,,
\end{equation}
where the sum goes over all possible non-negative integers $n_\gamma$ and $n_g$ which satisfy $2n_g+n_\gamma=N$. The number of such independent tensors for various values of $N$ is given in table \ref{tab:1spinsytemsize}.
\begin{table}[H]
        \centering
    \begin{tabular}{ c c c c c c c c c c c c c c} 
    \Xhline{3\arrayrulewidth}
     $N$ & 1  & 2 & 3 & 4 & 5 & 6 & 7 & 8 & 9 & 10 & 11 & 12 \\
     \hline
    $\dim (V^{(N)})$ &
    1 & 2 & 4 & 10 & 26 & 76 & 232 & 764 & 2,620 & 9,496 & 35,696 & 140,152 \\
    \Xhline{3\arrayrulewidth}
    \end{tabular}   
    \caption{Table showing the number of independent tensors for tensors with one fermion line and $N$ external Lorentz indices.}
    \label{tab:1spinsytemsize}      
\end{table}

Having defined the space of basis tensors, we can now construct an orbit partition formula in the same manner as in \cref{sec:noSpin}. As a consequence of the orthogonality relation, \cref{eq:gammaOrthog}, we can define the projector, $P_{n_\gamma}(\sigma)$ for a given basis element $t_{n_\gamma}(\sigma)$, as a linear combination of tensors with the same $n_g$ and $n_\gamma$. By construction $P_{n_\gamma}(\sigma)$ lies entirely in $V_{n_\gamma}^{(N)}$ and is therefore orthogonal under traces with any $t_{n_{\gamma}'}(\sigma)$ with $n_{\gamma}'\neq n_\gamma$.
As before, the number of unknown coefficients in the projector is restricted by the symmetries of the base element. Denoting the index symmetries of an element $t_{n_\gamma}$ by $H_{n_\gamma}(\sigma)$, we have for all $h\in H_{n_\gamma}(\sigma)$:
\begin{equation}\label{eq:1spin-sign}
    t_{n_\gamma}(h \circ \sigma) = s_{n_\gamma}(h,\sigma) \, t_{n_\gamma}(\sigma)\,,\qquad
\end{equation}
where we allow the symmetry relation to include an arbitrary sign
\begin{equation*}
s_{n_\gamma}(h,\sigma)=\pm 1\,,
\end{equation*}
which is essentially determined by whether $h \circ \sigma$ leads to an even or odd permutation of the $n_\gamma$ antisymmetric indices in $t_{n_\gamma}(\sigma)$.

Let us now return to the construction of the orbit partition formula for the projector $P_{n_\gamma}(\sigma)$. Following the procedure introduced in \cref{sec:noSpin}
we partition the tensors into orbits under the action of $H_{n_\gamma}(\sigma)$.
The structure of these orbits is discussed in detail in \cref{sec:1spin-orbits}. This leads us to the following ansatz for the projector
\begin{equation}
\label{eq:1spin-projector-ansatz}
P_{n_\gamma}(\sigma)=\sum_{k} c_{n_\gamma}^{k} T_{n_\gamma}^{k}(\sigma)\,.
\end{equation}
As before, the invariant sums $T^k_{n_\gamma}(\sigma)$ are constructed to have the same index symmetry as $t_{n_\gamma}(\sigma)$. 
This necessitates the use of the sign $s_{n_\gamma}$, defined in \cref{eq:1spin-sign}, in order to account for antisymmetric indices in $t_{n_\gamma}(\sigma)$.
The invariant sum can be written as
\begin{equation}\label{eq:1spin-invariant-sum}
    T^k_{n_\gamma}(\sigma) = \sum_{\tau \in C^k_{n_\gamma}(\sigma)} \!\!\!\!
    s_{n_\gamma}(\tau\circ \pi^{-1},\sigma) \, t_{n_\gamma}(\tau)\,,
\end{equation}
where $\pi$ is any permutation belonging to $C^k_{n_\gamma}(\sigma)$ that we fix for the entire sum, and which functions as a reference permutation for the orbit. Since both $\tau$ and $\pi$ belong to the same orbit, it is clear that $\tau=h\circ\pi$ for some element $h\in H_{n_\gamma}(\sigma)$, which makes the use of $s_{n_\gamma}$ in \cref{eq:1spin-invariant-sum} consistent with its definition in \cref{eq:1spin-sign}.
The choice of $\pi$ is arbitrary, and choosing a different $\pi\in C^k_{n_\gamma}(\sigma)$ will lead to at most an overall sign that factors out of the invariant sum.
These signs are discussed in more detail in \cref{sec:method}.

\subsection{Orbits}
\label{sec:1spin-orbits}
In contrast to the pure metric case, the extra complication which arises here is that indices can now be distributed between the metric and antisymmetric gamma parts. However, due to \cref{eq:1spin-grading} one can focus on one $n_\gamma$ at a time.

We consider a base element corresponding to
$$t_{n_\gamma}(e)= g^{\mu_1\mu_2}\dots g^{\mu_{2n_g-1}\mu_{2n_g}}\Gamma^{\mu_{2n_g+1}\dots\mu_{N}}.$$
As in the pure metric case, a procedure to identify different orbits is to map the tensors $t_{n_\gamma}(\sigma)$ to graphs. We extend our graphical notation to include antisymmetric gammas as follows. 
\begin{itemize}
    \item An antisymmetric product of gamma matrices in the base element, $t(e)$, is shown as a large dashed (pink) blob with its Lorentz indices shown as dashed (pink) lines connecting to the appropriate index/vertex.
    \begin{equation*}
      \begin{axopicture}{(100,30)(-50,-15)}
     \Text(0, 0)[rc]{$t(e)=\Gamma^{\mu_1\mu_2\mu_3}\dots \quad \to$}
     \pinkline{25}{0}{60}{0}
     \Line[color=RubineRed,width=1.5,dash,dsize=2](40,15)(60,15)
     \Line[color=RubineRed,width=1.5,dash,dsize=2](40,-15)(60,-15)
     \Bezier[color=RubineRed,width=1.5,dash,dsize=2](25,0)(25,10)(35,15)(40,15)
     \Bezier[color=RubineRed,width=1.5,dash,dsize=2](25,0)(25,-10)(35,-15)(40,-15)
     \pinkspin{25}{0} 
     \Vertex(60,15){1.5}
     \Vertex(60,0){1.5}
     \Vertex(60,-15){1.5}
     \Text(65,17)[lc]{1}
     \Text(65,0)[lc]{2}
     \Text(65,-17)[lc]{3}
     \Text(80,0)[lc]{$\dots$}
     \end{axopicture}
      \end{equation*}
  \item An antisymmetric product of gamma matrices in $t(\sigma)$ is shown as a large blob with its Lorentz indices shown as solid black lines connecting to the appropriate index/vertex.
  \begin{equation*}
    \begin{axopicture}{(100,30)(-50,-15)}
   \Text(0, 0)[rc]{$\Gamma^{\mu_1\mu_2\mu_3}\dots \quad \to$}
   \Line[width=1](25,0)(60,0)
   \Line[width=1](40,15)(60,15)
   \Line[width=1](40,-15)(60,-15)
   \Bezier[width=1](25,0)(25,10)(35,15)(40,15)
   \Bezier[width=1](25,0)(25,-10)(35,-15)(40,-15)
   \BCirc(25,0){10}
   \Vertex(60,15){1.5}
   \Vertex(60,0){1.5}
   \Vertex(60,-15){1.5}
   \Text(65,17)[lc]{1}
   \Text(65,0)[lc]{2}
   \Text(65,-17)[lc]{3}
   \Text(80,0)[lc]{$\dots$}
   \end{axopicture}
    \end{equation*}
\end{itemize}

To demonstrate the different features which can occur let us consider the base element $g^{\mu_1\mu_2}g^{\mu_3\mu_4}\Gamma^{\mu_5\mu_6}$. The five possible topologies are presented in \cref{tab:Pgraph-1spin}. Again, these topologies are invariant under the action of the stabiliser group and therefore label the orbits. A metric can either form part of a cycle of metric tensors, as in the case without fermion lines in \cref{sec:puremetricpartitions}, or can be in the \emph{spin-component}, by which we refer the connected component containing the two spin blobs. Any diagram can then be classified by two numbers. The  first, $n_{1}$, is the number of solid lines in cycles. The second, $n_2$, is the number of solid lines, which correspond to metrics, in the spin-component. In this way the tuple $(n_1,n_2)$ forms a partition of $n_g$.

\begin{table}[ht]
    \centering
    \begin{tabular}{ c c c c}
    \Xhline{3\arrayrulewidth}
    \quad \begin{minipage}{0.05\textwidth}\centering$\lambda_g$\end{minipage} & \begin{minipage}{0.05\textwidth}\centering$\lambda_\gamma$\end{minipage} & \begin{minipage}{0.25\textwidth}\centering Tensor\end{minipage}  &  \begin{minipage}{0.45\textwidth}\centering Graph\end{minipage} \\
        \hline
        (1,1)  & (0) & $g^{\mu_1\mu_2}g^{\mu_3\mu_4}\Gamma^{\mu_5\mu_6}$  &\multicolumn{1}{m{0.4\textwidth}}{
        \begin{axopicture}(130,70)(0,-30)
            \pinkline{0}{20}{40}{20}
            \pinkline{0}{-20}{40}{-20}
            \Bezier[width=1](0,20)(5,40)(35,40)(40,20)
            \Bezier[width=1](0,-20)(5,0)(35,0)(40,-20)            
            \Text(-5,25){1}
            \Text(45,25){2}
            \Text(-5,-25){3}
            \Text(45,-25){4}            
            \Line[color=RubineRed,width=1.5,dash,dsize=2](60,5)(90,5)
            \Line[color=RubineRed,width=1.5,dash,dsize=2](60,-5)(90,-5)
            \Line[width=1](90,5)(120,5)
            \Line[width=1](90,-5)(120,-5)
            \pinkspin{60}{0}
            \BCirc(120,0){10}
            \Vertex(90,5){1.5}
            \Text(90,15){$5$}
            \Vertex(90,-5){1.5}
            \Text(90,-15){$6$}
        \end{axopicture}}
    \\
        \hline
        (2)  & (0)  &   $g^{\mu_1\mu_3}g^{\mu_2\mu_4}\Gamma^{\mu_5\mu_6}$  &
    \multicolumn{1}{m{0.4\textwidth}}{
        \begin{axopicture}(130,70)(0,-30)
            \pinkline{0}{20}{40}{20}
            \pinkline{0}{-20}{40}{-20}
            \Line[width=1](0,20)(0,-20)
            \Line[width=1](40,20)(40,-20)            
            \Text(-5,25){1}
            \Text(45,25){2}
            \Text(-5,-25){3}
            \Text(45,-25){4}           
            \Line[color=RubineRed,width=1.5,dash,dsize=2](60,5)(90,5)
            \Line[color=RubineRed,width=1.5,dash,dsize=2](60,-5)(90,-5)
            \Line[width=1](90,5)(120,5)
            \Line[width=1](90,-5)(120,-5)           
            \pinkspin{60}{0}
            \BCirc(120,0){10}           
            \Vertex(90,5){1.5}
            \Text(90,15){$5$}
            \Vertex(90,-5){1.5}
            \Text(90,-15){$6$}
        \end{axopicture}}
    \\
        \hline
        (1) & (1) &   $g^{\mu_1\mu_2}g^{\mu_3\mu_5}\Gamma^{\mu_4\mu_6}$  & 
    \multicolumn{1}{m{0.45\textwidth}}{
        \begin{axopicture}(130,60)(0,-20)
            \pinkline{0}{20}{40}{20}
            \Bezier[width=1](0,20)(5,40)(35,40)(40,20)             
            \Text(-5,25){1}
            \Text(45,25){2}                
            \Line[color=RubineRed,width=1.5,dash,dsize=2](60,5)(90,5)
            \Line[color=RubineRed,width=1.5,dash,dsize=2](60,-5)(110,-5)
            \Line[width=1](90,5)(110,5)
            \Line[color=RubineRed,width=1.5,dash,dsize=2](110,5)(130,5)
            \Line[width=1](130,5)(160,5)
            \Line[width=1](110,-5)(160,-5)            
            \pinkspin{60}{0}
            \BCirc(160,0){10}            
            \Vertex(90,5){1.5}
            \Text(90,15){$5$}
            \Vertex(110,5){1.5}
            \Text(110,15){$3$}
            \Vertex(130,5){1.5}
            \Text(130,15){$4$}           
            \Vertex(110,-5){1.5}
            \Text(110,-15){$6$}
        \end{axopicture}}
    \\
    \hline
    (0) & (1,1) &   $g^{\mu_1\mu_6}g^{\mu_3\mu_5}\Gamma^{\mu_2\mu_4}$  & 
    \multicolumn{1}{m{0.4\textwidth}}{
    \begin{axopicture}(130,50)(50,-20)
        \Line[color=RubineRed,width=1.5,dash,dsize=2](60,5)(90,5) 
        \Line[color=RubineRed,width=1.5,dash,dsize=2](60,-5)(90,-5) 
        \Line[width=1](90,5)(110,5) 
        \Line[width=1](90,-5)(110,-5) 
        \Line[color=RubineRed,width=1.5,dash,dsize=2](110,5)(130,5) 
        \Line[color=RubineRed,width=1.5,dash,dsize=2](110,-5)(130,-5) 
        \Line[width=1](130,5)(160,5) 
        \Line[width=1](130,-5)(160,-5) 
        \pinkspin{60}{0}
        \BCirc(160,0){10}
        \Vertex(90,5){1.5}
        \Text(90,15){$5$}
        \Vertex(110,5){1.5}
        \Text(110,15){$3$}
        \Vertex(130,5){1.5}
        \Text(130,15){$4$}
        \Vertex(90,-5){1.5}
        \Text(90,-15){$6$}
        \Vertex(110,-5){1.5}
        \Text(110,-15){$1$}
        \Vertex(130,-5){1.5}
        \Text(130,-15){$2$}
    \end{axopicture}}
    \\ 
    \hline
    (0) & (2) &   $g^{\mu_1\mu_5}g^{\mu_2\mu_3}\Gamma^{\mu_4\mu_6}$  & 
    \multicolumn{1}{m{0.43\textwidth}}{
    \begin{axopicture}(130,50)(-5,-20)
        \Line[color=RubineRed,width=1.5,dash,dsize=2](0,5)(30,5) 
        \Line[color=RubineRed,width=1.5,dash,dsize=2](0,-5)(70,-5) 
        \Line[width=1](30,5)(50,5) 
        \Line[color=RubineRed,width=1.5,dash,dsize=2](50,5)(70,5) 
        \Line[width=1](70,5)(90,5) 
        \Line[color=RubineRed,width=1.5,dash,dsize=2](90,5)(110,5) 
        \Line[width=1](110,5)(140,5) 
        \Line[width=1](70,-5)(140,-5) 
        \pinkspin{0}{0}
        \BCirc(140,0){10}
        \Vertex(30,5){1.5}
        \Text(30,15){5}
        \Vertex(50,5){1.5}
        \Text(50,15){1}
        \Vertex(70,5){1.5}
        \Text(70,15){2}
        \Vertex(90,5){1.5}
        \Text(90,15){3}
        \Vertex(110,5){1.5}
        \Text(110,15){4}
        \Vertex(70,-5){1.5}
        \Text(70,-15){6}
    \end{axopicture}}
    \\
    \Xhline{3\arrayrulewidth}
        \end{tabular}
    \caption{The table lists examples of tensors from each of the five non-vanishing orbits for $N=6$, $n_g=2$ and $n_\gamma=2$. The left column lists the cycle structure of the non-spin part of the diagram in the same manner as \cref{sec:puremetricpartitions}. The second column specifies the graph structure of the spin-component, listing the number of metrics or solid lines on each path between the antisymmetric gamma blobs.}
    \label{tab:Pgraph-1spin}
\end{table} 

To label the topology we introduce two partitions:
\begin{equation}
    \lambda_g, \quad \lambda_\gamma, \quad \text{where}\quad \lambda_g \vdash n_1, \quad  \lambda_\gamma \vdash n_2\,,
\end{equation}
where $\lambda_\gamma$ is a partition of $n_2$ with at most $n_\gamma$ parts. The partition $\lambda_g$ describes the cycle structure of the purely metric components of the graphs, in the same way as in \cref{sec:puremetricpartitions}, while the $\lambda_\gamma$ describes how many solid (metric) lines are on each path between the spin blobs. Examples of $\lambda_g$ and $\lambda_\gamma$ are given in \cref{tab:Pgraph-1spin}.

There are, however, a few more diagrams that can be drawn. Consider the tensor $t_{n_\gamma}(\tau)=g^{\mu_1\mu_2}g^{\mu_5\mu_6}\Gamma^{\mu_3\mu_4}$.
In the present example we wish to construct an invariant sum involving this tensor that is antisymmetric on the indices $\mu_5$ and $\mu_6$.
However, such a quantity would identically be zero:
\begin{equation*}
    g^{\mu_1\mu_2}g^{\mu_5\mu_6}\Gamma^{\mu_3\mu_4} - g^{\mu_1\mu_2}g^{\mu_6\mu_5}\Gamma^{\mu_3\mu_4}=0\,.
\end{equation*}
Similarly, symmetrising this tensor over $\mu_3$ and $\mu_4$ gives 
\begin{equation*}
    g^{\mu_1\mu_2}g^{\mu_5\mu_6}\Gamma^{\mu_3\mu_4} + g^{\mu_1\mu_2}g^{\mu_5\mu_6}\Gamma^{\mu_4\mu_3}=0\,.
\end{equation*}
Similar cancellations happen between the remaining tensors in this orbit.
Therefore, the only invariant sum that can be constructed from the orbit of $t_{n_\gamma}(\tau)$ is identically zero. An orbit with this property is very easy to spot diagrammatically as the diagram will contain at least one closed cycle that starts and ends on the same blob. For the base term $g^{\mu_1\mu_2}g^{\mu_3\mu_4}\Gamma^{\mu_5\mu_6}$ the 3 possible diagrams of this type are shown in \cref{tab:1spin-zero-diagrams}.
\begin{table}
    \centering
    \begin{tabular}[h]{c c}
        \Xhline{3\arrayrulewidth}\vspace{-5ex}\\
        Tensor& \quad Graph \\\hline\vspace{-2.5ex}\\
        $g^{\mu_2\mu_4}g^{\mu_5\mu_6}\Gamma^{\mu_1\mu_3}$&
        \multicolumn{1}{m{0.45\textwidth}}{
        \begin{axopicture}(100,60)(-60,-30)
            \pinkline{0}{0}{30}{15}
            \pinkline{0}{0}{30}{-15}
            \Line[width=1](30,15)(30,-15)
            \pinkspin{0}{0}
            \Line[width=1](50,10)(50,-10)
            \pinkline{50}{10}{70}{20}
            \pinkline{50}{-10}{70}{-20}
            \Line[width=1](70,20)(90,0)
            \Line[width=1](70,-20)(90,0)
            \BCirc(90,0){10}
            \Text(70,25)[cb]{1}
            \Text(50,15)[rb]{2}

            \Text(70,-25)[ct]{3}
            \Text(50,-15)[rt]{4}

            \Text(30,20)[cb]{5}
            \Text(30,-20)[ct]{6}
        \end{axopicture}}\\\hline
        $g^{\mu_1\mu_5}g^{\mu_2\mu_6}\Gamma^{\mu_3\mu_4}$&
        \multicolumn{1}{m{0.45\textwidth}}{
        \begin{axopicture}(100,70)(-60,-35)
            \pinkline{0}{0}{20}{20}
            \pinkline{0}{0}{20}{-20}
            \Line[width=1](20,20)(40,10)
            \Line[width=1](20,-20)(40,-10)
            \pinkline{40}{10}{40}{-10}
            \pinkspin{0}{0}
            \pinkline{60}{15}{60}{-15}
            \Line[width=1](60,15)(90,0)
            \Line[width=1](60,-15)(90,0)
            \BCirc(90,0){10}
            \Text(20,25)[cb]{5}
            \Text(20,-25)[ct]{6}
            \Text(40,15)[lb]{1}
            \Text(40,-15)[lt]{2}
            \Text(60,20)[cb]{3}
            \Text(60,-20)[ct]{4}
        \end{axopicture}}\\\hline
        $g^{\mu_1\mu_2}g^{\mu_5\mu_6}\Gamma^{\mu_3\mu_4}$ & 
        \multicolumn{1}{m{0.45\textwidth}}{
        \begin{axopicture}(100,70)(-20,-35)
            \pinkline{0}{0}{40}{0}
            \Bezier[width=1](0,0)(5,20)(35,20)(40,0) 
            \pinkline{70}{0}{100}{15}
            \pinkline{70}{0}{100}{-15}
            \Line[width=1](100,15)(100,-15)
            \pinkspin{70}{0}
            \pinkline{120}{15}{120}{-15} 
            \Line[width=1](120,15)(150,0)
            \Line[width=1](120,-15)(150,0)
            \BCirc(150,0){10}
            \Text(0,-5)[tr]{1}
            \Text(40,-5)[tl]{2}
            \Text(100,20)[cb]{5}
            \Text(100,-20)[ct]{6}
            \Text(120,20)[cb]{3}
            \Text(120,-20)[ct]{4}
        \end{axopicture} }      \vspace{-2ex}
        \\
        \Xhline{3\arrayrulewidth}
    \end{tabular}
    \caption{The table lists examples of tensors in orbits for $N=6$, $n_g=2$ and $n_\gamma=2$ that result in an invariant sum that is identically zero.} \label{tab:1spin-zero-diagrams}
\end{table}

Enumerating the possible graph topologies for each orthogonal subspace $V^{(N)}_{n_\gamma}$ shows a similarly drastic reduction in the size of the problem in line with the results of \cref{sec:noSpin}. The number of graph topologies for a given $n_\gamma$ is given by
\begin{equation}\label{eq:1spin-lambda}
    \Lambda^{(N)}_{n_\gamma} = 
    \sum_{n_1+n_2=n_g} p(n_1)\,p(n_\gamma;n_2),
\end{equation}
where $p(m;n)$ is the number of integer partitions of $n$ with at most $m$ parts.

For each $n_\gamma$ the projector can be found independently and the number of topologies dictates the size of the system that needs to be solved. For $N=1,\dots,10$ this enumeration is performed in \cref{tab:1spinreducedsystemsize}.
\begin{table}[h]
    \centering
    \begin{tabularx}{0.95\textwidth}{ c *{10}{Y}}
    \Xhline{3\arrayrulewidth}
        $N$ & 1  & 2 & 3 & 4 & 5 & 6 & 7 & 8 & 9 & 10 \\
        \hline
    $\dim V^{(N)}$ & \textbf{1} & \textbf{2} & \textbf{4} & \textbf{10} & \textbf{26} & \textbf{76} & \textbf{232} & \textbf{764} & \textbf{2,620} & \textbf{9,496}  \\
        $\Lambda^{(N)}_{0}$ &-&1&-&2&-&3&-&5&- &7 \\
        $\Lambda^{(N)}_{1}$ &1&-&2&-&4&-&7&-&12&- \\
        $\Lambda^{(N)}_{2}$ &-&1&-&2&-&5&-&9&- &17\\
        $\Lambda^{(N)}_{3}$ &-&-&1&-&2&-&5&-&10&- \\
        $\Lambda^{(N)}_{4}$ &-&-&-&1&-&2&-&5&- &10\\
        $\Lambda^{(N)}_{5}$ &-&-&-&-&1&-&2&-&5 &- \\
        $\Lambda^{(N)}_{6}$ &-&-&-&-&-&1&-&2&- &5 \\
        $\Lambda^{(N)}_{7}$ &-&-&-&-&-&-&1&-&2 &- \\
        $\Lambda^{(N)}_{8}$ &-&-&-&-&-&-&-&1&- &2 \\
        $\Lambda^{(N)}_{9}$ &-&-&-&-&-&-&-&-&1 &- \\
        $\Lambda^{(N)}_{10}$&-&-&-&-&-&-&-&-&- &1 \\
    \Xhline{3\arrayrulewidth}
    \end{tabularx}
    \caption{The table compares the total number of one fermion line tensor structures, $\dim V^{(N)}$, for a given $N$ with the numbers of independent coefficients for the projectors, $\Lambda^{(N)}_{n_\gamma}$, for each possible $n_\gamma$.}
    \label{tab:1spinreducedsystemsize}
\end{table}


\subsection{Solution for the projectors}
Analogously to section \ref{sec:nospinsol}, the unknown coefficients $c^k_{n_\gamma}$ of \cref{eq:1spin-projector-ansatz} are obtained by solving the following system of equations:
\begin{equation}\label{eq:1spin-projector-solution}
    \tr\left(P_{n_\gamma}(\sigma) \cdot t_{n_\gamma}(\sigma_k)\right)= \delta_{\sigma,\sigma_k}\,, \qquad k=1,\dots, \Lambda^{(N)}_{n_\gamma}\,,
\end{equation}
with $\sigma$ fixed as the base permutation and $\sigma_k$ a representative from orbit $k$.
Here we understand the central dot, which represents contraction of the Lorentz indices, to mean 
\begin{equation}
    A \cdot B = 
    (A^{\muSet})_{ij} (B_{\muSet})_{jk}.
\end{equation}
Some coefficients for different $N$ are presented in \cref{tab:coeff-1spin}.


\begin{table}[h]
    \renewcommand*{\arraystretch}{2.5}
    \centering
    \begin{tabular}{c@{\hskip 3em}c@{\hskip 3em}c@{\hskip 3em}c}
        \Xhline{3\arrayrulewidth}\vspace{-10ex}\\
        $N$ & $n_\gamma$ & Graph & $c_{\lambda_g,\lambda_\gamma}$\\\hline
    \fbox{$N=1$} &$1$ & 

    {\scalebox{0.3}{
      \begin{axopicture}(80,20)
        \Pinkline{0}{0}{40}{0}
        \Line[width=4](40,0)(80,0)
        \Pinkspin{0}{0} 
        \Blackspin{80}{0}        
      \end{axopicture}}}
                
                &$c_{(),()}= \frac{1}{D\trId}$
           
    \\[1.5ex] \hline
    \fbox{$N=2$}&$0$ &
    {\scalebox{0.3}{
      \begin{axopicture}(120,20)
        \Pinkline{0}{0}{40}{0}
        \Bezier[width=4](0,0)(5,20)(35,20)(40,0)
        \Pinkspin{80}{0} 
        \Blackspin{120}{0} 
      \end{axopicture}}}
        &$c_{(1),()}= \frac{1}{D\trId}$
    \\ 
 &$2$ &
 \Gape[0.5ex][2ex]{{\scalebox{0.3}{
    \begin{axopicture}(80,40)
        \Pinkbezier{0}{0}{10}{10}{10}{20}{40}{20}
        \Pinkbezier{0}{0}{10}{-10}{10}{-20}{40}{-20}
        \Bezier[width=4](40,20)(70,20)(70,10)(80,0)
        \Bezier[width=4](40,-20)(70,-20)(70,-10)(80,0)
        \Pinkspin{0}{0} 
        \Blackspin{80}{0}  
      \end{axopicture}}}}
            
                & $c_{(),()}= \frac{1}{2D(D-1)\trId}$  
                    
    \\[1.5ex] \hline
   \fbox{$N=3$} &$1$ & 
        {\scalebox{0.3}{
            \begin{axopicture}(160,20)
            \Pinkline{0}{0}{40}{0}
            \Bezier[width=4](0,0)(5,20)(35,20)(40,0)
            \Pinkline{80}{0}{120}{0}
        \Line[width=4](120,0)(160,0)
            \Pinkspin{80}{0} 
            \Blackspin{160}{0} 
            \end{axopicture}}}
            
                 &$c_{(1),()} = \frac{-1}{D(D+2)(D-1)\trId}$,\\
            &&
            {\scalebox{0.3}{
            \begin{axopicture}(150,20)
            \Pinkline{0}{0}{40}{0}
            \Line[width=4](40,0)(70,0)
            \Pinkline{70}{0}{110}{0}
            \Line[width=4](110,0)(150,0)
            \Pinkspin{0}{0} 
            \Blackspin{150}{0} 
            \end{axopicture}}}&   $ c_{(),(1)} = \frac{D+1}{D(D+2)(D-1)\trId}$ 
            \\
    &$3$ & 
   \Gape[0.5ex][4ex]{{\scalebox{0.3}{
    \begin{axopicture}(80,80)
        \Pinkbezier{0}{0}{0}{10}{10}{40}{40}{40}
        \Pinkbezier{0}{0}{0}{-10}{10}{-40}{40}{-40}
        \Bezier[width=4](40,40)(70,40)(80,10)(80,0)
        \Bezier[width=4](40,-40)(70,-40)(80,-10)(80,0)
        \Pinkline{0}{0}{40}{0}
        \Line[width=4](40,0)(80,0)
        \Pinkspin{0}{0} 
        \Blackspin{80}{0}  
      \end{axopicture}}}  }     
                &$c_{(),()} =  \frac{1}{D(D-2)(D-1)\trId}$ 
            
    \\[1.5ex] \hline
    \fbox{$N=4$} &$0$ &

    {\scalebox{0.3}{
      \begin{axopicture}(120,20)
        \Pinkline{0}{20}{40}{20}
        \Bezier[width=4](0,20)(5,40)(35,40)(40,20)
        \Pinkline{0}{-20}{40}{-20}
        \Bezier[width=4](0,-20)(5,0)(35,0)(40,-20)
        \Pinkspin{80}{0} 
        \Blackspin{120}{0} 
      \end{axopicture}}}
            
                 &$c_{(1,1),()}= \frac{D+1}{D(D+2)(D-1)\trId},$\\
            &&
            {\scalebox{0.3}{
                \begin{axopicture}(120,20)
                  \Pinkline{0}{20}{40}{20}
                  \Line[width=4](0,20)(0,-20)
                  \Line[width=4](40,20)(40,-20)
                  \Pinkline{0}{-20}{40}{-20}
                  \Pinkspin{80}{0} 
                  \Blackspin{120}{0} 
                \end{axopicture}}}
            
            &$c_{(2),()} = \frac{-1}{D(D+2)(D-1)\trId}$ 
            \\
    &$2$ & 
    {\scalebox{0.3}{
    \begin{axopicture}(160,40)(-80,0)
        \Pinkline{-80}{0}{-40}{0}
            \Bezier[width=4](-80,0)(-75,20)(-45,20)(-40,0)
        \Pinkbezier{0}{0}{10}{10}{10}{20}{40}{20}
        \Pinkbezier{0}{0}{10}{-10}{10}{-20}{40}{-20}
        \Bezier[width=4](40,20)(70,20)(70,10)(80,0)
        \Bezier[width=4](40,-20)(70,-20)(70,-10)(80,0)
        \Pinkspin{0}{0} 
        \Blackspin{80}{0}  
      \end{axopicture}}}       
            
                & $c_{(1),()} =\frac{1}{D(D - 1) (D - 2) (D + 2)
                \trId}$,\\
            &&  
            {\scalebox{0.3}{
                \begin{axopicture}(150,40)
                   \Pinkbezier{0}{0}{10}{10}{10}{20}{40}{20}
                   \Line[width=4](40,20)(70,20)
                   \Pinkline{70}{20}{110}{20}
                    \Bezier[width=4](110,20)(140,20)(140,10)(150,0)
                  \Pinkbezier{0}{0}{10}{-20}{10}{-20}{70}{-20}
                  
                   \Bezier[width=4](70,-20)(140,-20)(140,-20)(150,0)
                   \Pinkspin{0}{0} 
                   \Blackspin{150}{0}  
                 \end{axopicture}}}
            
            &$c_{(),(1)} =\frac{1}{(D - 1) (D - 2) (D + 2)\trId}$ 
           \\[1ex]
    &$4$ &        
    {\scalebox{0.3}{
        \begin{axopicture}(80,40)
            \Pinkbezier{0}{0}{0}{30}{10}{60}{40}{60}
            \Pinkbezier{0}{0}{10}{0}{10}{20}{40}{20}
            \Pinkbezier{0}{0}{10}{0}{10}{-20}{40}{-20}
            \Pinkbezier{0}{0}{0}{-30}{10}{-60}{40}{-60}
            \Bezier[width=4](40,60)(70,60)(80,30)(80,0)
            \Bezier[width=4](40,20)(70,20)(70,0)(80,0)
            \Bezier[width=4](40,-20)(70,-20)(70,0)(80,0)
            \Bezier[width=4](40,-60)(70,-60)(80,-30)(80,0)
            \Pinkspin{0}{0} 
            \Blackspin{80}{0}  
          \end{axopicture}}}
               &$c_{(),()}=\frac{1}{ D (D - 1) (D - 2) (D - 3)
                \trId}$  
           \\[4ex] \Xhline{3\arrayrulewidth}
    \end{tabular}
    \caption{The table shows values for the unknown coefficients of \cref{eq:1spin-projector-ansatz} for the projectors for values of $N$ from 1 to 4. Each value of $n_\gamma$ in the table corresponds to a different projector. The value for each $c_{\lambda_g,\lambda_\gamma}$ is shown next to the graph for the corresponding orbit. The coefficients are labelled using the partition structure of the graphs that was discussed in \cref{sec:1spin-orbits}.}
    \label{tab:coeff-1spin}
    \end{table}
    A way to make the computation of the projectors more efficient is to write the projectors in terms of a the basis tensors 
    %
    %
    \begin{equation}
        \tilde t_{n_{\gamma}}^{\mu_{1} \dots \mu_{N}}(\sigma) =
        P^{\mu_{\sigma(1)}\dots\mu_{\sigma(2n_g)}}
        \Gamma^{\mu_{\sigma(2 n_{g}+1)} \dots \mu_{\sigma(N)}}\,,
    \end{equation}
    where the symbol $P$ denotes the pure-Lorentz projector defined in \cref{eq:nospin-projector-ansatz,eq:nospin-projector-definingproperty}.
    We then make a different ansatz for the one fermion line projector with $t$ replaced by $\tilde t$:
    \begin{equation}
    \label{eq:1spin-dual-projector-ansatz}
        \tilde P_{n_\gamma}(\sigma)=\sum_{k} \tilde c_k \sum_{\tau \in C^k_{n_\gamma}(\sigma)}
        s_{n_\gamma}(\tau\circ \pi_k^{-1},\sigma) \tilde t_{n_\gamma}(\tau).
    \end{equation}
    As in \cref{eq:1spin-invariant-sum}, $\pi_k$ is a reference permutation corresponding to the orbit $k$.
    Being dual to one another, the metric-only projectors have the same symmetry properties as the metric tensors.
    It follows that the sum in \cref{eq:1spin-dual-projector-ansatz} is over the same orbits $C^k_{n_\gamma}(\sigma)$ as the sum in \cref{eq:1spin-invariant-sum}.
    
    The advantage of this approach becomes apparent when we consider index contractions with the $\tilde P$s in \cref{eq:1spin-projector-solution}, which leads to contractions between metric tensors and $P$s.
    The evaluation of these contractions can be greatly sped up by using the following defining property of the $P$s:
    \begin{equation}
    \label{eq:Pprojectionidentity}
        g_{\mu_1\mu_2}\dots g_{\mu_{2n_g-1}\mu_{2n_g}} P^{\mu_{\sigma(1)}\dots \mu_{\sigma(2n_g)}}
        = \begin{cases}
        1 & \sigma = e \,,\\
        0 & \mathrm{otherwise}\,,
        \end{cases}\\ 
    \end{equation}
    which means that all of the contractions are automatically either zero or one. This simplification has allowed us to construct one fermion line projectors with up to $N=15$ external indices. Results are included in the ancillary files.

    \paragraph{Checks.} We have checked the valididty of the projectors by explicitely checking orthonormality with randomly chosen representatives from each orbit.
    Further checks will be presented in section \ref{sec:testing}.

\subsection{Optimisations} \label{sec:1spin-optimise}

A further level of simplifications can be achieved by taking into account the symmetry or antisymmetry properties of the integrand of the loop integral.
For example, when reducing a loop integral we may encounter a numerator of the form
\begin{equation}
\label{eq:1spinoptiexample}
    N^{\mu_1\mu_2\dots\mu_5}= k^{\mu_1}k^{\mu_2}k^{\mu_3}\Gamma^{k\mu_4\mu_5},
\end{equation}
where we again used the Schoonschip-notation, $k_\mu T^{\dots \mu \dots }= T^{\dots k \dots }$.
Our aim is to reduce this numerator so that the tensor integral becomes a sum of scalar integrals.
The numerator is clearly symmetric under exchanges of $\mu_1$, $\mu_2$ and $\mu_3$, and antisymmetric under exchanges of $\mu_4$ and $\mu_5$. These symmetry properties must be respected by the ansatz of possible tensor structures which may appear after integration. So, by simple inspection, many traces can be immediately set to zero. Taking this into account can quite drastically reduce the basis of possible required tensors, and their corresponding projectors, thereby leading to a faster and more efficient tensor reduction. Now let $P^{\mu_1\mu_2;\mu_3\mu_4\mu_5}$ be the projector for
$g_{\mu_1\mu_2}\Gamma_{\mu_3\mu_4\mu_5}$. An example of a trace that is zero by inspection  is
\[
    \tr( P^{\mu_4\mu_5;\mu_1\mu_2\mu_3} \, k^{\mu_1} k^{\mu_2} k^{\mu_3} \Gamma^{k\mu_4\mu_5} ) =0\,.
\]
Such terms can be systematically not included in the ansatz.

Yet another set of constraints for the ansatz comes from what we shall call the \emph{$\Gamma$-index rule} which holds as long as one works explicitly in the $\Gamma$-basis. It states that any Lorentz index on a $\Gamma$ in the integrand must also be on a $\Gamma$ in the ansatz. The proof goes as follows. Consider the general integral
\begin{align}
    I^\muSet &= \int \measure{k}{L} \mathcal{K}^{\mu_1\dots \mu_k }_{\alpha_1\dots \alpha_{m}} \Gamma^{\alpha_1\dots \alpha_{m}\mu_{k+1}\dots \mu_N } \\
    &=\Gamma^{\alpha_1\dots \alpha_{m}\mu_{k+1}\dots \mu_N }  \int \measure{k}{L} \mathcal{K}^{\mu_1\dots \mu_k }_{\alpha_1\dots \alpha_{m}}\\
     &=\Gamma^{\alpha_1\dots \alpha_{m}\mu_{k+1}\dots \mu_N }  \big(c g^{\mu_1\dots}\dots g^{\dots\mu_k} g^{\alpha_1\dots} \dots g^{\dots \alpha_{m}}+\dots\big)
\end{align}
with $\mathcal{K}^{\mu_1\dots \mu_k }_{\alpha_1\dots \alpha_{m}}$ a tensor integrand depending on the loop momenta. Performing the tensor reduction (without fermion lines) one obtains a sum over metric tensors which have to be contracted into the $\Gamma$. The $\Gamma$-index rule thus follows.
For the example in \cref{eq:1spinoptiexample} this means that the ansatz would be the following:
\begin{equation}
\label{eq:antisymbasisexX}
A(g^{\mu_1\mu_2}\Gamma^{\mu_3\mu_4\mu_5}+g^{\mu_2\mu_3}\Gamma^{\mu_1\mu_4\mu_5} +g^{\mu_3\mu_1}\Gamma^{\mu_2\mu_4\mu_5})\ .
\end{equation}
In cases where there are several loop momenta there will be additional restrictions but accounting for these adds their own computational difficulties.
Thus, the coefficient $A$ for the numerator in \cref{eq:1spinoptiexample} in the ansatz, \cref{eq:antisymbasisexX}, is given by
\begin{equation}
A=\tr P^{\mu_1\mu_2;\mu_3\mu_4\mu_5}N_{\mu_1\mu_2\dots\mu_5}=\tr P^{kk;k\mu_4\mu_5}\Gamma_{k\mu_4\mu_5}\,.
\end{equation}
Hence, the formalism allows one to neatly take into account the symmetry properties of the integral.

\section{$N$ Lorentz Indices and two Fermion Lines}
\label{sec:2fermion lines}

We now consider basis tensors for integrals involving two fermion lines.
We can form a basis for the spinor part by decomposing onto antisymmetric gammas on each fermion line separately. The generic structure of the spinor part is then of the form
\begin{equation}
    (\Gamma^\ixset{\mu}{n} \otimes \Gamma^\ixset{\nu}{m})_{ik;jl} =
    \left( \Gamma^\ixset{\mu}{n} \right)_{ij}
    \left( \Gamma^\ixset{\nu}{m} \right)_{kl}\,,
\end{equation}
multiplied by a product of metric tensors. We will use the tensor product notation to avoid writing explicit spinor indices. At two fermion lines we also must account for troublesome contracted Lorentz indices which we refer to as \emph{crossings}.
For instance, the set of vacuum tensors with two external indices includes
\begin{equation*}
    \Gamma^{\mu_1\mu_2\nu_1} \otimes \Gamma_{\nu_1}, \quad
    \Gamma^{\mu_1\mu_2\nu_1\nu_2} \otimes \Gamma_{\nu_1\nu_2}, \quad
    \Gamma^{\mu_1\mu_2\nu_1\nu_2\nu_3} \otimes \Gamma_{\nu_1\nu_2\nu_3},
\end{equation*}
with the contracted $\nu$-indices being the crossings. While in 4 dimensions no more than $4$ Lorentz indices can appear in a $\Gamma$, a complication arising in conventional dimensional regularization is that we can have, in principle, arbitrarily many antisymmetrised indices. The number of crossings is therefore unbounded, and so the number of vacuum basis tensors with a fixed number of external indices is infinite, i.e.\ we have an infinite-dimensional vector space. However, in any practical calculation the number of gamma matrices appearing in a given integral is always finite. 

We can further simplify the problem by transforming the integrand into the antisymmetric basis, this can be always be achieved efficiently via identity \cref{eq:gammatoGamma}, to obtain an integrand of the generic form
\begin{equation}
\label{eq:2spinintegral}
I^{\mu_1\dots\mu_N}_{N_1,N_2}=\int  d^Dk_1\dots d^Dk_L \,\mathcal{K}^{\mu_1\dots\mu_N}_{\alpha_1,\dots,\alpha_{N_1};\beta_1,\dots,\beta_{N_2}}\, \Gamma_1^{\alpha_1,\dots,\alpha_{N_1}}\otimes \Gamma_2^{\beta_1,\dots,\beta_{N_2}}\,,
\end{equation}
with $\mathcal{K}$ a tensor depending on the loop momenta. At this point, the orthogonality relation \cref{eq:gammaOrthog} can be used to ensure that only tensor products of $\Gamma$-matrices, i.e. of form $\Gamma_1\otimes \Gamma_2$, with the $\Gamma_i$ of the same rank as those appearing in the integrand of \cref{eq:2spinintegral}, can occur in the ansatz. In fact, as we will show below, these orthogonal subspaces are finite-dimensional, implying also that the projectors are built from a finite number of terms.

The structure of a tensor in the two fermion line basis can be specified by 4 numbers: the number of metric tensors $n_g$, the number of external indices on the first fermion line $n_{\gamma_1}$, the number of external indices on the second fermion line $n_{\gamma_2}$, and the number of crossings $n_c$.
These four numbers are combined into the tensor data tuple $\vec{n}=(n_g,n_{\gamma_1},n_{\gamma_2},n_c)$.
These numbers satisfy the obvious relation
$N=2n_g+n_{\gamma_1}+n_{\gamma_2}$.
There are many tensors that correspond to this data; they can be written generally as 
\begin{equation}
\label{eq:2spin-general-tensor}
    \begin{split}
        t^{\muSet}_{\vec{n}}(\sigma)=&
        \, g^{\mu_{\sigma(1)} \mu_{\sigma(2)}} \dots g^{\mu_{\sigma(2 n_{g}-1)} \mu_{\sigma(2 n_{g})}}\\
        &\times\Gamma^{\mu_{\sigma(2 n_{g}+1)} \dots \mu_{\sigma(2n_g+n_{\gamma_1})}\nu_1\dots\nu_{n_c}}
        \otimes
        \Gamma\indices{^{\mu_{\sigma(2 n_{g}+n_{\gamma_1}+1)} \dots \mu_{\sigma(n)}}_{\nu_1\dots\nu_{n_c}}},
    \end{split}
\end{equation}
where, as before, $\sigma$ is a permutation of the external indices belonging to the permutation group $S_N$.
Many of these permutations will correspond to identical tensors (up to a possible sign) in a way we shall now specify.
The symmetry of the external indices of the tensor \cref{eq:2spin-general-tensor} is described by the product group
\begin{equation}
    H = (S_2)^{n_g} \times S_{n_g} \times S_{n_{\gamma_1}} \times S_{n_{\gamma_2}},
\end{equation}
which we use as before to partition the full group of index permutations $S_N$ into cosets that live in the quotient set $S_N/H=S_{\vec{n}}$. 
The number of unique tensors with a given structure $\vec{n}$ is equal to the number of cosets:
\begin{equation}
|S_{\vec{n}}| =
\frac{n!}{2^{n_g}\, n_g!\,n_{\gamma_1}!\,n_{\gamma_2}!}=
\binom{n}{n_{\gamma_1}+n_{\gamma_2}}\binom{n_{\gamma_1}+n_{\gamma_2}}{n_{\gamma_1}}(2n_g-1)!!.
\end{equation}
With these considerations the tensor reduction of \cref{eq:2spinintegral} is given by
\begin{equation}
I^{\mu_1\dots\mu_N}_{N_1,N_2}=\sum_{\vec{n}} \sum_{\sigma\in S_{\vec{n}}} t^{\muSet}_{\vec{n}}(\sigma) I_{\vec{n}}(\sigma)
\end{equation}
where we sum over all independent tensors whose tensor data $\vec{n}$ is consistent with $N_1$ and $N_2$. 
We will give a concise expression for this sum in the following \cref{sec:2spin-orthoclasses}.

\subsection{Construction of the orthogonal subspaces}\label{sec:2spin-orthoclasses}
For a given two fermion line tensor in the antisymmetric basis let us define $N_1=n_{\gamma_1}+n_c$ and $N_2=n_{\gamma_2}+n_c$ as the number of indices on the two respective fermion lines, where we count all indices whether crossing or external. It follows from the orthogonality relation \cref{eq:gammaOrthog} that two tensors with different $N_1$ and/or different $N_2$ are orthogonal. We will say that tensors are in the same orthogonality class if they have the same $N_1$ and $N_2$. In the following we construct a basis for a given orthogonal subspace. The tensors $t_{\vec{n}'}$ with data $\vec{n}' = (n'_g, n'_{\gamma_1}, n'_{\gamma_2}, n'_c)$ belonging to a particular orthogonality class must then satisfy the following constraints:
\begin{equation}\label{eq:tensorreq}
    N = 2n_g'+n_{\gamma_1}'+n_{\gamma_2}', \qquad
    N_1 =n_{\gamma_1}'+n_c',\qquad
    N_2=n_{\gamma_2}'+n_c',
\end{equation}
 where $n_{\gamma_1}',\ n_{\gamma_2}',\ n_g',\ n_c'\geq 0$.

Thanks to orthogonality the equations, \eqref{eq:tensorreq}, in fact have a finite number of solutions. To determine these we now derive upper and lower bounds on each of the $n_{\gamma_1}',\ n_{\gamma_2}',\ n_g',\ n_c'$. First, let us recast the equations as follows:
\begin{align}
    N - N_1-N_2 &= 2n_g'-2n_c',\label{eq:Ncon}\\
    N_1 &=n_{\gamma_1}'+n_c',\label{eq:n1con}\\
    N_2&=n_{\gamma_2}'+n_c'.\label{eq:n2con}
\end{align}
The maximum values of $n_c'$ can be determined from \cref{eq:n1con,eq:n2con}.
Since $n_{\gamma_1}',\ n_{\gamma_2}'\geq 0$ we must have $n_c'\leq \min{(N_1,N_2)}$.
A lower bound comes from \cref{eq:Ncon} by writing $n_c' = n_c-c$:
\begin{equation}
    N - N_1-N_2 = 2n_g'-2(n_c-c)\,.
\end{equation}
For this condition to hold we must have $n_g'=n_g-c$. It thus follows that $c\le\min(n_g,n_c)$. Writing $n_c-n_g=(N_1+N_2-N)/2= n_c^{\min}$ we then derive
\begin{equation}
\max(n_c^{\min},0)\leq n_c'\leq \min{(N_1,N_2)}\,.
\end{equation}
It should be noted that for $n_c^{\min}$ to be an integer, the sum $N_1+N_2-N$ must be even. It is actually impossible to construct a tensor where this is not the case. Fixing $n'_c$ uniquely determines $ n_{\gamma_1}',n_{\gamma_2}'\text{and}\ n_g'$ through the equations \eqref{eq:tensorreq}. As $n'_c$ is now bounded to a finite range and uniquely determines the rest of the tensor data $\vec n'$, this not only provides a convenient way to generate the set of basis tensors for the construction of the orthogonal subspace, but also proves that their number is finite.

It is interesting to discuss the decomposition of the vector space $V^{(N)}$ in the case of the two fermion lines into the orthogonal subspaces. By grading on the pair of numbers $(N_1, N_2)$, the space of basis tensors decomposes into a direct sum
\begin{equation}\label{eq:2spin-grading}
    V^{(N)} = \bigoplus_{(N_1, N_2) \in X_N} V^{(N)}_{N_1, N_2}\,,
\end{equation}
where each of the summands are mutually orthogonal subspaces. The indexing set is defined as
\begin{equation}
    X_N = \{ (N_1, N_2) \in \mathbb N \times \mathbb N \ | \
    N_1 + N_2 + N \text{ even}, \
    |N_1 - N_2| \leq N\}.
\end{equation}
The even condition is evident from \cref{eq:Ncon} as the right-hand-side is clearly even. The magnitude condition can be seen by combining the expressions $N=N_1-N_2 +2n_g'+2n_{\gamma_2}'$ and $N=N_2-N_1 +2n_g'+2n_{\gamma_1}'$, which can be obtained from \cref{eq:tensorreq}. As before, in order to construct a projector for a base term in a particular subspace, one only has to consider tensors that belong to the same subspace $V^{(N)}_{N_1,N_2}$.

We denote the set of all tensor data inside a given $V^{(N)}_{N_1,N_2}$ as $\{\vec{n}\}_{N_1,N_2}$. The size of any one of these orthogonal subspaces is given by
\begin{equation}
\begin{split}
&|V^{(N)}_{N_1,N_2}|= \sum_{\{\vec n\}_{N_1,N_2}} |S_{\vec n}|\\
&\;\;=\sum_{n'_c=\max(n_c^{\min},0)}^{\min(N_{1},N_{2})}\binom{N}{N_{1}+N_{2}-2n_c'}\binom{N_{1}+N_{2}-2n_c'}{N_{1}-n_c'}(N-N_1-N_2+2n_c'-1)!!\,.
\end{split}
\end{equation}
This enumeration is done in table \ref{tab:2spinCount2n} for a tensor with $N=4$ and various values of $N_1,\ N_2$. For higher values of $N$ see the tables in appendix \ref{sec:2spintable}.
    \begin{table}[H]
        \centering
    \begin{tabular}{c c c c c c}
        \Xhline{3\arrayrulewidth}
        \vspace{-3ex}\\
        \fbox{$|V^{(N=4)}_{N_1,N_2}|$} &$N_1=1$ &$N_1=2$ &$N_1=3$ &$N_1=4$&$N_1=5$\\
        $N_2=1$& \textbf{15} & \textbf{-} &\textbf{10} &\textbf{-} &\textbf{1} \\
         $N_2=2$& \textbf{-} & \textbf{21} &\textbf{-} &\textbf{10} &\textbf{-} \\
         $N_2=3$& \textbf{10} & \textbf{-} &\textbf{21} &\textbf{-} &\textbf{10} \\
         $N_2=4$&\textbf{-} & \textbf{10} &\textbf{-} &\textbf{21} &\textbf{-} \\
         $N_2=5$&\textbf{1} & \textbf{-} &\textbf{10} &\textbf{-} &\textbf{21}\vspace{1ex}\\
         \Xhline{3\arrayrulewidth}
    \end{tabular}
    \caption{A table enumerating the number of independent tensors in the orthogonal subspace for $N=4$ and a range of values for $N_1,\ N_2$.}
    \label{tab:2spinCount2n}
\end{table}
It should be noted that in each orthogonality class several projectors will need to be constructed, one for each $\vec{n}$ within the class.

\subsection{Construction of the projectors}

As an example we will consider the construction of the projector for the following tensor:
\begin{equation}\label{eq:2spin-example}
    t_{(0,2,1,0)}(e) =
    \Gamma^{\mu_1\mu_2} \otimes \Gamma^{\mu_3}\,,
\end{equation}
which belongs to the subspace $V^{(3)}_{2,1}$.
From the above, we see that this subspace has a basis that includes not only tensors with data $(0,2,1,0)$, but also tensors with data $(1,1,0,1)$.
The complete basis of vacuum tensors for this subspace is
\begin{equation}
        \left. 
        \begin{array}{c}
        \Gamma^{\mu_1\mu_2} \otimes \Gamma^{\mu_3},  \\
        \Gamma^{\mu_1\mu_3} \otimes \Gamma^{\mu_2},  \\
        \Gamma^{\mu_2\mu_3} \otimes \Gamma^{\mu_1},  \\
        \end{array}
        \right\}   \quad (0,2,1,0) 
        \left. \qquad \quad
        \begin{array}{c}
        g^{\mu_1\mu_2} \Gamma^{\mu_3\nu} \otimes \Gamma_\nu, \\
        g^{\mu_1\mu_3} \Gamma^{\mu_2\nu} \otimes \Gamma_\nu, \\
        g^{\mu_2\mu_3} \Gamma^{\mu_1\nu} \otimes \Gamma_\nu. \\
        \end{array}
        \right\}  \quad (1,1,0,1)
\end{equation}
This accounts for \emph{all} of the basis tensors that are not orthogonal to \cref{eq:2spin-example}.

As before, we wish to partition the relevant set of basis tensors into orbits under the action of the stabiliser of the base term, which we call $H_{\vec n}(\sigma)$.
We observe that two tensors with different tensor data will never appear in the same orbit, because there is no permutation of indices that transforms one into the other.
Each orbit is labelled with an integer $k$ and the tensor data of the tensors in orbit $k$ is denoted $\vec n_k$.
As before the minimal set of permutations that generate the orbit $k$ is labelled by $C^k(\sigma, \vec n)$.

Having partitioned the set of basis tensors in this way, we can then construct invariant sums that possess the right symmetry properties. This allows us to write down the orbit partition formula for a base term $t_{\vec n}(\sigma)$ as follows:
\begin{equation}\label{eq:2spin-projector}
    P_{\vec n}(\sigma) = 
    \sum_{k} c^{k} 
    T^k(\sigma,\vec n). 
\end{equation}
In the two fermion line case the invariant sums take the form
\begin{equation}\label{eq:2spin-inv-sum}
    T^k(\sigma, \vec n) = 
    \sum_{\tau \in C^k(\sigma, \vec n)}
    s_{\vec n}(\tau\circ\pi^{-1},\sigma) t_{\vec n_k} (\tau).
\end{equation}
As before, $\pi$ is any permutation belonging to $C^k(\sigma, \vec n)$ that we fix for the entire sum. 
The signs $s_{\vec n}(\tau\circ\pi^{-1},\sigma)$ are defined for each $h\in H_{\vec n}(\sigma)$ by the equation
\begin{equation}\label{eq:2spin-sign}
    t_{\vec n}(h\circ \sigma) = s_{\vec n}(h,\sigma) t_{\vec n}(\sigma),
\end{equation}
in the same manner as \cref{eq:1spin-sign}.
We refer to \cref{sec:method} for a more in-depth discussion of these signs.

In our example, \cref{eq:2spin-example}, the stabiliser group is very simple, consisting of only two elements:
\begin{equation}
    H_{(0,2,1,0)}(e) = \{\ e,\ (12)\ \}.
\end{equation}
According to \cref{eq:2spin-sign}, we associate $e$ with the sign $+$ and $(12)$ with the sign $-$. 
For this example it is easy to construct the invariant sums by hand:
\begin{align}
    T^1 &= \Gamma^{\mu_1\mu_2} \otimes \Gamma^{\mu_3} \label{eq:2spin-t1}\\ 
    T^2 &= \Gamma^{\mu_1\mu_3} \otimes \Gamma^{\mu_2} -
    \Gamma^{\mu_2\mu_3} \otimes \Gamma^{\mu_1} \label{eq:2spin-t2}\\ 
    T^3 &= g^{\mu_1\mu_3} \Gamma^{\mu_2\nu} \otimes \Gamma_\nu
    -g^{\mu_2\mu_3} \Gamma^{\mu_1\nu} \otimes \Gamma_\nu \label{eq:2spin-t3}
\end{align}
All three have the desired antisymmetry on the exchange of $\mu_1$ and $\mu_2$, and have the same form as the general invariant sum given in \cref{eq:2spin-inv-sum}. It is impossible to construct an nonzero invariant sum with this property using $g^{\mu_1\mu_2} \Gamma^{\mu_3\nu} \otimes \Gamma_\nu$; we will say that this is a \emph{vanishing orbit}. Using the above notation we have $\vec n_1 = (0,2,1,0)$, $\vec n_2=\vec n_3=(1,1,0,0)$.
With \cref{eq:2spin-t1,eq:2spin-t2,eq:2spin-t3} in hand, the ansatz for the projector is
\begin{equation}
    P_{(0,2,1,0)}(e) = c^1 T^1 + c^2 T^2 + c^3 T^3\ .
\end{equation}
Projectors for the other tensors with data $(0,2,1,0)$ are related to this one by index permutation.
However, we cannot obtain a projector for a $(1,1,0,1)$ tensor in this manner.
Instead, we must repeat the above procedure, constructing orbits based on the symmetry of
\begin{equation}
    t_{(1,1,0,1)(e)} = g^{\mu_1\mu_2} \Gamma^{\mu_3\nu} \otimes \Gamma_\nu.
\end{equation}
Although this projector will be built from the same set of tensors, the invariant sums now must be \emph{symmetric} on $\mu_1$ and $\mu_2$.
Although the stabiliser group is the same as in the previous example, we now associate both elements with a plus sign, reflecting the symmetry (rather than antisymmetry) between $\mu_1$ and $\mu_2$. This leads to the following invariant sums:
 \begin{align}
     \hat{T}^{1} &= g^{\mu_1\mu_2} \Gamma^{\mu_3\nu} \otimes \Gamma_\nu \,,\\ 
      \hat{T}^{2} &= \Gamma^{\mu_1\mu_3} \otimes \Gamma^{\mu_2} +
      \Gamma^{\mu_2\mu_3} \otimes \Gamma^{\mu_1} \,,\\ 
     \hat{T}^3 &= g^{\mu_1\mu_3} \Gamma^{\mu_2\nu} \otimes \Gamma_\nu
     +g^{\mu_2\mu_3} \Gamma^{\mu_1\nu} \otimes \Gamma_\nu \,,
 \end{align}
such that the projector is
\begin{equation}
    P_{(1,1,0,1)}(e) = \hat c^1 \hat T^1 + \hat c^2 \hat T^2 + \hat c^3 \hat T^3\, .
\end{equation}
Note $t_{(0,2,1,0)(e)}$ does not appear here since it belongs to a vanishing orbit. Although the simple stabiliser group of this example allowed us to write down the invariant sums by inspection, for more complicated symmetry groups we will once again rely on a graphical method to construct them.

\subsection{Orbits} \label{sec:2spin-orbits}
\begin{figure}[H]
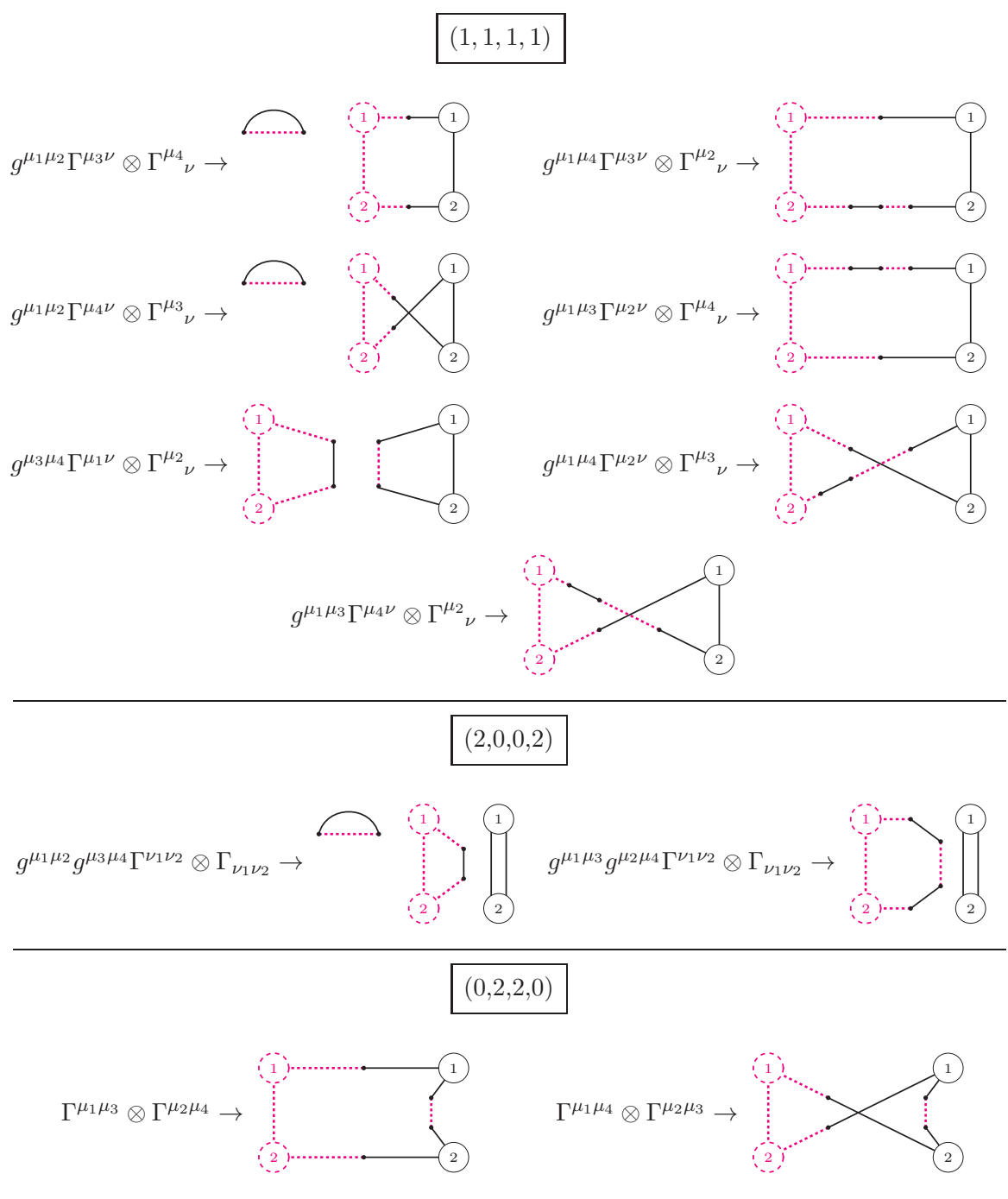

    \centering
    \rule{\textwidth}{3\arrayrulewidth}\\\vspace{1ex}
    \fbox{$(1,1,1,1)$}
    
    \scalebox{1}{
    \begin{minipage}{\textwidth}
        \centering
    \SetScale{0.65}
    \begin{minipage}[t]{0.49\textwidth}
        \raggedright

\begin{axopicture}(220,100)(-150,-35)
    \Text(-10,0)[rc]{$g^{\mu_1\mu_2} \Gamma^{\mu_3\nu} \otimes \Gamma\indices{^{\mu_4}_\nu}\rightarrow$}
    \pinkline{0}{20}{40}{20}
    \Bezier[width=1](0,20)(5,40)(35,40)(40,20)

    \pinkline{80}{30}{110}{30}
    \pinkline{80}{-30}{110}{-30}
    \pinkline{80}{30}{80}{-30}

    \Line[width=1](110,30)(140,30)
    \Line[width=1](110,-30)(140,-30)
    \Line[width=1](140,30)(140,-30)

    \pinkspin{80}{30}
    \BCirc(140,30){10}
    \pinkspin{80}{-30}
    \BCirc(140,-30){10}

    \textRubineRed
    \Text(80,30){\tiny 1}
    \Text(80,-30){\tiny 2}
    \textBlack
    \Text(140,30){\tiny 1}
    \Text(140,-30){\tiny 2}

\end{axopicture}

\begin{axopicture}(220,100)(-150,-35)
    \Text(-10,0)[rc]{$g^{\mu_1\mu_2} \Gamma^{\mu_4\nu} \otimes \Gamma\indices{^{\mu_3}_\nu}\rightarrow$}
    \pinkline{0}{20}{40}{20}
    \Bezier[width=1](0,20)(5,40)(35,40)(40,20)

    \pinkline{80}{30}{100}{10}
    \pinkline{80}{-30}{100}{-10}
    \pinkline{80}{30}{80}{-30}

    \Line[width=1](100,-10)(140,30)
    \Line[width=1](100,10)(140,-30)
    \Line[width=1](140,30)(140,-30)

    \pinkspin{80}{30}
    \BCirc(140,30){10}
    \pinkspin{80}{-30}
    \BCirc(140,-30){10}

    \textRubineRed
    \Text(80,30){\tiny 1}
    \Text(80,-30){\tiny 2}
    \textBlack
    \Text(140,30){\tiny 1}
    \Text(140,-30){\tiny 2}

\end{axopicture}

\begin{axopicture}(220,100)(-80,-35)
    
    \Text(60,0)[rc]{$g^{\mu_3\mu_4} \Gamma^{\mu_1\nu} \otimes \Gamma\indices{^{\mu_2}_\nu}\rightarrow$}
    \pinkline{80}{30}{130}{15}
    \pinkline{80}{-30}{130}{-15}
    \pinkline{80}{30}{80}{-30}

    \Line[width=1](130,-15)(130,15)
    
    \Line[width=1](210,30)(160,15)
    \Line[width=1](210,-30)(160,-16)
    \pinkline{160}{-15}{160}{15}
    
    \Line[width=1](210,30)(210,-30)

    \pinkspin{80}{30}
    \BCirc(210,30){10}
    \pinkspin{80}{-30}
    \BCirc(210,-30){10}

    \textRubineRed
    \Text(80,30){\tiny 1}
    \Text(80,-30){\tiny 2}
    \textBlack
    \Text(210,30){\tiny 1}
    \Text(210,-30){\tiny 2}

\end{axopicture}

\end{minipage}
\begin{minipage}[t]{0.49\textwidth}
    \raggedright
\begin{axopicture}(220,100)(-185,-35)
    \Text(-20,0)[rc]{$g^{\mu_1\mu_4} \Gamma^{\mu_3\nu} \otimes \Gamma\indices{^{\mu_2}_{\nu}}\rightarrow$}
    \pinkline{0}{30}{60}{30}
    \pinkline{0}{-30}{40}{-30}
    \pinkline{60}{-30}{80}{-30}
    \pinkline{0}{-30}{0}{30}

    \Line[width=1](60,30)(120,30)
    \Line[width=1](80,-30)(120,-30)
    \Line[width=1](40,-30)(60,-30)
    \Line[width=1](120,30)(120,-30)

    \pinkspin{0}{30}
    \BCirc(120,30){10}
    \pinkspin{0}{-30}
    \BCirc(120,-30){10}

    \textRubineRed
    \Text(0,30){\tiny 1}
    \Text(0,-30){\tiny 2}
    \textBlack
    \Text(120,30){\tiny 1}
    \Text(120,-30){\tiny 2}
\end{axopicture}

\begin{axopicture}(160,100)(-185,-35)
    \Text(-20,0)[rc]{$g^{\mu_1\mu_3} \Gamma^{\mu_2\nu} \otimes \Gamma\indices{^{\mu_4}_{\nu}}\rightarrow$}
    \pinkline{0}{-30}{60}{-30}
    \pinkline{0}{30}{40}{30}
    \pinkline{60}{30}{80}{30}
    \pinkline{0}{-30}{0}{30}

    \Line[width=1](60,-30)(120,-30)
    \Line[width=1](80,30)(120,30)
    \Line[width=1](40,30)(60,30)
    \Line[width=1](120,30)(120,-30)

    \BCirc(120,30){10}
    \pinkspin{0}{30}
    \pinkspin{0}{-30}
    \BCirc(120,-30){10}

    \textRubineRed
    \Text(0,30){\tiny 1}
    \Text(0,-30){\tiny 2}
    \textBlack
    \Text(120,30){\tiny 1}
    \Text(120,-30){\tiny 2}

\end{axopicture}


\begin{axopicture}(160,100)(-185,-35)
    \Text(-20,0)[rc]{$g^{\mu_1\mu_4} \Gamma^{\mu_2\nu} \otimes \Gamma\indices{^{\mu_3}_{\nu}}\rightarrow$}
    \pinkline{0}{-30}{20}{-20}
    \pinkline{0}{30}{40}{10}
    \pinkline{40}{-10}{80}{10}
    \pinkline{0}{-30}{0}{30}

    \Line[width=1](40,10)(120,-30)
    \Line[width=1](80,10)(120,30)
    \Line[width=1](20,-20)(40,-10)
    \Line[width=1](120,30)(120,-30)

    \BCirc(120,30){10}
    \pinkspin{0}{30}
    \pinkspin{0}{-30}
    \BCirc(120,-30){10}

    \textRubineRed
    \Text(0,30){\tiny 1}
    \Text(0,-30){\tiny 2}
    \textBlack
    \Text(120,30){\tiny 1}
    \Text(120,-30){\tiny 2}

\end{axopicture}

\end{minipage}


 \begin{axopicture}(160,100)(-100,-35)
    \Text(-20,0)[rc]{$g^{\mu_1\mu_3} \Gamma^{\mu_4\nu} \otimes \Gamma\indices{^{\mu_2}_{\nu}}\rightarrow$}
    \pinkline{0}{30}{20}{20}
    \pinkline{0}{-30}{40}{-10}
    \pinkline{40}{10}{80}{-10}
    \pinkline{0}{30}{0}{-30}

    \Line[width=1](40,-10)(120,30)
    \Line[width=1](80,-10)(120,-30)
    \Line[width=1](20,20)(40,10)
    \Line[width=1](120,30)(120,-30)

    \BCirc(120,30){10}
    \pinkspin{0}{30}
    \pinkspin{0}{-30}
    \BCirc(120,-30){10}

    \textRubineRed
    \Text(0,30){\tiny 1}
    \Text(0,-30){\tiny 2}
    \textBlack
    \Text(120,30){\tiny 1}
    \Text(120,-30){\tiny 2}
\end{axopicture}\\
\rule{\textwidth}{\arrayrulewidth}\\\vspace{1ex}
\fbox{(2,0,0,2)}

\begin{minipage}[t]{0.49\textwidth}
    \raggedright
\begin{axopicture}(220,100)(-190,-35)
    \Text(0,0)[rc]{$g^{\mu_1\mu_2} g^{\mu_3\mu_4} \Gamma^{\nu_1\nu_2} \otimes \Gamma_{\nu_1\nu_2}\rightarrow$}
    \pinkline{10}{20}{50}{20}
    \Bezier[width=1](10,20)(15,40)(45,40)(50,20)

    \pinkline{80}{30}{106.667}{10}
    \pinkline{80}{-30}{106.667}{-10}
    \pinkline{80}{30}{80}{-30}

    \Line[width=1](106.667,-10)(106.667,10)
    \Line[width=1](135,30)(135,-30)
    \Line[width=1](125,30)(125,-30)

    \pinkspin{80}{30}
    \BCirc(130,30){10}
    \pinkspin{80}{-30}
    \BCirc(130,-30){10}

    \textRubineRed
    \Text(80,30){\tiny 1}
    \Text(80,-30){\tiny 2}
    \textBlack
    \Text(130,30){\tiny 1}
    \Text(130,-30){\tiny 2}

\end{axopicture}
\end{minipage}
\begin{minipage}[t]{0.49\textwidth}
    \raggedright
\begin{axopicture}(220,100)(-155,-35)
 
    \Text(60,0)[rc]{$g^{\mu_1\mu_3} g^{\mu_2\mu_4} \Gamma^{\nu_1\nu_2} \otimes \Gamma_{\nu_1\nu_2}\rightarrow$}
    \pinkline{80}{30}{110}{30}
    \pinkline{80}{-30}{110}{-30}
    \pinkline{80}{30}{80}{-30}

    \pinkline{130}{-15}{130}{15}

    \Line[width=1](110,30)(130,15)
    \Line[width=1](110,-30)(130,-15)
    
    \Line[width=1](145,30)(145,-30)
    \Line[width=1](155,30)(155,-30)

    \pinkspin{80}{30}
    \BCirc(150,30){10}
    \pinkspin{80}{-30}
    \BCirc(150,-30){10}

    \textRubineRed
    \Text(80,30){\tiny 1}
    \Text(80,-30){\tiny 2}
    \textBlack
    \Text(150,30){\tiny 1}
    \Text(150,-30){\tiny 2}

\end{axopicture}
\end{minipage}\\
\rule{\textwidth}{\arrayrulewidth}\\\vspace{1ex}
\fbox{(0,2,2,0)}

\begin{minipage}[t]{0.49\textwidth}
    \raggedright

$\begin{axopicture}(160,100)(-170,-35)
    \Text(-20,0)[rc]{$\Gamma^{\mu_1\mu_3} \otimes \Gamma^{\mu_2\mu_4}\rightarrow$}
    \pinkline{0}{30}{60}{30}
    \pinkline{0}{-30}{60}{-30}
    \pinkline{105}{10}{105}{-10}
    \pinkline{0}{30}{0}{-30}

    \Line[width=1](60,30)(120,30)
    \Line[width=1](60,-30)(120,-30)
    \Line[width=1](120,30)(105,10)
    \Line[width=1](120,-30)(105,-10)

    \BCirc(120,30){10}
    \pinkspin{0}{30}
    \pinkspin{0}{-30}
    \BCirc(120,-30){10}

    \textRubineRed
    \Text(0,30){\tiny 1}
    \Text(0,-30){\tiny 2}
    \textBlack
    \Text(120,30){\tiny 1}
    \Text(120,-30){\tiny 2}

\end{axopicture}$
\end{minipage}
\begin{minipage}[t]{0.49\textwidth}
    \raggedright

\begin{axopicture}(160,100)(-170,-35)
    \Text(-20,0)[rc]{$\Gamma^{\mu_1\mu_4} \otimes \Gamma^{\mu_2\mu_3}\rightarrow$}
    \pinkline{0}{30}{40}{10}
    \pinkline{0}{-30}{40}{-10}
    \pinkline{105}{10}{105}{-10}
    \pinkline{0}{30}{0}{-30}

    \Line[width=1](40,-10)(120,30)
    \Line[width=1](40,10)(120,-30)
    \Line[width=1](120,30)(105,10)
    \Line[width=1](120,-30)(105,-10)

    \BCirc(120,30){10}
    \pinkspin{0}{30}
    \pinkspin{0}{-30}
    \BCirc(120,-30){10}

    \textRubineRed
    \Text(0,30){\tiny 1}
    \Text(0,-30){\tiny 2}
    \textBlack
    \Text(120,30){\tiny 1}
    \Text(120,-30){\tiny 2}
\end{axopicture}\vspace{3ex}
\end{minipage}

\end{minipage}}
\rule{\textwidth}{3\arrayrulewidth}
\caption{The figure shows representative tensors from each of the 11 non-vanishing orbits for the tensor $g^{\mu_1\mu_2} \Gamma^{\mu_3\nu} \otimes \Gamma\indices{^{\mu_4}_\nu}$. We note the appearances of tensors with data $(2,0,0,2)$ and $(0,2,2,0)$ as well.}
\label{fig:2spin-orbits}
\end{figure}

In order to construct the orbits in a systematic way we employ our graphical notation.
To account for the extra fermion line we simply extend the notation by drawing another blob.
We label the blobs 1 and 2 for clarity, as indicated in the following:
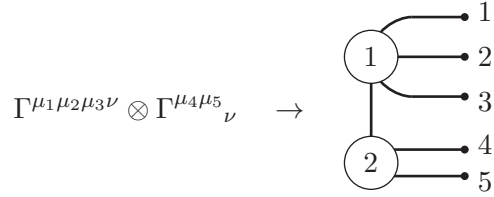
\begin{equation*}
    \begin{axopicture}(150,70)(-60,-30)
        \Text(0,0)[rc]{$\Gamma\indices{^{\mu_1\mu_2\mu_3\nu}}\otimes\Gamma\indices{^{\mu_4\mu_5}_\nu}\quad \rightarrow$}
        \Line[width=1](25,20)(60,20)
        \Line[width=1](25,20)(25,-20)
   \Line[width=1](40,35)(60,35)
   \Line[width=1](40,5)(60,5)
   \Bezier[width=1](25,20)(25,30)(35,35)(40,35)
   \Bezier[width=1](25,20)(25,10)(35,5)(40,5)
   \BCirc(25,20){10} 
   \Text(25,20){1}
   \Vertex(60,35){1.5}
   \Vertex(60,20){1.5}
   \Vertex(60,5){1.5}
   \Text(65,37)[lc]{1}
   \Text(65,20)[lc]{2}
   \Text(65,3)[lc]{3}
   
   \Line[width=1](25,-15)(60,-15)
   \Line[width=1](25,-25)(60,-25)
   \BCirc(25,-20){10} 
   \Text(25,-20){2}
   \Vertex(60,-15){1.5}
   \Vertex(60,-25){1.5}
   \Text(65,-13)[lc]{4}
   \Text(65,-27)[lc]{5}
    \end{axopicture}
\end{equation*}
Note that the contracted index $\nu$ leads to a line between the two blobs. We proceed as before by drawing diagrams for each independent non-orthogonal tensor.
To demonstrate this with a more complicated example we construct a projector for the base term 
\begin{equation}
\label{eq:twospinexample}
    t_{\vec{n}}(e)=g^{\mu_1\mu_2} \Gamma^{\mu_3\nu} \otimes \Gamma\indices{^{\mu_4}_\nu},\quad \text{where} \quad 
    \vec n=(n_g,n_{\gamma_1},n_{\gamma_2},n_{c})=(1,1,1,1)\,.
\end{equation}
Using the notation of \cref{eq:2spin-grading}, we say this tensor lives in the subspace $V^{(4)}_{2,2}$, the size (or dimension) of which is $|V^{(4)}_{2,2}|=21$. The possible tensor data for tensors living in the $N_1=2=N_2$ subspace are
\begin{equation}
\{(0,2,2,0),\,(1,1,1,1),\,(2,0,0,2)\}.
\end{equation}
We therefore must also draw diagrams for tensors of the form $\Gamma^{\mu_1\mu_2} \otimes \Gamma^{\mu_3\mu_4}$ and of the form $g^{\mu_1\mu_2}g^{\mu_3\mu_4} \Gamma^{\nu_1\nu_2} \otimes \Gamma_{\nu_1\nu_2}$. There are 11 topologies which correspond to nonvanishing orbits. These are presented in figure \ref{fig:2spin-orbits}, alongside an example tensor that corresponds to the structure. As before, the number of graphs dictates the size of the system of equations we need to solve. The complexity of these graphs is more involved than those of \cref{sec:noSpin,sec:1fermion line} and cannot be compactly summarised in terms of a cycle or partition structure. We therefore find that the best way to label these orbits is simply with the graphs themselves.

\begin{figure}
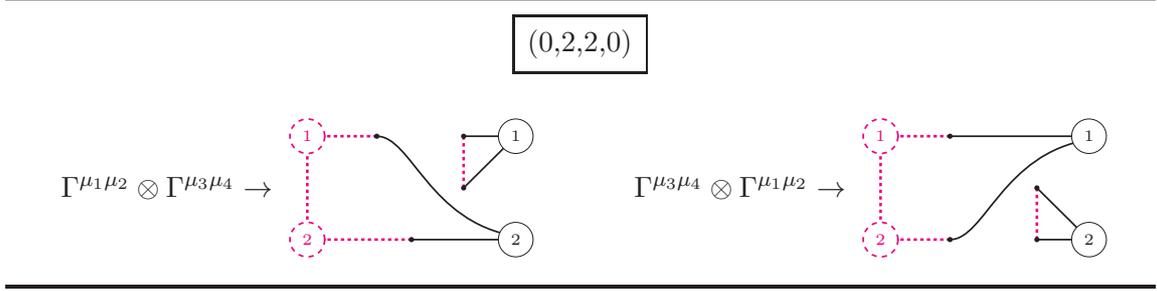

    \centering
    \SetScale{0.65}
    \rule{\textwidth}{3\arrayrulewidth}\\\vspace{1ex}
    \fbox{(0,2,2,0)}\\
    \begin{minipage}[t]{0.49\textwidth}
    \begin{axopicture}(160,100)(-170,-35)
        \Text(-20,0)[rc]{$\Gamma^{\mu_1\mu_2} \otimes \Gamma^{\mu_3\mu_4}\rightarrow$}
        \pinkline{0}{30}{40}{30}
        \pinkline{0}{-30}{60}{-30}
        
        \pinkline{0}{30}{0}{-30}
        \pinkline{90}{30}{90}{0}
    
        \Bezier[width=1](40,30)(60,30)(70,-20)(120,-28)
        \Line[width=1](60,-30)(120,-30)

        \Line[width=1](90,30)(120,30)
        \Line[width=1](90,0)(120,30)
        
        \BCirc(120,30){10}
        \pinkspin{0}{30}
        \pinkspin{0}{-30}
        \BCirc(120,-30){10}
    
        \textRubineRed
        \Text(0,30){\tiny 1}
        \Text(0,-30){\tiny 2}
        \textBlack
        \Text(120,30){\tiny 1}
        \Text(120,-30){\tiny 2}
    \end{axopicture}
    \end{minipage}
    \begin{minipage}[t]{0.49\textwidth}
        \begin{axopicture}(160,100)(-170,-35)
            \Text(-20,0)[rc]{$\Gamma^{\mu_3\mu_4} \otimes \Gamma^{\mu_1\mu_2}\rightarrow$}
            \pinkline{0}{30}{40}{30}
            \pinkline{0}{-30}{40}{-30}
            
            \pinkline{0}{30}{0}{-30}
            \pinkline{90}{-30}{90}{0}
        
            \Bezier[width=1](40,-30)(60,-30)(70,20)(120,28)
            \Line[width=1](40,30)(120,30)
    
            \Line[width=1](90,-30)(120,-30)
            \Line[width=1](90,0)(120,-30)
            
            \BCirc(120,30){10}
            \pinkspin{0}{30}
            \pinkspin{0}{-30}
            \BCirc(120,-30){10}
        
            \textRubineRed
            \Text(0,30){\tiny 1}
            \Text(0,-30){\tiny 2}
            \textBlack
            \Text(120,30){\tiny 1}
            \Text(120,-30){\tiny 2}
        \end{axopicture}
        \end{minipage}\\
    \rule{\textwidth}{3\arrayrulewidth}
    \caption{ The figure shows the 2 orbits -- for the projector of $g^{\mu_1\mu_2} \Gamma^{\mu_3\nu} \otimes \Gamma\indices{^{\mu_4}_\nu}$ -- whose
    invariant sum vanishes.}
    \label{fig:2spin-zero-orbits}
\end{figure}

As was the case in \cref{sec:1fermion line}, certain diagrams should be excluded from consideration, because they generate an invariant sum that is identically zero. These diagrams contain a closed cycle that starts and ends on the same blob. For this example only tensors with data $(0,2,2,0)$ can generate this sort of diagram as the blobs must have at least 2 external lines. There are only two of these diagrams, and they are presented in \cref{fig:2spin-zero-orbits}. 

The diagrams in \cref{fig:2spin-orbits} provide the invariant sums entering the projectors for any tensor with tensor data $(1,1,1,1)$. We need, however, to repeat the exercise for projectors of tensors in the same orthogonal subspace $V^{(4)}_{2,2}$ which have different tensor data, i.e. those with data $(2,0,0,2)$ and $(0,2,2,0)$. Their projectors will be built from the same basis of tensors, but the orbit partitioning will be different. 

The number of nonvanishing orbits, which we denote $\Lambda_{\vec{n}}$, can be exactly enumerated from $\vec{n}$ by exhaustively constructing all possible diagrams. This enumeration is done for $N=4$ in table \ref{tab:2spinCount2nClass}. A number of tables for other values of $N$ are presented in appendix \ref{sec:2spintable2}.
    \begin{table}[h]
        \centering
        \begin{tabular}{c c c c c c}
            \Xhline{3\arrayrulewidth}
            \vspace{-3ex}\\
            \fbox{\begin{minipage}{5em}\centering $|V^{(N=4)}_{N_1,N_2}|$\\ $\{\Lambda_{\vec{n}}\}$
            \end{minipage}} &$N_1=1$ &$N_1=2$ &$N_1=3$ &$N_1=4$&$N_1=5$\\
            
            $N_2=1$& \makecell{\textbf{15}\\9,4 }& \makecell{\textbf{-} }&\makecell{\textbf{10}\\3,3 }&\makecell{\textbf{-} }&\makecell{\textbf{1}\\1 }\\
             
             $N_2=2$& \makecell{\textbf{-} }& \makecell{\textbf{21}\\6,11,5 }&\makecell{\textbf{-} }&\makecell{\textbf{10}\\3,3 }&\makecell{\textbf{-} }\\
             
             $N_2=3$& \makecell{\textbf{10}\\3,3 }& \makecell{\textbf{-} }&\makecell{\textbf{21}\\6,11,5 }&\makecell{\textbf{-} }&\makecell{\textbf{10}\\3,3 }\\
             
             $N_2=4$&\makecell{\textbf{-} }& \makecell{\textbf{10}\\3,3 }&\makecell{\textbf{-} }&\makecell{\textbf{21}\\6,11,5}&\makecell{\textbf{-} }\\
             
             $N_2=5$&\makecell{\textbf{1}\\1 }& \makecell{\textbf{-} }&\makecell{\textbf{10}\\3,3 }&\makecell{\textbf{-} }&\makecell{\textbf{21}\\6,11,5 }\vspace{1ex}\\
             \Xhline{3\arrayrulewidth}
        \end{tabular}
    \caption{A table enumerating the number of tensors which are in the same orthogonal subspace as a tensor with $N=4$ and various values of $N_1,\ N_2$.
    For each $N_1,\,N_2$ several $\vec n$ are possible, and each corresponds to a different projector $P_{\vec{n}}$. E.g. for $N_1=N_2=2$ the possible $\vec{n}$ are
    $\{(2,0,0,2),(1,1,1,1),(0,2,2,0)\}$. For each of these projectors the table lists the number of independent coefficients $\Lambda_{\vec n}$ required to determine the projector $P_{\vec{n}}$, e.g.  $\Lambda_{(2,0,0,2)}=6,\ \Lambda_{(1,1,1,1)}=11,\ \Lambda_{(0,2,2,0)}=5$.}
    \label{tab:2spinCount2nClass}
\end{table}
\subsection{Solution for the projectors}
We again must calculate each $c^k$ in \cref{eq:2spin-projector}  by solving the following set of simultaneous equations:
\begin{equation}
    \tr\left(P_{\vec n}(\sigma) \cdot t_{\vec n_k}(\sigma_k)\right)= \delta_{\sigma,\sigma_k}\delta_{\vec n,\vec n_k}\,, \qquad k=1,\dots, \Lambda_{\vec n}\,,
\end{equation}
where the trace of a two fermion line quantity is understood to be 
\begin{equation}
    \tr (A_1 \otimes A_2) = \tr(A_1)\tr(A_2).
\end{equation}
The coefficients for the example considered in \cref{sec:2spin-orbits} are presented in \cref{tab:2spin-results}. We have employed the method to compute projectors for all tensors in the range $N<8$, $N_{1,2}<6$. Results for these are included in an ancillary file.
\begin{table}[h]
    \centering 
    \begin{tabular}{c c c c}
        \Xhline{3\arrayrulewidth}
        $\vec n$ & $k$ & Graph & $c_k$ \\
        \hline
        \fbox{$(1,1,1,1)$}& 1 & 
            \multicolumn{1}{m{0.15\textwidth}}{\scalebox{0.3}{ $\begin{axopicture}(150,100)(-30,-40)
                \Pinkline{0}{20}{40}{20}
                \Bezier[width=4](0,20)(5,40)(35,40)(40,20)
            
                \Pinkline{80}{30}{110}{30}
                \Pinkline{80}{-30}{110}{-30}
                \Pinkline{80}{30}{80}{-30}
            
                \Line[width=4](110,30)(140,30)
                \Line[width=4](110,-30)(140,-30)
                \Line[width=4](140,30)(140,-30)       
                \Pinkspin{80}{30}
                \Blackspin{140}{30}
                \Pinkspin{80}{-30}
                \Blackspin{140}{-30}
            
            
            \end{axopicture}$}} & $\frac{D^3+3D^2-3D-3}{D(D-1)(D-2)(D-3)(D+2)(D+1)(D+4)\trId^2}$
        \\
        & 2 &
        \multicolumn{1}{m{0.15\textwidth}}{\scalebox{0.3}{
        \begin{axopicture}(220,100)(-30,-35)
            \Pinkline{0}{20}{40}{20}
            \Bezier[width=4](0,20)(5,40)(35,40)(40,20)
        
            \Pinkline{80}{30}{100}{10}
            \Pinkline{80}{-30}{100}{-10}
            \Pinkline{80}{30}{80}{-30}
        
            \Line[width=4](100,-10)(140,30)
            \Line[width=4](100,10)(140,-30)
            \Line[width=4](140,30)(140,-30)
        
            \Pinkspin{80}{30}
            \Blackspin{140}{30}
            \Pinkspin{80}{-30}
            \Blackspin{140}{-30}
        
        \end{axopicture}}}& $\frac{-D^2-2D+1}{D(D-1)(D-2)(D-3)(D+2)(D+1)(D+4)\trId^2}$\\
        & 3 &
        \multicolumn{1}{m{0.15\textwidth}}{\scalebox{0.3}{
        \begin{axopicture}(220,100)(40,-35)
            \Pinkline{80}{30}{130}{15}
            \Pinkline{80}{-30}{130}{-15}
            \Pinkline{80}{30}{80}{-30}

            \Line[width=4](130,-15)(130,15)
    
            \Line[width=4](210,30)(160,15)
            \Line[width=4](210,-30)(160,-16)
            \Pinkline{160}{-15}{160}{15}
    
            \Line[width=4](210,30)(210,-30)

            \Pinkspin{80}{30}
            \Blackspin{210}{30}
            \Pinkspin{80}{-30}
            \Blackspin{210}{-30}
        
        \end{axopicture}}}& $\frac{3D+5}{D(D-1)(D-2)(D-3)(D+2)(D+1)(D+4)\trId^2} $\\
        & 4 &
        \multicolumn{1}{m{0.15\textwidth}}{\scalebox{0.3}{
        \begin{axopicture}(220,100)(-40,-35)
            \Pinkline{0}{30}{60}{30}
            \Pinkline{0}{-30}{40}{-30}
            \Pinkline{60}{-30}{90}{-30}
            \Pinkline{0}{-30}{0}{30}

            \Line[width=4](60,30)(120,30)
            \Line[width=4](90,-30)(120,-30)
            \Line[width=4](40,-30)(60,-30)
            \Line[width=4](130,30)(130,-30)

            \Pinkspin{0}{30}
            \Blackspin{130}{30}
            \Pinkspin{0}{-30}
            \Blackspin{130}{-30}
            
        \end{axopicture}}}& $ \frac{-D^2-3D-3}{D(D-1)(D-2)(D-3)(D+2)(D+1)(D+4)\trId^2}$\\
        & 5 &
        \multicolumn{1}{m{0.15\textwidth}}{\scalebox{0.3}{
        \begin{axopicture}(220,100)(-40,-35)
            \Pinkline{0}{-30}{60}{-30}
            \Pinkline{0}{30}{40}{30}
            \Pinkline{60}{30}{90}{30}
            \Pinkline{0}{-30}{0}{30}
        
            \Line[width=4](60,-30)(120,-30)
            \Line[width=4](90,30)(120,30)
            \Line[width=4](40,30)(60,30)
            \Line[width=4](130,30)(130,-30)

            \Blackspin{130}{30}
            \Pinkspin{0}{30}
            \Pinkspin{0}{-30}
            \Blackspin{130}{-30}
            
        \end{axopicture}}}& $ \frac{-D^2-3D-3}{D(D-1)(D-2)(D-3)(D+2)(D+1)(D+4)\trId^2}$\\
        & 6 &
        \multicolumn{1}{m{0.15\textwidth}}{\scalebox{0.3}{
        \begin{axopicture}(220,100)(-40,-35)
            \Pinkline{0}{-30}{20}{-20}
            \Pinkline{0}{30}{40}{10}
            \Pinkline{40}{-10}{80}{10}
            \Pinkline{0}{-30}{0}{30}
        
            \Line[width=4](40,10)(130,-30)
            \Line[width=4](80,10)(130,30)
            \Line[width=4](20,-20)(40,-10)
            \Line[width=4](130,30)(130,-30)

            \Blackspin{130}{30}
            \Pinkspin{0}{30}
            \Pinkspin{0}{-30}
            \Blackspin{130}{-30}
        \end{axopicture}}}& $\frac{2D+1}{D(D-1)(D-2)(D-3)(D+2)(D+1)(D+4)\trId^2} $\\
        & 7 &
        \multicolumn{1}{m{0.15\textwidth}}{\scalebox{0.3}{
        \begin{axopicture}(220,100)(-40,-35)
            \Pinkline{0}{30}{20}{20}
            \Pinkline{0}{-30}{40}{-10}
            \Pinkline{40}{10}{80}{-10}
            \Pinkline{0}{30}{0}{-30}
        
            \Line[width=4](40,-10)(130,30)
            \Line[width=4](80,-10)(130,-30)
            \Line[width=4](20,20)(40,10)
            \Line[width=4](130,30)(130,-30)

            \Blackspin{130}{30}
            \Pinkspin{0}{30}
            \Pinkspin{0}{-30}
            \Blackspin{130}{-30}
        \end{axopicture}}}& $\frac{2D+1}{D(D-1)(D-2)(D-3)(D+2)(D+1)(D+4)\trId^2} $\\ \hline
        \fbox{$(2,0,0,2)$} &8&
        \multicolumn{1}{m{0.15\textwidth}}{\scalebox{0.3}{
            \begin{axopicture}(220,100)(-30,-35)
            \Pinkline{0}{20}{40}{20}
            \Bezier[width=4](0,20)(5,40)(35,40)(40,20)

            \Pinkline{80}{30}{106.667}{10}
            \Pinkline{80}{-30}{106.667}{-10}
            \Pinkline{80}{30}{80}{-30}

            \Line[width=4](106.667,-10)(106.667,10)
            \Line[width=4](145,30)(145,-30)
            \Line[width=4](135,30)(135,-30)

            \Blackspin{140}{30}
            \Pinkspin{80}{30}
            \Pinkspin{80}{-30}
            \Blackspin{140}{-30}
        \end{axopicture}}}& $ \frac{-D^2-2D+1}{D(D-1)(D-2)(D-3)(D+2)(D+1)(D+4)\trId^2}$\\
        &9&
        \multicolumn{1}{m{0.15\textwidth}}{\scalebox{0.3}{
            \begin{axopicture}(220,100)(40,-35)
                \Pinkline{80}{30}{110}{30}
                \Pinkline{80}{-30}{110}{-30}
                \Pinkline{80}{30}{80}{-30}
                \Pinkline{150}{-15}{150}{15}
            
                \Line[width=4](110,30)(150,15)
                \Line[width=4](110,-30)(150,-15)
                \Line[width=4](215,30)(215,-30)
                \Line[width=4](205,30)(205,-30)

            \Blackspin{210}{30}
            \Pinkspin{80}{30}
            \Pinkspin{80}{-30}
            \Blackspin{210}{-30}
        \end{axopicture}}}& $\frac{2D+1}{D(D-1)(D-2)(D-3)(D+2)(D+1)(D+4)\trId^2}$\\\hline
        \fbox{$(0,2,2,0)$} & 10 &
        \multicolumn{1}{m{0.15\textwidth}}{\scalebox{0.3}{
            \begin{axopicture}(220,100)(-40,-35)
                \Pinkline{0}{30}{60}{30}
                \Pinkline{0}{-30}{60}{-30}
                \Pinkline{105}{10}{105}{-10}
                \Pinkline{0}{30}{0}{-30}

                \Line[width=4](60,30)(130,30)
                \Line[width=4](60,-30)(130,-30)
                \Line[width=4](130,30)(105,10)
                \Line[width=4](130,-30)(105,-10)

            \Blackspin{130}{30}
            \Pinkspin{0}{30}
            \Pinkspin{0}{-30}
            \Blackspin{130}{-30}
        \end{axopicture}}}& $\frac{-1}{(D-1)(D-2)(D-3)(D+2)(D+1)\trId^2}$\\
        & 11 &
        \multicolumn{1}{m{0.15\textwidth}}{\scalebox{0.3}{
            \begin{axopicture}(220,100)(-40,-35)
                \Pinkline{0}{30}{40}{10}
                \Pinkline{0}{-30}{40}{-10}
                \Pinkline{110}{10}{110}{-10}
                \Pinkline{0}{30}{0}{-30}
        
                \Line[width=4](40,-10)(130,30)
                \Line[width=4](40,10)(130,-30)
                \Line[width=4](130,30)(110,10)
                \Line[width=4](130,-30)(110,-10)

            \Blackspin{130}{30}
            \Pinkspin{0}{30}
            \Pinkspin{0}{-30}
            \Blackspin{130}{-30}
        \end{axopicture}}}& $\frac{1}{D(D-1)(D-2)(D-3)(D+2)(D+1)\trId^2}$\\
        \Xhline{3\arrayrulewidth}
    \end{tabular}
    \caption{The table shows the solutions for the unknown coefficients, corresponding to the eleven orbit diagrams of \cref{fig:2spin-orbits}, for the projector of $g^{\mu_1\mu_2} \Gamma^{\mu_3\nu} \otimes \Gamma\indices{^{\mu_4}_\nu}$ considered as an example in \cref{sec:2spin-orbits}. For clarity, we reproduce the orbit diagram next to its corresponding coefficient. The tensor data of tensors in that orbit is shown in a box in the left column.}
\label{tab:2spin-results}
\end{table}

\paragraph{Checks.} We have checked the valididty of the projectors by explicitely checking orthonormality with randomly chosen representatives from each orbit.
Further checks will be presented in section \ref{sec:testing}.

\subsection{Optimisations}
We now briefly discuss how to further optimise the reduction of an integral by reducing the size of the basis of tensors in the ansatz. Given that we are working with integrands in the antisymmetric basis, we can apply the $\Gamma$-index rule (as described in \cref{sec:1spin-optimise}) separately on each fermion line.
Consider a numerator in a two fermion line reduction that has the following form:
\begin{equation}
    N^{\mu_1\dots\mu_7} = k_1^{\mu_1}k_1^{\mu_2}k_2^{\mu_3}\Gamma^{k_1k_2\mu_4\mu_5}\otimes\Gamma^{k_1\mu_6\mu_7}.
\end{equation}
This $N$ is in the $V^{(7)}_{4,3}$ subspace which has a dimension of 665. This means a naive ansatz would have us apply 665 projectors, however after applying the $\Gamma$-index rule we find only 6 possible basis elements in the ansatz:
\begin{equation}
    \begin{split}
      N^{\mu_1\dots\mu_7} \to\phantom{+} &A_1 g^{\mu_1\mu_2} \Gamma^{\mu_3\mu_4\mu_5\nu}\otimes\Gamma\indices{^{\mu_6\mu_7}_\nu}+A_2\Gamma^{\mu_1\mu_2\mu_4\mu_5}\otimes\Gamma\indices{^{\mu_3\mu_6\mu_7}}\\
        +&A_3 g^{\mu_1\mu_3} \Gamma^{\mu_2\mu_4\mu_5\nu}\otimes\Gamma\indices{^{\mu_6\mu_7}_\nu}+A_4\Gamma^{\mu_1\mu_3\mu_4\mu_5}\otimes\Gamma\indices{^{\mu_2\mu_6\mu_7}}\\
        +&A_5 g^{\mu_2\mu_3} \Gamma^{\mu_1\mu_4\mu_5\nu}\otimes\Gamma\indices{^{\mu_6\mu_7}_\nu}+A_6\Gamma^{\mu_2\mu_3\mu_4\mu_5}\otimes\Gamma\indices{^{\mu_1\mu_6\mu_7}}\,.
    \end{split}
\end{equation}
These are the only elements in the basis of $V_{4,3}^{(7)}$ that have $\mu_4,\mu_5$ on the first fermion line and $\mu_6,\mu_7$ on the second, in accordance with the $\Gamma$-index rule. Thus, we have achieved a simplification of this particular example by a factor of about 100. 

Similarly to the single spin case described in \cref{sec:1spin-optimise}, one can use the symmetry properties of the projector to further reduce the number of projectors which actually have to be applied. In particular the symmetry under    
$\mu_1$ and $\mu_2$ leads to relations among the $A_i$-coefficients:
\begin{equation}
    \begin{split}
A_1=P_1^{\mu_1\dots\mu_7} N_{\mu_1\dots\mu_7},\qquad &A_2=0,\\
A_3=A_5=P_3^{\mu_1\dots\mu_7} N_{\mu_1\dots\mu_7},\qquad   &A_4=A_6=P_4^{\mu_1\dots\mu_7} N_{\mu_1\dots\mu_7}\,,
    \end{split}
\end{equation}
where the $P_i$ are the projectors of the tensors accompanying the $A_i$-coefficients. Thus, we gain another factor of 2 from symmetry in this example.

\section{External momenta}
\label{sec:external}
\label{sec:Etensor}
While vacuum tensor Feynman integrals are important in their own right for instance in the context of renormalization, of course most problems in Feynman integral computations involve external momenta. Fortunately there is a straightforward way to extend our method to this more general case, which will be implemented as part of the \OPITER{} \FORM{} program \cite{opiter}. In the following we shall consider an $L$-loop tensor integral depending on $E$ independent, external momenta $p_i$:
\begin{equation}
\label{ref:Etensor}
\int d^Dk_1\dots d^Dk_L \frac{k_{i_1}^{\mu_1}\dots k_{i_N}^{\mu_N}} {f(k_1,\dots,k_L,p_1,\dots,p_E)}.
\end{equation}
The denominator $f$ is a scalar function of all the momenta and perhaps some other masses and scales. We will largely ignore it since it does not participate in the tensor reduction. It is now convenient to follow the approach originally due to Van Neerven and one of the authors  \cite{vanNeerven:1983vr,Binoth:2002xg,Ellis:2011cr} to split the $D$-dimensional loop momentum space, $V$, into a subspace $V_{\parallel}$ spanned by the external momenta and its transverse complement $V_{\perp}$, such that $V=V_{\parallel}\oplus V_{\perp}$. The metric then decomposes as follows:
\begin{align}
    g^{\mu\nu} &= g^{\mu\nu}_\perp + g^{\mu\nu}_\parallel.
\end{align}
We have made the usual assumption of conventional dimensional regularization: $D>E$. The explicit forms of these metric tensors are given by:
\begin{align}
g^{\mu\nu}_\parallel &=\sum_{i,j} p_i^\mu H_{ij} p_j^\nu, \\
g^{\mu\nu}_\perp &= g^{\mu\nu}-\sum_{i,j} p_i^\mu H_{ij} p_j^\nu,
\end{align}
with $H=G^{-1}$ the inverse of the Gram matrix $G_{ij}=p_i\cdot p_j$. This allows us to split the loop momenta into transverse and parallel components as follows:
\begin{align}
    k^\mu = k^\mu_\perp + k^\mu_\parallel\,,\qquad \text{where}\qquad  
    k^\mu_\perp= g^{\mu\nu}_\perp k_\nu  ,\qquad k^\mu_\parallel=g^{\mu\nu}_\parallel k_\nu   \,.
\end{align}
This also leads to the following useful identities:
\begin{equation}
    \begin{array}{cc}
    k_{i\perp}\cdot k_{j\perp}=k_{i\perp}\cdot k_j=k_i\cdot k_{j\perp},
    \qquad\qquad &\qquad \qquad k_\perp \cdot p_i =0,\\
   (g_\perp)^\mu_{\phantom{\mu}\mu} = D_\perp= D-E,\qquad
   \qquad &\qquad \qquad g^{\mu\nu}_\perp (p_i)_\nu = 0\,.
    \end{array}
\end{equation}
Being spanned by the external momenta, the $k_{i\parallel}$ have Lorentz indices only on external momenta, and hence are already fully reduced.
Thus, the general problem of eq. (\ref{ref:Etensor}) is reduced to the form
\begin{equation}
\label{ref:ETtensor}
\int d^Dk_1\dots d^Dk_L \frac{(k_{i_1})_\perp^{\mu_1}\dots (k_{i_n})_\perp^{\mu_n}} {f(k_1,\dots,k_L,p_1,\dots,p_E)}\,.
\end{equation}
After the integration of \cref{ref:ETtensor}, the indices $\mu_1,\dots,\mu_n$ can only be carried on transverse metrics $g_\perp$.
Note that this also means that transverse integrals of type \cref{ref:ETtensor} with odd $n$ vanish.
The reduction of a tensor integral onto symmetric metric tensors was exactly the case considered already in \cref{sec:noSpin}.
We arrive at the very useful result, that the projectors of \cref{sec:noSpin} are suitable for the integral \cref{ref:ETtensor}, if we make the replacements $g\to g_\perp$ and $D\to D_\perp$.

We now consider how to implement a similar reduction on integrals that have one or more fermion lines.
To achieve this we will also need to decompose the gamma matrices, which proceeds as follows:
\begin{align}
    \gamma^\mu = \gamma^\mu_\perp + \gamma^\mu_\parallel\,,\qquad \text{where}\qquad   \gamma^\mu_\perp= g^{\mu\nu}_\perp \gamma^\nu  ,\qquad \gamma^\mu_\parallel=g^{\mu\nu}_\parallel \gamma^\nu   \,,
\end{align}
such that
\begin{equation}
    \begin{array}{cc}
    \gamma^{\mu}_{\perp}(p_i)_\mu =0,
    \qquad\qquad&\qquad\qquad\gamma^{\mu} (p_i)_\mu = \slashed{p}_i= \gamma^{\mu}_{\parallel}(p_i)_\mu ,\\
    \{ \gamma^{\mu}, \gamma^{\nu}_{\perp}\} =   \{ \gamma^{\mu}_{\perp}, \gamma^{\nu}_{\perp}\} = 2 g^{\mu\nu}_\perp, 
    \qquad\qquad&\qquad\qquad\{ \gamma^{\mu}_{\perp}, \gamma^{\nu}_{\parallel}\} = 0.
\end{array}
\end{equation}
We also define 
\begin{equation}
\label{eq:GAMMAperp}
    \Gamma_{\perp}^{\mu_1\dots\mu_p}=\gamma_{\perp}^{[\mu_1}\dots\gamma_{\perp}^{\mu_p]}= \frac{1}{p!}\,\delta^{\mu_1\dots\mu_p}_{\nu_1\dots\nu_p}\gamma_{\perp}^{\nu_1}\dots\gamma_{\perp}^{\nu_p}
\end{equation}
as a transverse equivalent of the antisymmetric gamma matrix of \cref{eq:antisymmetric-gamma}.
Once again we will only need to reduce tensors composed of transverse elements, so they can only depend on the transverse metric $g_\perp$ and the transverse gamma $\Gamma_\perp$.
As a result we can use the projectors from section \ref{sec:1fermion line} and \ref{sec:2fermion lines} after replacing $ \Gamma\rightarrow \Gamma_\perp,\ $ $g \rightarrow g_\perp$, and $D \rightarrow D_\perp$. 

\subsection{Self Energy integrals}\label{sec:selfenergy}
As an explicit example we shall now consider the important class of self-energy integrals, sometimes also known as propagator integrals or p-integrals, which depend on a single off-shell external momentum $Q$, i.e.\  $E=1$. This class of integrals forms a basis for multi-loop renormalization constants via  the $R^*$ operation \cite{Chetyrkin:1982nn,Chetyrkin:1984xa,Smirnov:1986me}. We normalise the momenta such that $Q^2=1$, so that
\begin{align}
    g^{\mu\nu}_\perp&= g^{\mu\nu}-Q^\mu Q^\nu,\\
    k^\mu &= k^\mu_\perp + k\cdot Q\, Q^\mu.
\end{align}
It is instructive to see the decomposition of a simple example into transverse and perpendicular parts.
We begin by decomposing all objects individually into parallel and perpendicular parts:
\begin{equation}
    \begin{split}
        k^\mu k^\nu  &= k^\mu_\perp  k^\nu_\perp + k\cdot Q \, (k^\mu_\perp  Q^\nu +k^\nu_\perp  Q^\mu) + (k\cdot Q)^2\,Q^\mu Q^\nu\\
    &\to k^\mu_\perp  k^\nu_\perp  + (k\cdot Q)^2\,Q^\mu Q^\nu,
    \end{split}
\end{equation}
where we have dropped the middle term as it has an odd number of transverse momenta, and therefore vanishes under integration over the loop momentum. The last term has no need for further reduction. Applying the relevant projector to the first term, we get
\begin{equation}
    \begin{split}
    k^\mu k^\nu  \rightarrow \quad& \left[g^{\mu\nu}_\perp (P_\perp)_{\alpha\beta}\right] k^\alpha_\perp k^\beta_\perp + (k\cdot Q)^2\,Q^\mu Q^\nu\\
    &= \left[ g^{\mu\nu}_\perp\frac{1}{D-1} (g_{\perp})_{\alpha\beta}\right]k^\alpha_\perp  k^\beta_\perp + (k\cdot Q)^2\,Q^\mu Q^\nu\\
    &= k_\perp \cdot k_\perp\frac{1}{D-1} g^{\mu\nu}_\perp + (k\cdot Q)^2\,Q^\mu Q^\nu
    \end{split}
\end{equation}
Here, $P_\perp^{\mu\nu}$ is the projector for $g^{\mu\nu}_\perp$.
Finally, we can substitute the definitions of the transverse quantities to arrive at a result in terms of the full loop momenta and the external momenta:
\begin{equation}
    k^\mu k^\nu \rightarrow  \frac{1}{D-1} \left(
    (k\cdot Q)^2-k\cdot k\right)g^{\mu\nu} +\frac{1}{D-1}\left(
      D(k\cdot Q)^2-k\cdot k\right)Q^\mu Q^\nu.
\end{equation}

Let us now turn to the transverse decomposition of $\gamma$-matrices. We note the following identity in the self-energy case:
\begin{equation}
    \gamma^{\mu} = \gamma^{\mu}_{\perp} + \gamma^{\mu}_{\parallel}
    = \gamma^{\mu}_{\perp}  + \slashed{Q} Q^\mu \,,
\end{equation}
from which follows
\begin{align}
\label{eq:zeroGamma}
    \gamma_{\mu} Q^\mu = \slashed{Q}= \gamma^{\mu}_{\parallel}Q_\mu \,,\qquad \gamma^{\mu}_{\perp}Q_\mu =0\,.
\end{align}
In the self-energy case we also have the following useful identities, which assist in decomposing $\Gamma$-tensors:
\begin{align}
\label{eq:gammatotrans}
    \Gamma^{\mu_1...\mu_n} &= \Gamma^{\mu_1...\mu_n}_\perp + \sum_{m=1}^n\slashed{Q}\,Q^{\mu_m}\, \Gamma^{\mu_{1}...\widehat{\mu_m}...\mu_{n}}_\perp (-1)^{m+1},\\
\label{eq:transtogamma}
    \Gamma^{\mu_1...\mu_n}_\perp &= \Gamma^{\mu_1...\mu_n} + \sum_{m=1}^n\,Q^{\mu_m}\, \Gamma^{\mu_1...\widehat{\mu_m}...\mu_n,Q} (-1)^{m+n+1},
\end{align}
with $\Gamma_\perp$ defined via \cref{eq:GAMMAperp}. In the above formulae the hat denotes an index that has been omitted. Furthermore, we have the identities,
\begin{align}
\label{eq:shuffelQslash}
    \slashed{Q}\Gamma^{\mu_1...\mu_n}_\perp &= \Gamma^{\mu_1...\mu_n Q}(-1)^n,\\
\label{eq:zeroAntiGamma}
    \Gamma^{\mu_{1}...Q...\mu_{n}}_\perp &= 0.
\end{align}
Derivations for \cref{eq:transtogamma,eq:gammatotrans,eq:shuffelQslash} are presented in \cref{sec:transgammaproof} whilst \cref{eq:zeroAntiGamma} follows directly from \cref{eq:zeroGamma}.

\section{$\gamma$-factorisation vs.\ Fermionic Projectors}
\label{sec:testing}
Using the methods presented in  \cref{sec:1fermion line,sec:2fermion lines,sec:noSpin} we calculated projectors with up to 32 Lorentz indices for the pure Lorentz case, up to 15 Lorentz indices with the one fermion line projectors, and up to $7$ external indices with $N_{1,2}<6$ for the two fermion line case. We provide these projectors in \textsc{Form} procedures in the ancillary files.

\begin{table}[t]
   \begin{minipage}[t]{0.49\textwidth}
      \centering
   \begin{tabular}[t]{ccc}
   \multicolumn{3}{c}{$T(n)= k_1^{\nu_1}k_2^{\nu_2}\Gamma^{k_3\,k_4\,\mu_1\dots\mu_{n}}$}               \\ \Xhline{3\arrayrulewidth}
   \multicolumn{1}{c|}{\multirow{2}{*}{$n$}} & \multicolumn{2}{c}{run-time (sec)}       \\ \cline{2-3}
      \multicolumn{1}{c|}{}                   & \multicolumn{1}{c|}{\makecell{Fermionic \\ Projectors}}     & \begin{minipage}{7em}\vspace{1ex}\centering $\gamma$-factorisation  \vspace{.5ex}\end{minipage} \\ \hline
      \multicolumn{1}{c|}{1}  &  \multicolumn{1}{c|}{0.003}&    0.004     \\
      \multicolumn{1}{c|}{2}  &  \multicolumn{1}{c|}{0.005}&    0.002      \\
      \multicolumn{1}{c|}{3}  &  \multicolumn{1}{c|}{0.003}&    0.003     \\
      \multicolumn{1}{c|}{4}  &  \multicolumn{1}{c|}{0.009}&    0.005      \\
      \multicolumn{1}{c|}{5}  &  \multicolumn{1}{c|}{0.036}&    0.003     \\
      \multicolumn{1}{c|}{6}  &  \multicolumn{1}{c|}{0.268}&   0.004      \\
      \multicolumn{1}{c|}{7}  &  \multicolumn{1}{c|}{2.489}&  0.005    \\
      \multicolumn{1}{c|}{8}  &  \multicolumn{1}{c|}{26.753}& 0.004    \\\Xhline{3\arrayrulewidth}
   \end{tabular}
   \caption{Table of run-times (sec) for the family of tensors $T(n)= k_1^{\nu_1}\dots k_4^{\nu_4}\slashed{k}_1\dots \slashed{k}_4\gamma^{\mu_1}\dots\gamma^{\mu_{n}}$ in the vacuum case.}
   \label{tab:onespin-changindex}
   \end{minipage}
   \hfill
   \begin{minipage}[t]{0.49\textwidth}
      \centering
   \begin{tabular}[t]{ccc}
   \multicolumn{3}{c}{$T(n)= k_1^{\nu_1}\dots k_{n-1}^{\nu_{n-1}}\,k_{n+1}^{\nu_{n}}\Gamma^{\mu_1\,\mu_2\, k_{1}\,\dots\, k_{n}}$}               \\ \Xhline{3\arrayrulewidth}
   \multicolumn{1}{c|}{\multirow{2}{*}{$n$}} & \multicolumn{2}{c}{run-time (sec)}       \\ \cline{2-3}
      \multicolumn{1}{c|}{}                   & \multicolumn{1}{c|}{\makecell{Fermionic \\ Projectors}}     & \begin{minipage}{7em}\vspace{1ex}\centering $\gamma$-factorisation  \vspace{.5ex}\end{minipage} \\ \hline
   \multicolumn{1}{c|}{1}  &  \multicolumn{1}{c|}{0.027}&  0.017 \\
   \multicolumn{1}{c|}{2}  &  \multicolumn{1}{c|}{0.024}&  0.018 \\
   \multicolumn{1}{c|}{3}  &  \multicolumn{1}{c|}{0.025}&  0.022 \\
   \multicolumn{1}{c|}{4}  &  \multicolumn{1}{c|}{0.030}&  0.118 \\
   \multicolumn{1}{c|}{5}  &  \multicolumn{1}{c|}{0.062}&  6.503 \\
   \multicolumn{1}{c|}{6}  &  \multicolumn{1}{c|}{0.340}&  1036 \\
   \multicolumn{1}{c|}{7}  &  \multicolumn{1}{c|}{3.107}&  56h \\
   \multicolumn{1}{c|}{8}  &  \multicolumn{1}{c|}{33.42}&  \dots \\\Xhline{3\arrayrulewidth}
   \end{tabular}
   \caption{Table of run-times (sec) for the family of tensors $T(n)= k_1^{\nu_1}k_2^{\nu_2}\Gamma^{k_3\,k_4\mu_1\dots\mu_{n}}_4$ in the vacuum case. The bottom right entry of the table is omitted because the expected length of the calculation was deemed impractical.}
   \label{tab:onespin-changemom}
   \end{minipage}
   \end{table}

For tensor integrals with spinor indices there are two ways to apply these projectors. The first is using the full range of fermionic projectors and a suitable completeness relation to decompose the integrand directly into its basis of tensors with fermion lines. We refer to this as the \emph{fermionic projectors} method.
The second method, which we refer to as \emph{$\gamma$-factorisation},  proceeds by factorising out the gamma structures, via (schematically)
\begin{equation}
\int d^D k \, \dots  \slashed{k}\dots \,{=}\, \gamma_\mu \int d^D k \, \dots k^\mu\dots  \,,
\end{equation}
and the tensor reduction is then performed on the pure Lorentz structure only.

To compare these two methods the tensor decomposition is performed for several families of tensors using both of these methods. These comparisons also lead to a strong check for the correctness of the various projectors we computed, since in all cases reported below we always checked explicitely that the results obtained in the two appraoches agree.

In the following we consider vacuum integrals, i.e.\ no external momentum. Further below we also consider one self-energy example with dependence on $Q$. Some results for one fermion line are shown in \cref{tab:onespin-changindex,tab:onespin-changemom}, where the difference in the methods becomes apparent. The code was run with \verb|tform| using 8 threads on an AMD EPYC 7532 32-Core processor, and we have subtracted the projector-table read-in time. In the family $T(n)= k_1^{\nu_1}k_2^{\nu_2}\Gamma^{k_3\,k_4\,\mu_1\dots\mu_{n}}$ (\cref{tab:onespin-changindex}) the number of external momenta remains fixed and  the number of Lorentz indices on gamma matrices increases. The $\gamma$-factorisation method produces consistent timings as it only applies the same 4 index projectors each time since there are only ever 4 loop momenta. Conversely, the fermionic projector increases with size as $n$ increases so the contraction with the integrand gets more complicated. This becomes noticeable from $n\ge5$, from where the time starts to increase roughly linearly with a scale factor of $\sim10$. Compare this to the family $T(n)= k_1^{\nu_1}\dots k_{n-1}^{\nu_{n-1}}\,k_{n+1}^{\nu_{n}}\Gamma^{\mu_1\,\mu_2\, k_{1}\,\dots\, k_{n}}$ (\cref{tab:onespin-changemom}) where the number of loop momenta increases with $n$.  The $\gamma$-factorisation method loses out to the fermionic projectors for $n>4$ as each of the added momenta raises the rank of the tensor to be reduced, whereas the full projector method is only ever reducing over 4 Lorentz indices. However, at $n=4$ this is already a 5-loop example so for most practical applications the $\gamma$-factorisation is still faster.

The pattern of where each method is faster continues with two-spin tensors. Consider \cref{tab:twospin-changeindex} which shows the run-time for tensors in the $T(n,m)=k_1^{\nu_1}k_1^{\nu_2}\slashed{k}_1\gamma^{\mu_1}\dots\gamma^{\mu_{n}}\otimes \slashed{k}_1\gamma^{\rho_1}\dots\gamma^{\rho_{m}}$ family of tensors. $n$ and $m$ can be varied independently and both increase  the number of Lorentz indices on gamma matrices without changing the number of loop momenta.
\begin{table}[t]
   \centering
   \begin{tabular}{ccccccccc}
   \multicolumn{9}{c}{$T(n,m)=k_1^{\nu_1}k_1^{\nu_2}\slashed{k}_1\gamma^{\mu_1}\dots\gamma^{\mu_{n}}\otimes \slashed{k}_1\gamma^{\rho_1}\dots\gamma^{\rho_{m}}$}       \\ \Xhline{3\arrayrulewidth}
   \multicolumn{1}{c|}{\multirow{2}{*}{\begin{minipage}{4em}\centering run-time (sec)\end{minipage}}}& \multicolumn{4}{c|}{\makecell{Fermionic Projectors}}& \multicolumn{4}{c}{$\gamma$-factorisation }\\ \cline{2-9}
   \multicolumn{1}{c|}{}  & $n=1$      & $n=2$      & $n=3$& \multicolumn{1}{c|}{$n=4$} &$n=1$      & $n=2$      & $n=3$& $n=4$\\ \hline
   \multicolumn{1}{c|}{$m=1$} &  0.016  & 0.043  & 0.158  & \multicolumn{1}{c|}{0.754} & 0.005& 0.004 & 0.011 & 0.011 \\
   \multicolumn{1}{c|}{$m=2$} &  0.046 & 0.176   & 0.993  & \multicolumn{1}{c|}{1.087} & 0.007& 0.011 & 0.012 & 0.021\\
   \multicolumn{1}{c|}{$m=3$} &  0.195  & 1.137 & 0.963  & \multicolumn{1}{c|}{8.565}& 0.012& 0.011 & 0.019 & 0.042 \\
   \multicolumn{1}{c|}{$m=4$} &  1.205 & 1.147 & 9.147  & \multicolumn{1}{c|}{ 6.686}& 0.017& 0.021 & 0.045 & 0.094 \\\Xhline{3\arrayrulewidth}
   \end{tabular}
   \caption{Table of run-times (sec) for the family of tensors $T(n,m)=k_1^{\nu_1}k_1^{\nu_2}\slashed{k}_1\gamma^{\mu_1}\dots\gamma^{\mu_{n}}\otimes \slashed{k}_1\gamma^{\rho_1}\dots\gamma^{\rho_{m}}$ in the vacuum case.}
\label{tab:twospin-changeindex}
   \end{table}
   As expected the fermionic projector method handles all the tensors well, though the time increases with both $n$ and $m$. On the other hand the $\gamma$-factorisation method has a fairly consistently low time for all the tensors as it only ever reduces the same rank 4 tensor. 
   
   \Cref{tab:twospin-changemom} shows the run-time for tensors in the family $T(n,m)=k_1^{\nu_1}k_1^{\nu_2}\gamma^{\mu_1}\slashed{k}_1\dots \slashed{k}_n\otimes \gamma^{\mu_2}\slashed{k}_1\dots \slashed{k}_m$, which we consider for the self-energy case with $Q$-dependence. In this case increasing $n$ or $m$ increases the number of loop momenta while keeping the rank of the tensor fixed.
   \begin{table}[h]
      \centering
      \begin{tabular}{ccccccccc}
      \multicolumn{9}{c}{$T(n,m)=k^{\nu_1}k^{\nu_2}\gamma^{\mu_1}\slashed{k}_1\dots \slashed{k}_n\otimes \gamma^{\mu_2}\slashed{k}_1\dots \slashed{k}_m$}       \\ \Xhline{3\arrayrulewidth}
      \multicolumn{1}{c|}{\multirow{2}{*}{\begin{minipage}{4em}\centering run-time (sec)\end{minipage}}}& \multicolumn{4}{c|}{\makecell{Fermionic Projectors}}& \multicolumn{4}{c}{$\gamma$-factorisation }\\ \cline{2-9}
      \multicolumn{1}{c|}{}  & $n=1$      & $n=2$      & $n=3$       & \multicolumn{1}{c|}{$n=4$} & $n=1$      & $n=2$      & $n=3$       & $n=4$ \\ \hline
      \multicolumn{1}{c|}{$m=1$} & 0.063 & 0.086 & 0.221 & \multicolumn{1}{c|}{0.563}  & 0.012 & 0.011 & 0.021 & 0.046 \\
      \multicolumn{1}{c|}{$m=2$} & 0.083 & 0.389 & 0.828 & \multicolumn{1}{c|}{3.029}  & 0.013 & 0.017 & 0.039 & 0.732 \\
      \multicolumn{1}{c|}{$m=3$} & 0.209 & 0.834 & 4.912 & \multicolumn{1}{c|}{12.92}  & 0.022 & 0.038 & 0.534 & 3.922 \\
      \multicolumn{1}{c|}{$m=4$} & 0.520 & 2.976 & 12.63 & \multicolumn{1}{c|}{60.39}  & 0.065 & 0.770 & 4.217 & 116.0 \\
      \Xhline{3\arrayrulewidth}
      \end{tabular}
      \caption{Table of run-times (sec) for the family of tensors $T(n,m)=k^{\nu_1}k^{\nu_2}\gamma^{\mu_1}\slashed{k}_1\dots \slashed{k}_n\otimes \gamma^{\mu_2}\slashed{k}_1\dots \slashed{k}_m$ in the self-energy case with $Q$-dependence.}
   \label{tab:twospin-changemom}
      \end{table}
As expected both methods are able to handle all the tensors in question and the fermionic projectors eventually become faster than the  $\gamma$-factorisation. However, this only happens at $n=m=4$ which is again a 5-loop example.

\FloatBarrier

\section{Conclusions}
\label{sec:conclusions}

In this paper we provide conceptual and algorithmic improvements, as well as extensions, to the projector-based approach to tensor reduction introduced for multi-loop vacuum Feynman integrals in refs.~\cite{Ruijl:2018poj,Herzog:2017ohr}. The efficiency of this approach derives from exploiting the symmetry properties of the projector. 
By building an ansatz for the projector that manifests these symmetries, the system of linear equations for the projector coefficients is greatly simplified, in comparison to Passarino-Veltmann-like approaches.

One goal that we achieve in this work is to spell out the group theory behind this approach, which is governed by what we call the \emph{orbit partition formula}.
More precisely, we describe the orbits into which the basis of tensors, which span a given projector, are partitioned under the action of its symmetry group.
For the vacuum case we set up a correspondence between these orbits and the cycle structure of certain bi-chord diagrams.
This allows us to determine a complete and minimal set of unknown coefficients, and their associated invariant tensors in the projector ansatz.
Each of these unknown coefficients, as well as their associated invariants, corresponds to an integer partition of $N/2$, with $N$ the rank of the tensor.
These integer partitions correspond exactly to the cycle structure of the associated bi-chord diagrams.
We compute the coefficients for projectors up to rank 32 and provide various optimisations for their implementation in a \FORM{} program.
We also find a convenient representation for the pure Lorentz projectors in terms of totally symmetric tensors.

We extend the methodology to tensor Feynman integrals containing one and two fermion lines. We find that although the structure of the projectors is somewhat more complicated, especially in the case of 2 fermion lines, the method can be employed more or less identically, with the different orbits again corresponding to certain diagrams. A major difference to the pure Lorentz case is that not all tensor structures are
related by an index permutation and so we require several structurally distinct projectors.
For one fermion line we compute projectors with up to 15 Lorentz indices, while for two fermion lines we compute projectors with up to 7 Lorentz indices.

In the application to fermion lines, we find it convenient to work in a basis of totally antisymmetric gamma matrices, which we denote by $\Gamma$.
We also provide a new relation, \cref{eq:gammatoGamma}, which allows one to pass into the anti-symmetric basis in a highly efficient way, that is particularly well optimised for \FORM. A useful property is that $\Gamma$s of different lengths are orthogonal under the trace. This restricts the basis for each projector to tensors with $\Gamma$s of the same length.
Due to the antisymmetry, one must also take into account relative signs in the invariant sums. With fewer symmetry relations in the basis of tensors, the power of the orbit partition formula gets somewhat reduced, i.e.\ there is less reduction in the size of the linear system of equations. This effect increases with more fermion lines. A general description of the orbit partition formula approach which captures all these complications is also provided in \cref{sec:method}.

Working in $D$ spacetime dimensions there is a major complication when considering tensors with two or more fermion lines. Namely, the set of possible vacuum tensors becomes infinite, since there can be an arbitrary number of contracted Lorentz indices between different fermion lines. We show that this problem is naturally circumvented by the orthogonality of the $\Gamma$-basis, both for the construction of a particular projector, as well as for the reduction of any two-fermion-line integrand.


We go on to describe how to extend the use of these vacuum projectors to problems with any number of external momenta. This is achieved through the decomposition of the underlying vector space into spaces transverse and parallel to the external momenta. To illustrate this we include a detailed description for  the case of a single off-shell external momentum $Q$, as is encountered in the calculation of self-energy or propagator integrals. The extension to more momenta is conceptually relatively straightforward as described in \cref{sec:external}. For the pure Lorentz case this has recently also been discussed in ref.\ \cite{Anastasiou:2023koq}, where an efficient way to generate the combinatorics of the external momenta was presented in terms of Wick contractions. Combining this with our new methodologies could be a promising avenue to explore in the future.


In \cref{sec:testing} we compare two methods for applying these projectors to integrands with fermion lines. The first of these is the fermionic projector method, which employs the projectors constructed in \cref{sec:1fermion line,sec:2fermion lines}. The second is the $\gamma$-factorisation method, where any loop momenta contracted with a $\gamma$-matrix are factored out, and the reduction is performed on the loop momenta only. The latter method therefore only uses the projectors of \cref{sec:noSpin}. An implementation of both of these methods in \FORM{} can be found in the ancillary files. 

The testing revealed that the $\gamma$-factorisation method is faster in situations where there are few loop momenta in the spinor part of the integrand. Each momentum that is factored out of the fermion line increases the tensor rank of the integrand, thus increasing the cost of the tensor reduction.
We note that, in the antisymmetric basis, each loop momentum can only be contracted into a given fermion line once.
Therefore, a problem where many loop momenta are contracted into the fermion line(s) can only occur at high loop numbers.
We see this reflected in \cref{tab:onespin-changemom,tab:twospin-changemom} where the fermionic projector method only outpaces the $\gamma$-factorisation at high loop number (more than 4).
On the other hand, we see in \cref{tab:onespin-changindex,tab:twospin-changeindex} that adding more indices to the $\Gamma$ necessitates the use of a higher-rank fermionic projector, but makes no difference to the $\gamma$-factorisation.



A version of the $\gamma$-factorisation algorithm has already been implemented in a novel $R^*$-algorithm to be published in a forthcoming paper. For challenging applications, e.g.\ with higher-dimensional operators, one requires high-rank tensor reduction as it involves Taylor expansion to the order of the superficial degree of divergence of the diagram. Similarly, we envision that our algorithms will prove particularly useful in the application of asymptotic expansions to higher orders, via the method of regions, in momentum space \cite{Smirnov:1990rz,Smn94,BnkSmn97,SmnRkmt99,Gardi:2022khw,Ma:2023hrt,Herzog:2023sgb}. However, as demonstrated in ref.\ \cite{Anastasiou:2023koq}, the method can also prove useful for multi-loop amplitude calculations with several external momenta.

As the number of fermion lines increases, the amount of symmetry in the orthogonality class of a given integrand decreases. This was already evident at two fermion lines, where there are several groups of basis tensors not related by an index permutation. With less symmetry in the problem, the power of the orbit partition formula is reduced, and so the number of coefficients needed for the projector may at some point become intractable. However, one would not expect to encounter a three fermion line problem before reaching a six point fermionic diagram, and so three fermion line projectors may be of limited use. We may, however, always employ the  $\gamma$-factorisation method to these cases. It would also be interesting to explore a hybrid approach for three or more fermion lines by factorising the momenta out of one of the fermion lines and using fermionic projectors for the remainder.

A promising avenue to improve the effectiveness of the projectors is the use of integrand symmetry explored in \cref{sec:nospin_integrandsymm}. It identifies terms in the tensor ansatz whose scalar coefficients are the same, before the projectors are applied. It thereby minimises the application of the projectors to the integrand. Further work would be required to extend these ideas to integrands with fermion lines. A fully automated implementation of the projectors in a \textsc{Form} program to reduce tensor integrals of arbitrary loop, tensor rank, with arbitrary number of fermion lines and number of external momenta (the \OPITER{} program), which also incorporates these integrand symmetries, is the topic of a publication in preparation \cite{opiter}.

\section*{Acknowledgements}
FH is supported by the UKRI FLF ``Forest Formulas for the LHC'' Mr/S03479x/1.
JG, FH, AK and ST are supported by the STFC Consolidated Grant ``Particle
Physics at the Higgs Centre''.

\appendix
\section{General Method}
\label{sec:method}
In this appendix we will rephrase our method of tensor reduction into a general language that does not refer to specific examples or kinds of vacuum tensors.  
In doing so, we uncover a general method for constructing projectors that encompasses all three of the cases we encountered in earlier sections.


Our starting point is a set of tensors with $N$ indices, which we denote by $t_i$, that we take as a basis for the vector space $V^{(N)}$. 
It is assumed that we have already gone through the procedure of ``factoring out'' equivalent tensors as we did in \cref{sec:noSpin,sec:1fermion line,sec:2fermion lines}, so that none of the basis tensors are related to each other by, for example, moving metric tensors around or permuting the indices on a $\Gamma$.
Our ansatz for the value of the integral is a general linear combination of these elements $I = \sum_i A_i t_i$.
In this section we will adopt an index-free notation as in the equation above. To show index contraction in the absence of explicit indices, we will use a central dot. For two basis elements $t$ and $t'$ which each have $N$ indices, we define
\begin{equation}\label{eq:method-contraction-example}
    t\cdot t' = t^{\mu_1\dots\mu_N} t'_{\mu_1\dots\mu_N}.
\end{equation}
Should the basis tensors possess some other kind of group indices, the central dot is understood as a trace over those indices as well. In other words, the result of the central dot operation should always be a scalar.
We use the symbol $P^{(i)}$ to denote the projector (or dual vector) for $t_i$ and require them to have the property
\begin{equation}\label{eq:method-projectorproperty}
    P^{(i)} \cdot t_j = 
    \begin{cases}
        1 & i = j \\
        0 & \mathrm{otherwise}
    \end{cases}.
\end{equation}

\subsection{Permutations and index symmetry}
The key simplification of the orbit partition formula happens because of the symmetry properties of the indices of the basis tensors. We can represent permutations on the indices by defining a group action on the tensors by some element of the permutation group $S_N$.
Consider the following example using a one fermion line tensor.
Using cycle notation to represent the transposition of $\mu_1$ and $\mu_2$, we can write
\begin{equation}
    (12).g^{\mu_2\mu_3}\gamma^{\mu_1} = g^{\mu_1\mu_3}\gamma^{\mu_2}.
\end{equation}
We should take care to make the distinction between the central dot which denotes index contraction and the lower one which represents the group action. Furthermore, we note that the transposition $(12)$ acts by swapping the indices $\mu_1$ and $\mu_2$, whatever their position on the tensor. It does \emph{not} correspond to swapping the first and second indices of the tensor.

In previous sections we established an equivalence relation between tensors that are equal up to a sign under index permutation,
such as swapping indices on a metric, exchanging the position of metrics, or permuting indices on a $\Gamma$ tensor.
In the following let us denote by $X$ a set of inequivalent basis tensors.
The choice of these ``representative'' tensors is arbitrary, the only restriction being that we should pick one from each equivalence class. We want the group action to be closed on the set of basis tensors, i.e.\ acting with $\sigma\in S_N$ on $x\in X$ should produce an element of $X$. This can always be achieved by rewriting the result of a permutation in terms of the representative in $X$ and will be understood in what follows.


In the case where we have to do an odd permutation of gamma indices to close the group action, the group action carries an additional minus sign.
To give an example, we have the following:
\begin{equation}
    (34).g^{\mu_1\mu_2}\Gamma^{\mu_3\mu_4} = -g^{\mu_1\mu_2}\Gamma^{\mu_3\mu_4},
\end{equation}
because we require an odd permutation of indices on the gamma to bring the intermediate result $g^{\mu_1\mu_2}\Gamma^{\mu_4\mu_3}$ back into the set of basis tensors.

Let us now discuss the effect of permutations on the dot product defined in \eqref{eq:method-contraction-example}.
Since the contracted indices are dummy indices the result of the contraction is unchanged if we do the same permutation on both sets of indices.
For two tensors $a$ and $b$, we thus have
\begin{equation}
    a \cdot b = (\sigma . a) \cdot (\sigma . b),
\end{equation}
for some $\sigma\in S_N$. If we let $\sigma. b = c$, we arrive at the useful property 
\begin{equation}\label{eq:method-contractionproperty}
    a \cdot (\sigma^{-1}.c) = (\sigma.a) \cdot c,
\end{equation}
which shows that we can take an index permutation across the contraction by inverting it.


The index permutations which correspond to symmetries of a given basis element $t_i$ form a subgroup of $S_N$ called the stabiliser subgroup, which we denote $H(t_i)$. Elements of $H(t_i)$ leave $t_i$ invariant (up to a sign in the case where we swap antisymmetric indices):
\begin{equation}\label{eq:method-stabiliser}
    H(t_i) = \left\{ h \in S_N | h.t_i = \pm t_i \right\}.
\end{equation} 
It is important that we keep track of which elements of $H(t_i)$ change the sign
of $t_i$. To do this we denote the sign in equation \eqref{eq:method-stabiliser} by $s_i(h)$ such that:
\begin{equation}
    h.t_i = s_i(h) \ t_i\,.
\end{equation}
In effect, this associates every element of $H(t_i)$ to a plus or minus sign.
We also note that
\begin{equation}\label{eq:method-inverse-sign}
    s_i(h) = s_i(h^{-1}),
\end{equation}
which follows since $s_i$ is a group homomorphism, i.e. $s_i(h g)=s_i(h)s_i(g)$, and the identity permutation has sign $+1$. 

We now define the \emph{orbit} corresponding to a particular $t_j\in X$, which consists
of all the elements of $X$ that can be reached by acting on $t_j$ with elements
of the group $H(t_i)$. We represent the orbit generated by acting on $t_j$ with the
symbol $X^{(i)}_j$:
\begin{equation}
    X^{(i)}_j = \left\{ h.t_j | h \in H(t_i) \right\}.
\end{equation}
In contrast to the $C_k$ of the main body of the paper which were sets of permutations, $X_j$ is a set of tensors.
The group action partitions the set of $t_i$s; each $t_i$ will belong to
exactly one orbit. It follows that one can have $X^{(i)}_j=X^{(i)}_k$ for
distinct $t_j$ and $t_k$ if they belong to the same orbit under the action of
$H(t_i)$.
It is important to keep this in mind when building our ansatz for the projector, because it will lead to some superfluous numerical factors.

The terms in a particular $X^{(i)}_j$ can be summed to create a quantity with the symmetries of $t_i$, as we will now demonstrate.
Consider the following sum of the terms in $X^{(i)}_j$, which throughout the text we have called $T$:
\begin{equation}\label{eq:method-oij1}
    T^{(i)}_j = \frac{1}{K} \sum_{h\in H(t_i)} s_i(h) \ h.t_j.
\end{equation}
The sign function $s_i(h)$ appears since, if two terms are related
by an \emph{antisymmetry} of the element $t_i$, we want them to appear with a
relative minus sign in the sum.
This is the same sign that appeared in \cref{eq:1spin-invariant-sum,eq:2spin-inv-sum}, but there is no need for a ``reference'' permutation, since the sum in this particular definition of the orbit is over elements of the stabiliser group $H(\sigma)$.
The overcounting factor $K$ will be explained
shortly. We can see that the definition above has the correct symmetry
properties by acting on both sides with a permutation $h'\in H(t_i)$:
\begin{align*}
    h'. T^{(i)}_j &= \frac{1}{K} \sum_{h\in H(t_i)} s_i(h) \ (h'h).t_j \\
    &= \frac{1}{K} \sum_{h''\in H(t_i)} s_i((h')^{-1}h'') \ h''.t_j \\
    &= s_i(h')\frac{1}{K} \sum_{h''\in H(t_i)} s_i(h'') \ h''.t_j \\
    &= s_i(h') T^{(i)}_j .
\end{align*}
In the second line we have re-indexed the sum in terms of $h''=h'h$, and have then used the fact that $s_i(h^{-1})=s_i(h)$ and also that the sign function factorises, $s_i(h'h'')=s_i(h')s_i(h'')$.
This allows us see that $T^{(i)}_j$ has the same symmetries as $t_i$; both are stabilised under the action of $H(t_i)$.
Since we want the projector for some term to have the same overall symmetry properties as that term, these invariant sums will form the building blocks of the projector.

Next we turn to the overcounting factor $K$, which accounts for the fact that sometimes the sum in equation \eqref{eq:method-oij1} will have an irrelevant overall numeric factor.
This happens when multiple elements of $H(t_i)$ map $t_j$ to the same basis tensor, or, in other words, when some elements of $H(t_i)$ are also index symmetries of $t_j$.
Alternatively, one could note that if the length of the orbit $X^{(i)}_j$ is smaller than the order of $H(t_i)$, the sum necessarily generates the same term multiple times.

We can use this last fact to determine the value of $K$, since the length of $X^{(i)}_j$ is related to the order of $H(t_i)$ through the orbit stabiliser theorem.
In our case, the order of $H(t_i)$ is the length of $X^{(i)}_j$ multiplied by the number of elements of $H(t_i)$ which are also symmetries of $t_j$.
This is a subgroup of $H(t_i)$ which we call $H(t_i,t_j)$. We then have
\begin{equation}\label{eq:method-orbitstabiliser}
    \big| H(t_i) \big| = \big| X^{(i)}_j \big| \big| H(t_i,t_j) \big|.
\end{equation}
We can write $H(t_i,t_j)$ explicitly as 
\begin{align*}
    H(t_i,t_j) &= \{h\in H(t_i) : h.t_j = \pm t_j\} \\
    &= H(t_i) \cap H(t_j).
\end{align*}
Each element in the sum \eqref{eq:method-oij1} appears $K=|H(t_i)|/|X^{(i)}_j|$ times. Then, using equation \eqref{eq:method-orbitstabiliser} we can write
\begin{equation}\
\label{eq:method-oij2}
\begin{split}
T^{(i)}_j &= \frac{1}{|H(t_i,t_j)|} \sum_{h\in H(t_i)} s_i(h) \ h.t_j\,,\\
              &=\sum_{h\in C_j(t_i)} s_i(h) \ h.t_j\,,
\end{split}
\end{equation}
where in the second line we have written the sum using the set $C_j(t_i)$ which generates each tensor in the orbit exactly once.

Some basis tensors have symmetry properties which we can use to exclude them automatically from the projector. For instance, if we are requiring the projector to be symmetric on two indices, we cannot include terms in the projector ansatz that are antisymmetric on those indices. The invariant sums corresponding to these terms are identically zero.

\subsection{Solving the projector}\label{sec:appendixsolving}
As we have stated previously, every tensor that could appear in the projector is accounted for in one of these orbits.
Therefore, a linear combination of all the unique orbit sums is the most general ansatz that involves every relevant basis element while also obeying the correct symmetry properties.
To ensure that we account for all the unique orbit sums without encountering duplicates,
we employ a subset of the $t_i$ which consists of one member from each orbit. We denote this subset by using a capitalised index, $t_I$, and call them \emph{orbit representatives}.
The choice of representatives is somewhat arbitrary; it doesn't matter which elements we choose as representatives as long as we choose one from each orbit. Our minimal ansatz for the projector is then 
\begin{equation}
    P^{(i)} = \sum_J c_J T^{(i)}_J\,,
\end{equation}
with $c_J$ being a coefficient (function of $D$ and $\trId$ in our case) to be determined. This is the general case of what we have called the orbit partition formula in previous sections.

Next, we consider how this ansatz can be made to satisfy the requirements in \eqref{eq:method-projectorproperty}.
It is not necessary to account for the contraction of the projector with every possible basis tensor, and we will demonstrate this as follows.
Consider two basis elements $t, t' \in X^{(i)}_j$. Since these basis elements are in the same orbit, there exists a group element $h\in H(t_i)$ such that $h.t'=t$. 
It can then be shown that $t$ and $t'$ give the same result, up to a sign, when contracted into any orbit $T^{(i)}_k$:
\begin{equation}
\begin{split}
    t' \cdot T^{(i)}_k &= (h^{-1}.t)\cdot T^{(i)}_k \\
    &= t\cdot (h.T^{(i)}_k) \\
    &= s_i(h) \ t \cdot T^{(i)}_k.
\end{split}
\end{equation}
In the second line we have used the property \eqref{eq:method-contractionproperty}.
This then extends straightforwardly to the projector which is just a linear combination of orbits:
\begin{equation}
    t' \cdot P^{(i)} = s_i(h) \ t \cdot P^{(i)}.
\end{equation}
It follows that if the contraction of the projector with a representative element of some orbit is zero, the contraction with any other element of that orbit will automatically also be zero.

With this in mind, the minimal set of requirements that the projector has to satisfy is as follows:
\begin{equation}
    \begin{split}
        P^{(i)} \cdot t_J &= 
        \begin{cases}
            1 & t_i = t_J \\
            0 & \mathrm{otherwise}
        \end{cases} \\
        &= \delta^{(i)}_J.
    \end{split}
\end{equation}
The symbol $\delta^{(i)}_J$ can be viewed as a normal vector in the space of representatives that points in the $i$-direction.
Substituting in our ansatz for the projector, we get 
\begin{equation}
    \left( \sum_K c_K T^{(i)}_K \right) 
    \cdot t_J = \delta^{(i)}_J.
\end{equation}
This can be rewritten as a matrix equation,
\begin{equation}
    \sum_K M^{(i)}_{JK} c_K = \delta^{(i)}_J,\quad \text{where} \quad
    M^{(i)}_{JK} = t_J \cdot T^{(i)}_K \,.
\end{equation}
This is a system of $\Lambda$ equations in $\Lambda$ unknowns, where $\Lambda$ is the number of representatives (i.e.\ the number of orbits).
Inverting the matrix yields the unknown coefficients $c_J$:
\begin{equation}
    c_J = \sum_K \left( M^{(i)} \right)^{-1}_{JK} \delta^{(i)}_K.
\end{equation}
Once we have solved this system of equations, we can get further projectors for free by permuting the indices as follows. Given a permutation $\sigma\in S_N$ such that $\sigma.t_i=t_k$ then $P^{(k)}=\sigma.P^{(i)}$. This follows since:
\begin{align}
\label{eq:method-freeprojector}
    (\sigma.P^{(i)}) \cdot t_j &=
    P^{(i)} \cdot (\sigma^{-1}.t_j) \nn\\
    &= \begin{cases}
        1 & t_i = \sigma^{-1} . t_j \quad \Rightarrow \quad t_k=t_j \ ,\\
        0 & \mathrm{otherwise}
     \end{cases} \\
    &= P^{(k)} \cdot t_j \nn\ .
\end{align}
with the last line following from the second, since it uniquely defines the projector.

\section{Proof of the Gamma matrix reduction formula}
\label{sec:gammaproof}
We wish to prove the relationship: 
\begin{equation}
    \label{eq:gamma2GAMMA_proof}
    \gamma^{\mu_1}\dots\gamma^{\mu_n}=\sum_{k=0}^{n}\sum_{\pi\in \Sigma_n^k}\sgn(\pi)\Gamma^{\mu_{\pi{(1)}}\dots\mu_{\pi(k)}}\tr(\gamma^{\mu_{\pi(k+1)}}\dots\gamma^{\mu_{\pi(n)}})\,,
\end{equation}
where the sum over $\Sigma_{n}^k$ shuffles the first $k$ indices with the remaining $n-k$ indices over the two tensors presented in \cref{sec:1fermion line}.

In $D$ dimensions we can express the trace of an arbitrary string of gamma matrices as 
\begin{equation}\label[]{eq:trace1}
   \begin{split}
     \tr\left(\gamma_{\mu_1} \dots \gamma_{\mu_{2 m+1}}\right)&=0, \\ \tr\left(\gamma_{\mu_1} \dots \gamma_{\mu_{2 m}}\right)&=\frac{1}{2^m m !} \sum_{p \in {S}_{2 m}} \operatorname{sgn}(p)\left(\prod_{k=1}^m \delta_{\mu_{p(k)}, \mu_{p(m+k)}}\right)\trId,
   \end{split}
\end{equation}
where $\sgn(p)=\pm 1$, is the sign of the permutation. Each independent term in the sum is generated by several permutations, the  factor of $2^m m!$ corrects this overcounting. The sum over  $S_{2 m}$ then has $(2m-1)!! = \frac{(2m)!}{2^m m!}$. As we have seen the antisymmetric gamma matrices are orthogonal under traces following from equation \ref{eq:trace1} this orthogonality can be expressed as 
\begin{equation}
    \tr\left(\Gamma_{\nu_1\dots\nu_j}\Gamma^{\rho_1\dots\rho_k}\right)= k! \delta_{jk} \delta^{\rho_k}_{[\nu_1}\dots \delta^{\rho_1}_{\nu_k]}\trId.
\end{equation}
We now wish to express a generic product of gamma matrices in the antisymmetric basis,
\begin{equation}
    \gamma_{\mu_1} \dots \gamma_{\mu_{n}}= \sum^n_{k=0}c^{\nu_1\dots\nu_k}_{\mu_1\dots\mu_n}\Gamma_{\nu_k\dots\nu_1},
\end{equation}
where our prescription for ordering the indices follows \cite{Cvitanovic:1982bq}.
The coefficient $c$ is antisymmetric in its upper indices and can be determined from orthogonality,
\begin{align}
        \tr\left(\gamma_{\mu_1}\dots\gamma_{\mu_n}\Gamma^{\rho_1\dots\rho_j}\right)&=\sum^n_{k=0}c^{\nu_1\dots\nu_k}_{\mu_1\dots\mu_n}\tr\left(\Gamma^{}_{\nu_1\dots\nu_k}\Gamma^{\rho_1\dots\rho_j}\right) \nonumber\\
    &=\sum^n_{k=0}c^{\nu_1\dots\nu_k}_{\mu_1\dots\mu_n}k! \delta_{jk} \delta^{\rho_k}_{[\nu_1}\dots \delta^{\rho_1}_{\nu_k]}\trId\nonumber\\
    &=j!c^{\rho_j\dots\rho_1}_{\mu_1\dots\mu_n}\trId.
\end{align}
The coefficients vanish if $j>\min(n,D)$ or $ n+j$ is odd. Thus, our string of gamma matrices can be expressed as 
\begin{equation} \label{eq:untraced}
    \gamma_{\mu_1} \dots \gamma_{\mu_{n}}= \frac{1}{\trId}\sum_{k=0}^n\frac{1}{k!}\tr\left(\gamma_{\mu_1}\dots\gamma_{\mu_n}\Gamma^{\nu_1\dots\nu_k}\right)\Gamma_{\nu_1\dots\nu_k}.
\end{equation}
We now need to evaluate the trace, with some relabelling it becomes
    \begin{equation}\label{eq:traces}
        \begin{split}
            \tr&\left(\gamma_{\mu_1}\dots\gamma_{\mu_n}\Gamma^{\nu_1\dots\nu_k}\right)\Gamma_{\nu_1\dots\nu_k}=\tr\left(\gamma_{\mu_1}\dots\gamma_{\mu_{n+k}}\right)\Gamma^{\mu_{n+1}\dots\mu_{n+k}}\\
            =&\begin{cases}
                \displaystyle\frac{1}{2^m m !} \sum_{p \in S_{2 m}} \operatorname{sgn}(p)\left(\prod_{j=1}^m \delta_{\mu_{p(j)}, \mu_{p(m+j)}}\right)\Gamma^{\mu_{n+1}\dots\mu_{n+k}}\trId, & \text{$n+k$ even}\\
                0, & \text{$n+k$ odd}
            \end{cases} 
        \end{split}
    \end{equation} 
where now $2m=n+k$.
We now wish to reduce the sum over ${S}_{n+k}$ to one over only the independent terms. Each element $q \in{S}_{n+k}$ partitions the set $\{1,\dots,n+k\}$ into disjoint sets $\{q(1),....,q(2k)\}$ and $\{q(2k+1),...,q(n+k)\}$. This corresponds to the division of indices between the deltas and gamma in equation \ref{eq:traces}. We define an equivalence relation $p\sim q$ on permutations $p,q\in {S}_{n+k}$ when they produce the same partition. An equivalence class $[q]\in {S}_{n+k}/\sim={S}_{n+k}/({S}_{2k}\times{S}_{n-k})$ contains $(2k)!(n-k)!$ permutations and there are $\binom{n+k}{2k}$ classes. We define a signum of each partition as $\sgn([q])=\sgn(q_0)$ where $q_0\in[q]$ is the permutation for which $q_0(1)<\cdots<q_0(2k)$ and $q_0(2k+1)<\cdots<q_0(n+k)$.
After contracting the indices, \cref{eq:traces} reduces to (in the case $n+k$ is even)
 \begin{equation}
  \begin{split}
      \tr\left(\gamma_{\mu_1}\dots\gamma_{\mu_n}\Gamma^{\nu_1\dots\nu_k}\right)&\Gamma_{\nu_1\dots\nu_k}\\ &= r\hspace{-2em}\sum_{\substack{[q]\in\\{S}_{n+k}/({S}_{2k}\times{S}_{n-k})}}\hspace{-2em} \sgn([q])\tr(\gamma_{\mu_{q(1)}}\dots\gamma_{\mu_{q(n-k)}})\Gamma_{\mu_{q(n-k+1)}\dots\mu_{q(n)}},
  \end{split}
 \end{equation}
 where the sum is reduced to one over the equivalence classes $[q]$ and $r$ is a combinatorial factor. It is given by 
 \begin{equation}
    r=(2k)!(n-k)!\ \cdot\ \frac{2^{m-k}(m-k)!}{(n-k)!}\ \cdot\ \frac{2^k(k!)^2}{(2k)!}\ \cdot\ \binom{m}{k}\ \cdot\ \frac{1}{2^mm!}\ =k!.
 \end{equation} 
The $(2k)!(n-k)!$ is the number of permutations in each partition. $\frac{2^{m-k}(m-k)!}{(2(m-k))!}=\frac{2^{m-k}(m-k)!}{(n-k)!}$ is the normalisation of the permutations that create equivalent traces. The term $ \frac{2^k(k!)^2}{(2k)!}$ normalises the permutations that chose which indices end up in the antisymmetric tensor the extra factor of $k!$ accounting for the antisymmetry. Finally, the $ \frac{1}{2^mm!}$ is just the factor from equation \ref{eq:traces}. Inserting this into equation \ref{eq:untraced} we obtain 
\begin{equation}
    \gamma_{\mu_1} \dots \gamma_{\mu_{n}}=\frac{1}{\trId}\sum_{k=0}^{n} \sum_{\substack{[q]\in\\{S}_{n+k}/({S}_{2k}\times{S}_{n-k})}} \hspace{-2em}\sgn([q])\tr(\gamma_{\mu_{q(1)}}\dots\gamma_{\mu_{q(n-k)}})\Gamma_{\mu_{q(n-k+1)}\dots\mu_{q(n)}},
\end{equation}
noting that this sum is equivalent to the formulation in terms of shuffles (\cref{eq:gammatoGamma,eq:gamma2GAMMA_proof}).

\section{Further tables for the two fermion lines system sizes}
\subsection*{Number of non-orthogonal independent tensors}\label{sec:2spintable}
\begin{minipage}{.45\linewidth}
    \begin{table}[H]
        \centering
        \resizebox{0.65\width}{!}{
    \begin{tabular}{c c c c c c}
        \Xhline{3\arrayrulewidth}
        \vspace{-3ex}\\
        \fbox{$|V^{(N=2)}_{N_1,N_2}|$} &$N_1=1$ &$N_1=2$ &$N_1=3$ &$N_1=4$&$N_1=5$\\
        $N_2=1$& \begin{tabular}{@{}c@{}}\textbf{3} \end{tabular}& \begin{tabular}{@{}c@{}}\textbf{-} \end{tabular}&\begin{tabular}{@{}c@{}}\textbf{1} \end{tabular} &\begin{tabular}{@{}c@{}}\textbf{-} \end{tabular} &\begin{tabular}{@{}c@{}}\textbf{-} \end{tabular}\\
         $N_2=2$& \begin{tabular}{@{}c@{}}\textbf{-}\end{tabular} &\begin{tabular}{@{}c@{}}\textbf{3} \end{tabular}  & \begin{tabular}{@{}c@{}}\textbf{-} \end{tabular} & \begin{tabular}{@{}c@{}}\textbf{1} \end{tabular}&\begin{tabular}{@{}c@{}}\textbf{-}\end{tabular}\\
         $N_2=3$&\begin{tabular}{@{}c@{}}\textbf{1}\end{tabular} &\begin{tabular}{@{}c@{}}\textbf{-}\end{tabular} & \begin{tabular}{@{}c@{}}\textbf{3} \end{tabular}& \begin{tabular}{@{}c@{}}\textbf{-} \end{tabular}&\begin{tabular}{@{}c@{}}\textbf{1}\end{tabular}\\
         $N_2=4$&\begin{tabular}{@{}c@{}}\textbf{-}\end{tabular} & \begin{tabular}{@{}c@{}}\textbf{1}\end{tabular}& \begin{tabular}{@{}c@{}}\textbf{-}\end{tabular}& \begin{tabular}{@{}c@{}}\textbf{3} \end{tabular}&\begin{tabular}{@{}c@{}}\textbf{-}\end{tabular}\\
         $N_2=5$&\begin{tabular}{@{}c@{}}\textbf{-}\end{tabular} & \begin{tabular}{@{}c@{}}\textbf{-}\end{tabular}& \begin{tabular}{@{}c@{}}\textbf{1}\end{tabular}& \begin{tabular}{@{}c@{}}\textbf{-} \end{tabular}&\begin{tabular}{@{}c@{}}\textbf{3}\end{tabular}\vspace{1ex}\\
         \Xhline{3\arrayrulewidth}
    \end{tabular}}
    \caption{A table enumerating the number of independent tensors in the orthogonal subspace for $N=2$ and various values of $N_1,\ N_2$. }
    \label{tab:2spinCount3n}
\end{table}
\end{minipage}
\hfill
\begin{minipage}{.45\linewidth}

    \begin{table}[H]
        \centering
        \resizebox{0.65\width}{!}{
            \begin{tabular}{c c c c c c}
                \Xhline{3\arrayrulewidth}
                \vspace{-3ex}\\
        \fbox{$|V^{(N=3)}_{N_1,N_2}|$} &$N_1=1$ &$N_1=2$ &$N_1=3$ &$N_1=4$&$N_1=5$\\
        $N_2=1$& \textbf{-} & \textbf{6} &\textbf{-} &\textbf{1} &\textbf{-} \\
         
         $N_2=2$&  \textbf{6} & \textbf{-} &\textbf{6} &\textbf{-} &\textbf{1} \\
         
         $N_2=3$& \textbf{-} & \textbf{6} &\textbf{-} &\textbf{6} &\textbf{-} \\
         
         $N_2=4$& \textbf{1} & \textbf{-} &\textbf{6} &\textbf{-} &\textbf{6} \\
         
         $N_2=5$&\textbf{-} & \textbf{1} &\textbf{-} &\textbf{6} &\textbf{-}\vspace{1ex}\\
         \Xhline{3\arrayrulewidth}
    \end{tabular}}
    \caption{A table enumerating the number of independent tensors in the orthogonal subspace for $N=3$ and various values of $N_1,\ N_2$. }
    \label{tab:2spinCount4n}
\end{table}
\end{minipage}
\begin{minipage}{.45\linewidth}

    \begin{table}[H]
        \centering
        \resizebox{0.65\width}{!}{
     \begin{tabular}{c c c c c c}
        \Xhline{3\arrayrulewidth}
        \vspace{-3ex}\\
        \fbox{$|V^{(N=5)}_{N_1,N_2}|$} &$N_1=1$ &$N_1=2$ &$N_1=3$ &$N_1=4$&$N_1=5$\\
        $N_2=1$& \textbf{-} & \textbf{45} &\textbf{-} &\textbf{15} &\textbf{-} \\
         
         $N_2=2$&  \textbf{45} & \textbf{-} &\textbf{55} &\textbf{-} &\textbf{15} \\
         
         $N_2=3$& \textbf{-} & \textbf{55} &\textbf{-} &\textbf{55} &\textbf{-} \\
         
         $N_2=4$& \textbf{15} & \textbf{-} &\textbf{55} &\textbf{-} &\textbf{55} \\
         
         $N_2=5$&\textbf{-} & \textbf{15} &\textbf{-} &\textbf{55} &\textbf{-}\vspace{1ex}\\
         \Xhline{3\arrayrulewidth}
    \end{tabular}}
    \caption{A table enumerating the number of independent tensors in the orthogonal subspace for $N=5$ and various values of $N_1,\ N_2$. }
    \label{tab:2spinCount3n}
\end{table}
\end{minipage}
 \hfill
\begin{minipage}{.45\linewidth}
    \begin{table}[H]
        \centering
        \resizebox{0.65\width}{!}{
     \begin{tabular}{c c c c c c}
        \Xhline{3\arrayrulewidth}
        \vspace{-3ex}\\
        \fbox{$|V^{(N=6)}_{N_1,N_2}|$} &$N_1=1$ &$N_1=2$ &$N_1=3$ &$N_1=4$&$N_1=5$\\
        $N_2=1$& \textbf{105} & \textbf{-} &\textbf{105} &\textbf{-} &\textbf{21} \\
         
         $N_2=2$& \textbf{-} & \textbf{195} &\textbf{-} &\textbf{120} &\textbf{-} \\
         
         $N_2=3$& \textbf{105} & \textbf{-} &\textbf{215} &\textbf{-} &\textbf{120} \\
         
         $N_2=4$&\textbf{-} & \textbf{120} &\textbf{-} &\textbf{215} &\textbf{-} \\
         
         $N_2=5$&\textbf{21} & \textbf{-} &\textbf{120} &\textbf{-} &\textbf{215}\vspace{1ex}\\
         \Xhline{3\arrayrulewidth}
    \end{tabular}}
    \caption{A table enumerating the number of independent tensors in the orthogonal subspace for $N=6$ and various values of $N_1,\ N_2$. }
    \label{tab:2spinCount4n}
\end{table}
\end{minipage}
\begin{minipage}{.45\linewidth}
    \begin{table}[H]
        \resizebox{0.65\width}{!}{
        \centering
     \begin{tabular}{c c c c c c}
        \Xhline{3\arrayrulewidth}
        \vspace{-3ex}\\
        \fbox{$|V^{(N=7)}_{N_1,N_2}|$} &$N_1=1$ &$N_1=2$ &$N_1=3$ &$N_1=4$&$N_1=5$\\
        $N_2=1$& \textbf{-} & \textbf{420} &\textbf{-} &\textbf{210} &\textbf{-} \\
         
         $N_2=2$&  \textbf{420} & \textbf{-} &\textbf{630} &\textbf{-} &\textbf{231} \\
         
         $N_2=3$& \textbf{-} & \textbf{630} &\textbf{-} &\textbf{665} &\textbf{-} \\
         
         $N_2=4$& \textbf{210} & \textbf{-} &\textbf{665} &\textbf{-} &\textbf{665} \\
         
         $N_2=5$&\textbf{-} & \textbf{231} &\textbf{-} &\textbf{665} &\textbf{-}\vspace{1ex}\\
         \Xhline{3\arrayrulewidth}
    \end{tabular}}
    \caption{A table enumerating the number of independent tensors in the orthogonal subspace for $N=7$ and various values of $N_1,\ N_2$. }
    \label{tab:2spinCount3n}
\end{table}
\end{minipage}
\hfill
\begin{minipage}{.45\linewidth}
    \begin{table}[H]
        \centering
        \resizebox{0.65\width}{!}{
     \begin{tabular}{c c c c c c}
        \Xhline{3\arrayrulewidth}
        \vspace{-3ex}\\
        \fbox{$|V^{(N=8)}_{N_1,N_2}|$} &$N_1=1$ &$N_1=2$ &$N_1=3$ &$N_1=4$&$N_1=5$\\
        $N_2=1$& \textbf{945} & \textbf{-} &\textbf{1260} &\textbf{-} &\textbf{378} \\
         
         $N_2=2$& \textbf{-} & \textbf{2205} &\textbf{-} &\textbf{1680} &\textbf{-} \\
         
         $N_2=3$& \textbf{1260} & \textbf{-} &\textbf{2765} &\textbf{-} &\textbf{1736} \\
         
         $N_2=4$&\textbf{-} & \textbf{1680} &\textbf{-} &\textbf{2835} &\textbf{-} \\
         
         $N_2=5$&\textbf{378} & \textbf{-} &\textbf{1736} &\textbf{-} &\textbf{2835}\vspace{1ex}\\
         \Xhline{3\arrayrulewidth}
    \end{tabular}}
    \caption{A table enumerating the number of independent tensors in the orthogonal subspace for $N=8$ and various values of $N_1,\ N_2$. }
    \label{tab:2spinCount4n}
\end{table}
\end{minipage}
\begin{minipage}{.45\linewidth}
    \begin{table}[H]
        \resizebox{0.65\width}{!}{
        \centering
     \begin{tabular}{c c c c c c}
        \Xhline{3\arrayrulewidth}
        \vspace{-3ex}\\
        \fbox{$|V^{(N=9)}_{N_1,N_2}|$} &$N_1=1$ &$N_1=2$ &$N_1=3$ &$N_1=4$&$N_1=5$\\
        $N_2=1$& \textbf{-} & \textbf{4725} &\textbf{-} &\textbf{3150} &\textbf{-} \\
         
         $N_2=2$&  \textbf{4725} & \textbf{-} &\textbf{8505} &\textbf{-} &\textbf{3906} \\
         
         $N_2=3$& \textbf{-} & \textbf{8505} &\textbf{-} &\textbf{9765} &\textbf{-} \\
         
         $N_2=4$& \textbf{3150} & \textbf{-} &\textbf{9765} &\textbf{-} &\textbf{9891} \\
         
         $N_2=5$&\textbf{-} & \textbf{9891} &\textbf{-} &\textbf{3906} &\textbf{-} \\
     \Xhline{3\arrayrulewidth}
    \end{tabular}
    }
    \caption{A table enumerating the number of independent tensors in the orthogonal subspace for $N=9$ and various values of $N_1,\ N_2$. }
\end{table}
\end{minipage}
\hfill
\begin{minipage}{.45\linewidth}
    \begin{table}[H]
        \centering
        \resizebox{0.65\width}{!}{
     \begin{tabular}{c c c c c c}
        \Xhline{3\arrayrulewidth}
        \vspace{-3ex}\\
        \fbox{$|V^{(N=10)}_{N_1,N_2}|$} &$N_1=1$ &$N_1=2$ &$N_1=3$ &$N_1=4$&$N_1=5$\\
        $N_2=1$& \textbf{10395} & \textbf{-} &\textbf{17325} &\textbf{-} &\textbf{6930} \\
         
         $N_2=2$& \textbf{-} & \textbf{29295} &\textbf{-} &\textbf{26775} &\textbf{-} \\
         
         $N_2=3$& \textbf{17325} & \textbf{-} &\textbf{41895} &\textbf{-} &\textbf{29295} \\
         
         $N_2=4$&\textbf{-} & \textbf{26775} &\textbf{-} &\textbf{45045} &\textbf{-} \\
         
         $N_2=5$&\textbf{6930} & \textbf{-} &\textbf{29295} &\textbf{-} &\textbf{45297}\vspace{1ex}\\
         \Xhline{3\arrayrulewidth}
    \end{tabular}}
    \caption{A table enumerating the number of independent tensors in the orthogonal subspace for $N=10$ and various values of $N_1,\ N_2$. }
    \label{tab:2spinCount4n}
\end{table}
\end{minipage}
\subsection*{Number of orbits}\label{sec:2spintable2}


\begin{minipage}{.47\linewidth} 
    \begin{table}[H]
        \centering
        \resizebox{0.6\width}{!}{
            \begin{tabular}{c  c c c cc}
                \Xhline{3\arrayrulewidth}
                 \vspace{-3ex}\\
                \fbox{\begin{minipage}{5em}\centering $|V^{(N=2)}_{N_1,N_2}|$\\ $\{\Lambda_{\vec{n}}\}_{\vec n}$
                \end{minipage}} &$N_1=1$ &$N_1=2$ &$N_1=3$ &$N_1=4$&$N_1=5$\\
                $N_2=1$& \makecell{\textbf{3}\\3,2}& \textbf{-}&\makecell{\textbf{1}\\1}  &\textbf{-} &\textbf{-}\\
                 
                 $N_2=2$& \textbf{-} &\makecell{\textbf{3}\\3,2 }  & \makecell{\textbf{-} } & \makecell{\textbf{1}\\1 }&\makecell{\textbf{-}}\\
                 
                 $N_2=3$&\makecell{\textbf{1}\\1} &\makecell{\textbf{-}} & \makecell{\textbf{3}\\3,2 }& \makecell{\textbf{-} }&\makecell{\textbf{1}\\1}\\
                 
                 $N_2=4$&\makecell{\textbf{-}} & \makecell{\textbf{1}\\1}& \makecell{\textbf{-}}& \makecell{\textbf{3}\\3,2 }&\makecell{\textbf{-}}\\
               
                 $N_2=5$&\makecell{\textbf{-}} & \makecell{\textbf{-}}& \makecell{\textbf{1}\\1}& \makecell{\textbf{-} }&\makecell{\textbf{3}\\3,2}\vspace{1ex}\\
                 \Xhline{3\arrayrulewidth}
            \end{tabular}
    }
    \caption{A table enumerating the number of tensors which are not orthogonal to a tensor with $N=3$ and various values of $N_1,\ N_2$. The number of orbits for each projector is enumerated below the total number of tensors.}
    \label{tab:2spinCount3n}
\end{table}
\end{minipage}
\hfill
\begin{minipage}{.47\linewidth}

    \begin{table}[H]
        \centering
        \resizebox{0.6\width}{!}{
            \begin{tabular}{c c c c c c}
                \Xhline{3\arrayrulewidth}
                \vspace{-3ex}\\
                \fbox{\begin{minipage}{5em}\centering $|V^{(N=3)}_{N_1,N_2}|$\\ $\{\Lambda_{\vec{n}}\}_{\vec n}$
                \end{minipage}} &$N_1=1$ &$N_1=2$ &$N_1=3$ &$N_1=4$&$N_1=5$\\
                
                $N_2=1$& \makecell{\textbf{-} }& \makecell{\textbf{6}\\3,3 }&\makecell{\textbf{-} }&\makecell{\textbf{1}\\1 }&\makecell{\textbf{-} }\\
                 
                 $N_2=2$&  \makecell{\textbf{6}\\3,3 }& \makecell{\textbf{-} }&\makecell{\textbf{6}\\3,3 }&\makecell{\textbf{-} }&\makecell{\textbf{1}\\1 }\\
                 
                 $N_2=3$& \makecell{\textbf{-} }& \makecell{\textbf{6}\\3,3 }&\makecell{\textbf{-} }&\makecell{\textbf{6}\\3,3 }&\makecell{\textbf{-} }\\
                 
                 $N_2=4$& \makecell{\textbf{1}\\1 }& \makecell{\textbf{-} }&\makecell{\textbf{6}\\3,3 }&\makecell{\textbf{-} }&\makecell{\textbf{6}\\3,3 }\\
                 
                 $N_2=5$&\makecell{\textbf{-} }& \makecell{\textbf{1}\\1 }&\makecell{\textbf{-} }&\makecell{\textbf{6}\\3,3 }&\makecell{\textbf{-} }\vspace{1ex}\\
                 \Xhline{3\arrayrulewidth}
            \end{tabular}
    }
    \caption{A table enumerating the number of tensors which are not orthogonal to a tensor with $N=4$ and various values of $N_1,\ N_2$. The number of orbits for each projector is enumerated below the total number of tensors. }
    \label{tab:2spinCount4n}
\end{table}
\end{minipage}
\begin{minipage}{.47\linewidth}

    \begin{table}[H]
        \centering
        \resizebox{0.6\width}{!}{
    \begin{tabular}{c c c c c c}
        \Xhline{3\arrayrulewidth}
        \vspace{-3ex}\\
        \fbox{\begin{minipage}{5em}\centering $|V^{(N=5)}_{N_1,N_2}|$\\ $\{\Lambda_{\vec{n}}\}_{\vec n}$
        \end{minipage}}&$N_1=1$ &$N_1=2$ &$N_1=3$ &$N_1=4$&$N_1=5$\\
        
        $N_2=1$& \makecell{\textbf{-} }& \makecell{\textbf{45}\\11,7 }&\makecell{\textbf{-} }&\makecell{\textbf{15}\\3,3 }&\makecell{\textbf{-} }\\
         
         $N_2=2$&  \makecell{\textbf{45}\\11,7  }& \makecell{\textbf{-} }&\makecell{\textbf{55}\\6,13,8 }&\makecell{\textbf{-} }&\makecell{\textbf{15}\\3,3 }\\
         
         $N_2=3$& \makecell{\textbf{-} }& \makecell{\textbf{55}\\6,13,8 }&\makecell{\textbf{-} }&\makecell{\textbf{55}\\6,13,8 }&\makecell{\textbf{-} }\\
         
         $N_2=4$& \makecell{\textbf{15}\\3,3 }& \makecell{\textbf{-} }&\makecell{\textbf{55}\\6,13,8 }&\makecell{\textbf{-} }&\makecell{\textbf{55} \\6,13,8}\\
         
         $N_2=5$&\makecell{\textbf{-} }& \makecell{\textbf{15}\\3,3 }&\makecell{\textbf{-} }&\makecell{\textbf{55}\\6,13,8 }&\makecell{\textbf{-} }\vspace{1ex}\\
         \Xhline{3\arrayrulewidth}
    \end{tabular}}
    \caption{A table enumerating the number of tensors which are not orthogonal to a tensor with $N=5$ and various values of $N_1,\ N_2$. The number of orbits for each projector is enumerated below the total number of tensors.}
    \label{tab:2spinCount3n}
\end{table}
\end{minipage}
 \hfill
\begin{minipage}{.47\linewidth}
    \begin{table}[H]
        \centering
        \resizebox{0.6\width}{!}{
    \begin{tabular}{c c c c c c}
        \Xhline{3\arrayrulewidth}
        \vspace{-3ex}\\
        \fbox{\begin{minipage}{5em}\centering $|V^{(N=6)}_{N_1,N_2}|$\\ $\{\Lambda_{\vec{n}}\}_{\vec n}$
        \end{minipage}} &$N_1=1$ &$N_1=2$ &$N_1=3$ &$N_1=4$&$N_1=5$\\
        
        $N_2=1$& \makecell{\textbf{105}\\21,7 }& \makecell{\textbf{-} }&\makecell{\textbf{105}\\11,8 }&\makecell{\textbf{-} }&\makecell{\textbf{21}\\3,3 }\\
         
         $N_2=2$& \makecell{\textbf{-} }& \makecell{\textbf{195}\\24,30,9 }&\makecell{\textbf{-} }&\makecell{\textbf{120}\\6,13,9 }&\makecell{\textbf{-} }\\
         
         $N_2=3$& \makecell{\textbf{105}\\11,8 }& \makecell{\textbf{-} }&\makecell{\textbf{215}\\10,27,\\32,10 }&\makecell{\textbf{-} }&\makecell{\textbf{120}\\6,13,9 }\\
         
         $N_2=4$&\makecell{\textbf{-} }& \makecell{\textbf{120}\\6,13,9 }&\makecell{\textbf{-} }&\makecell{\textbf{215}\\10,27,\\32,10 }&\makecell{\textbf{-} }\\
         
         $N_2=5$&\makecell{\textbf{21}\\3,3 }& \makecell{\textbf{-} }&\makecell{\textbf{120}\\6,13,9 }&\makecell{\textbf{-} }&\makecell{\textbf{215}\\10,27,\\32,10 }\vspace{1ex}\\
         \Xhline{3\arrayrulewidth}
    \end{tabular}}
    \caption{A table enumerating the number of tensors which are not orthogonal to a tensor with $N=6$ and various values of $N_1,\ N_2$. The number of orbits for each projector is enumerated below the total number of tensors.}
    \label{tab:2spinCount4n}
\end{table}
\end{minipage}
\begin{minipage}{.47\linewidth}
    \begin{table}[H]
        \resizebox{0.6\width}{!}{
        \centering
    \begin{tabular}{c c c c c c}
        \Xhline{3\arrayrulewidth}
        \vspace{-3ex}\\
        \fbox{\begin{minipage}{5em}\centering $|V^{(N=7)}_{N_1,N_2}|$\\ $\{\Lambda_{\vec{n}}\}_{\vec n}$
        \end{minipage}} &$N_1=1$ &$N_1=2$ &$N_1=3$ &$N_1=4$&$N_1=5$\\
        
        $N_2=1$& \makecell{\textbf{-} }& \makecell{\textbf{420}\\30,14 }&\makecell{\textbf{-} }&\makecell{\textbf{210}\\11,8 }&\makecell{\textbf{-} }\\
         
         $N_2=2$&  \makecell{\textbf{420}\\30,14 }& \makecell{\textbf{-} }&\makecell{\textbf{630}\\27,40,17 }&\makecell{\textbf{-} }&\makecell{\textbf{231}\\6,13,9 }\\
         
         $N_2=3$& \makecell{\textbf{-} }& \makecell{\textbf{630}\\27,40,17 }&\makecell{\textbf{-} }&\makecell{\textbf{665}\\10,30,\\42,18 }&\makecell{\textbf{-} }\\
         
         $N_2=4$& \makecell{\textbf{210}\\11,8 }& \makecell{\textbf{-} }&\makecell{\textbf{665}\\10,30,\\42,18  }&\makecell{\textbf{-} }&\makecell{\textbf{665}\\10,30,\\42,18  }\\
         
         $N_2=5$&\makecell{\textbf{-} }& \makecell{\textbf{231}\\6,13,9 }&\makecell{\textbf{-} }&\makecell{\textbf{665}\\10,30,\\42,18  }&\makecell{\textbf{-} }\vspace{1ex}\\
         \Xhline{3\arrayrulewidth}
    \end{tabular}}
    \caption{A table enumerating the number of tensors which are not orthogonal to a tensor with $N=7$ and various values of $N_1,\ N_2$. The number of orbits for each projector is enumerated below the total number of tensors.}
    \label{tab:2spinCount3n}
\end{table}
\end{minipage}
\hfill
\begin{minipage}{.47\linewidth}
    \begin{table}[H]
        \centering
        \resizebox{0.6\width}{!}{
    \begin{tabular}{c c c c c c}
        \Xhline{3\arrayrulewidth}
        \vspace{-3ex}\\
        \fbox{\begin{minipage}{5em}\centering $|V^{(N=8)}_{N_1,N_2}|$\\ $\{\Lambda_{\vec{n}}\}_{\vec n}$
        \end{minipage}} &$N_1=1$ &$N_1=2$ &$N_1=3$ &$N_1=4$&$N_1=5$\\
        
        $N_2=1$& \makecell{\textbf{945}\\42,12 }& \makecell{\textbf{-} }&\makecell{\textbf{1260}\\32,17 }&\makecell{\textbf{-} }&\makecell{\textbf{378}\\11,8 }\\
         
         $N_2=2$& \makecell{\textbf{-} }& \makecell{\textbf{2205}\\75,67,17 }&\makecell{\textbf{-} }&\makecell{\textbf{1680}\\27,42,20 }&\makecell{\textbf{-} }\\
         
         $N_2=3$& \makecell{\textbf{1260}\\32,17 }& \makecell{\textbf{-} }&\makecell{\textbf{2765}\\46,91,\\76,19 }&\makecell{\textbf{-} }&\makecell{\textbf{1736}\\10,30,\\44,21 }\\
         
         $N_2=4$&\makecell{\textbf{-} }& \makecell{\textbf{1680}\\27,42,20 }&\makecell{\textbf{-} }&\makecell{\textbf{2835}\\15,50,94,\\78,20 }&\makecell{\textbf{-} }\\
         
         $N_2=5$&\makecell{\textbf{378}\\11,8 }& \makecell{\textbf{-} }&\makecell{\textbf{1736}\\10,30,\\44,21 }&\makecell{\textbf{-} }&\makecell{\textbf{2835}\\15,50,94,\\78,20  }\vspace{1ex}\\
         \Xhline{3\arrayrulewidth}
    \end{tabular}}
    \caption{A table enumerating the number of tensors which are not orthogonal to a tensor with $N=8$ and various values of $N_1,\ N_2$.The number of orbits for each projector is enumerated below the total number of tensors. }
    \label{tab:2spinCount4n}
\end{table}
\end{minipage}
\begin{minipage}{.47\linewidth}
    \begin{table}[H]
        \resizebox{0.6\width}{!}{
        \centering
    \begin{tabular}{c c c c c c}
        \Xhline{3\arrayrulewidth}
        \vspace{-3ex}\\
        \fbox{\begin{minipage}{5em}\centering $|V^{(N=9)}_{N_1,N_2}|$\\ $\{\Lambda_{\vec{n}}\}_{\vec n}$
        \end{minipage}} &$N_1=1$ &$N_1=2$ &$N_1=3$ &$N_1=4$&$N_1=5$\\
        
        $N_2=1$& \makecell{\textbf{-} }& \makecell{\textbf{4725}\\67,26 }&\makecell{\textbf{-} }&\makecell{\textbf{3150}\\32,18 }&\makecell{\textbf{-} }\\
         
         $N_2=2$&  \makecell{\textbf{4725}\\67,26 }& \makecell{\textbf{-} }&\makecell{\textbf{8505}\\91,99,34 }&\makecell{\textbf{-} }&\makecell{\textbf{3906}\\27,42,21 }\\
         
         $N_2=3$& \makecell{\textbf{-} }& \makecell{\textbf{8505}\\91,99,34 }&\makecell{\textbf{-} }&\makecell{\textbf{9765}\\50,108,\\109,37 }&\makecell{\textbf{-} }\\
         
         $N_2=4$& \makecell{\textbf{3150}\\32,18 }& \makecell{\textbf{-} }&\makecell{\textbf{9765}\\50,108,\\109,37 }&\makecell{\textbf{-} }&\makecell{\textbf{9891}\\15,54,111,\\111,38 }\\
         
         $N_2=5$&\makecell{\textbf{-} }& \makecell{\textbf{3906}\\27,42,21 }&\makecell{\textbf{-} }&\makecell{\textbf{9891}\\15,54,111,\\111,38  }&\makecell{\textbf{-} }\vspace{1ex}\\
         \Xhline{3\arrayrulewidth}
    \end{tabular}}
    \caption{A table enumerating the number of tensors which are not orthogonal to a tensor with $N=9$ and various values of $N_1,\ N_2$. The number of orbits for each projector is enumerated below the total number of tensors.}
\end{table}
\end{minipage}
\hfill
\begin{minipage}{.47\linewidth}
    \begin{table}[H]
        \centering
        \resizebox{0.6\width}{!}{
    \begin{tabular}{c c c c c c}
        \Xhline{3\arrayrulewidth}
        \vspace{-3ex}\\
        \fbox{\begin{minipage}{5em}\centering $|V^{(N=10)}_{N_1,N_2}|$\\ $\{\Lambda_{\vec{n}}\}_{\vec n}$
        \end{minipage}} &$N_1=1$ &$N_1=2$ &$N_1=3$ &$N_1=4$&$N_1=5$\\
        
        $N_2=1$& \makecell{\textbf{10395}\\78,19 }& \makecell{\textbf{-} }&\makecell{\textbf{17325}\\76,34 }&\makecell{\textbf{-} }&\makecell{\textbf{6930}\\32,18 }\\
         
         $N_2=2$& \makecell{\textbf{-} }& \makecell{\textbf{29295}\\183,136,\\28 }&\makecell{\textbf{-} }&\makecell{\textbf{26775}\\94,109,43 }&\makecell{\textbf{-} }\\
         
         $N_2=3$& \makecell{\textbf{17325}\\76,34 }& \makecell{\textbf{-} }&\makecell{\textbf{41895}\\164,242,\\163,33 }&\makecell{\textbf{-} }&\makecell{\textbf{29295}\\50,111,\\119,46 }\\
         
         $N_2=4$&\makecell{\textbf{-} }& \makecell{\textbf{26775}\\94,109,\\43 }&\makecell{\textbf{-} }&\makecell{\textbf{45045}\\75,187,258,\\172,35 }&\makecell{\textbf{-} }\\
         
         $N_2=5$&\makecell{\textbf{6930}\\32,18 }& \makecell{\textbf{-} }&\makecell{\textbf{29295}\\50,111,\\119,46 }&\makecell{\textbf{-} }&\makecell{\textbf{45297}\\21,80,191,\\261,174,36 }\vspace{1ex}\\
         \Xhline{3\arrayrulewidth}
    \end{tabular}}
    \caption{A table enumerating the number of tensors which are not orthogonal to a tensor with $N=10$ and various values of $N_1,\ N_2$. The number of orbits for each projector is enumerated below the total number of tensors. }
    \label{tab:2spinCount4n}
\end{table}
\end{minipage}

\section{Proof of the transverse Gamma formula}
\label{sec:transgammaproof}
We now present proofs of \cref{eq:gammatotrans,eq:transtogamma,eq:shuffelQslash} used to express $\Gamma$ in terms of $\Gamma_\perp$ and vice versa.
We begin with \cref{eq:gammatotrans}, starting on the left-hand-side:
\begin{align}
    \Gamma^{\mu_1\dots \mu_n} &= \Gamma^{\nu_1\dots \nu_n} g^{\mu_1\nu_1}\cdots g^{\mu_n\nu_n}\\
                            &= \Gamma^{\nu_1\dots \nu_n} \left(g^{\mu_1\nu_1}_\perp + Q^{\mu_1}Q^{\nu_1}\right)\cdots \left(g^{\mu_n\nu_n}_\perp + Q^{\mu_n}Q^{\nu_n}\right)
\end{align}
Now we note that if two or more of $Q$s are contracted into the $\Gamma$ the term will vanish by antisymmetry, so the only terms that survive are 
\begin{equation}\label{eq:splittransgamma}
    \begin{split}
         \Gamma^{\mu_1\dots \mu_n} =& \Gamma^{\nu_1\dots \nu_n}g^{\mu_1\nu_1}_\perp\cdots g^{\mu_n\nu_n}_\perp\\
                                        &+ Q^{\mu_1} \Gamma^{Q\,\nu_2\dots \nu_n} g^{\mu_2\nu_2}_\perp\cdots g^{\mu_n\nu_n}_\perp\\
                                        & + \cdots\\
                                        & + Q^{\mu_n} \Gamma^{\nu_1\dots \nu_{n-1}\,Q} g^{\mu_1\nu_1}_\perp\cdots g^{\mu_{n-1}\nu_{n-1}}_\perp.
    \end{split}
\end{equation}
The first term is just $\Gamma^{\mu_1\dots \mu_n}_\perp$. We now need to simplify the other terms. Consider one such term and expand out the antisymmetrised gamma, which results in 
\begin{equation}\label{eq:QinGamma}
    \begin{split}
        \Gamma^{\nu_1\dots \nu_{n-1}\,Q} g^{\mu_1\nu_1}_\perp\cdots &g^{\mu_{n-1}\nu_{n-1}}_\perp \\&= Q^{\nu_n} \left(\,\frac{1}{n!}\sum_{\sigma \in S_n}\sgn(\sigma)\gamma^{\nu_{\sigma(1)}}\dots\gamma^{\nu_{\sigma(n)}}\right)g^{\mu_1\nu_1}_\perp\cdots g^{\mu_{n-1}\nu_{n-1}}_\perp .
    \end{split}
\end{equation}
Take the right-hand-side of \cref{eq:QinGamma} and decompose the sum over $\sigma\in S_n$ in to one over $S_{n-1}$ and a shuffle in of the $\gamma^{\mu_n}$ with the appropriate sign, so it becomes 
\begin{equation}
    \begin{split}
        Q^{\nu_n} \left(\,\frac{1}{n!}\sum_{k=1}^{n}\sum_{\sigma \in S_{n-1}}\sgn(\sigma)\gamma^{\nu_{\sigma(1)}}\dots\gamma^{\nu_{\sigma(k-1)}}\gamma^{\nu_n}\gamma^{\nu_{\sigma(k)}}\dots\gamma^{\nu_{\sigma(n-1)}}(-1)^{(k+n)}\right)&
        \\&\hspace{-5em}\times g^{\mu_1\nu_1}_\perp\cdots g^{\mu_{n-1}\nu_{n-1}}_\perp.
    \end{split}
\end{equation}
The contraction with the $Q^{\nu_n}$ and the $g_\perp$s leaves us with
\begin{equation}
    \frac{1}{n!}\sum_{k=1}^{n}\sum_{\sigma \in S_{n-1}}\sgn(\sigma)\gamma^{\mu_{\sigma(1)}}_\perp\dots\gamma^{\mu_{\sigma(k-1)}}_\perp\slashed{Q}\gamma^{\mu_{\sigma(k)}}_\perp\dots\gamma^{\mu_{\sigma(n-1)}}_\perp(-1)^{(k+n)}.
\end{equation}
Anticommuting the $\slashed{Q}$ past the first $k-1$ $\gamma_\perp$s results in an additional factor of $(-1)^{k-1}$, leaving us with 
\begin{equation}
    \frac{1}{n!}\sum_{k=1}^{n}\sum_{\sigma \in S_{n-1}}\sgn(\sigma)\slashed{Q}\gamma^{\mu_{\sigma(1)}}_\perp\dots\gamma^{\mu_{\sigma(k-1)}}_\perp\gamma^{\mu_{\sigma(k)}}_\perp\dots\gamma^{\mu_{\sigma(n-1)}}_\perp(-1)^{(2k+n-1)}.
\end{equation}
The sum over $k$ is now trivial along with the power of $2k$ in the sign, so the expression simplifies:
\begin{equation}
    \frac{n}{n!}\sum_{\sigma \in S_{n-1}}\sgn(\sigma)\slashed{Q}\gamma^{\mu_{\sigma(1)}}_\perp\dots\gamma^{\mu_{\sigma(n-1)}}_\perp(-1)^{(n+1)}.
\end{equation} 
Finally, we apply again the definition of $\Gamma$ to arrive at the simple expression
\begin{equation}\label{eq:QinGamma-simplified}
    \Gamma^{\nu_1\dots \nu_{n-1}\,Q} g^{\mu_1\nu_1}_\perp\cdots g^{\mu_{n-1}\nu_{n-1}}_\perp =\slashed{Q}\Gamma^{\mu_1\dots \mu_{n-1}}_\perp(-1)^{(n+1)}.
\end{equation}
This result generalises to the following:
\begin{equation}
    \Gamma^{\nu_1\dots \nu_{k-1}\,Q\,\nu_{k}\dots \nu_{n-1}}g^{\mu_1\nu_1}_\perp\cdots g^{\mu_{n-1}\nu_{n-1}}_\perp = (-1)^{k+1}\slashed{Q}\Gamma^{\nu_1\dots \nu_{n-1}}_\perp.
\end{equation}
Finally, we can apply this result to \cref{eq:splittransgamma} and recover 
\begin{equation}\label{eq:gammaproof-gammatotrans}
    \Gamma^{\mu_1...\mu_n} = \Gamma^{\mu_1...\mu_n}_\perp + \sum_{m=1}^n\slashed{Q}\,Q^{\mu_m}\, \Gamma^{\mu_{1}...\widehat{\mu_m}...\mu_{n}}_\perp (-1)^{m+1}.
\end{equation}
To find the reverse transformation we expand out $\Gamma_\perp$ in terms of $Q$ and metric tensors 
\begin{equation}
    \begin{split}
        \Gamma^{\mu_1...\mu_n}_\perp  &= \Gamma^{\nu_1...\nu_n} g^{\mu_1\nu_1}_\perp\cdots g^{\mu_{n}\nu_{n}}_\perp\\
            &=\Gamma^{\nu_1\dots \nu_n} \left(g^{\mu_1\nu_1} - Q^{\mu_1}Q^{\nu_1}\right)\cdots \left(g^{\mu_n\nu_n} - Q^{\mu_n}Q^{\nu_n}\right).
    \end{split}
\end{equation}
Now take the right-hand-side and expand the brackets:
\begin{equation}
    \begin{split}
        \Gamma^{\mu_1\dots \mu_n}-& Q^{\mu_1} \Gamma^{Q\,\nu_2\dots \nu_n} g^{\mu_2\nu_2}\cdots g^{\mu_n\nu_n}
            - \cdots\\
            \dots-& Q^{\mu_k} \Gamma^{\nu_1\dots \nu_{k-1}\,Q\,\nu_{k+1}\dots \nu_{n}} g^{\mu_1\nu_1}\cdots g^{\mu_{k-1}\nu_{k-1}} g^{\mu_{k+1}\nu_{k+1}}\cdots g^{\mu_{n-1}\nu_{n-1}}- \cdots\\
             \dots-& Q^{\mu_n} \Gamma^{\nu_1\dots \nu_{n-1}\,Q} g^{\mu_1\nu_1}\cdots g^{\mu_{n-1}\nu_{n-1}}.\\
    \end{split}
\end{equation}
The previous expression can be expressed more simply as a sum which allows us to anticommute the $Q$ to the end of the $\Gamma$ and picking up an appropriate sign:
\begin{equation}
    \begin{split}
        \Gamma^{\mu_1\dots \mu_n}\ - &\sum_{k=1}^{n} Q^{\mu_k} \Gamma^{\mu_1\dots\mu_{k-1}Q\mu_{k+1}\dots\mu_{n}}\\
        &= \Gamma^{\mu_1\dots \mu_n}\ - \sum_{k=1}^{n} Q^{\mu_k} \Gamma^{\mu_1\dots\mu_{k-1}\mu_{k+1}\dots\mu_{n}\,Q}(-1)^{n-k}
        .
    \end{split}
\end{equation}
We are left with the result
\begin{equation}
    \Gamma^{\mu_1...\mu_n}_\perp = \Gamma^{\mu_1...\mu_n} + \sum_{m=1}^n\,Q^{\mu_m}\, \Gamma^{\mu_1...\widehat{\mu_m}...\mu_n,Q} (-1)^{m+n+1}.
\end{equation}
Finally, we wish to express $\slashed{Q}\Gamma^{\mu_1\dots\mu_n}_\perp$ in terms of non-transverse quantities, which we achieve by working backwards from $\Gamma^{\mu_1\dots\mu_n\,Q}$:
    \begin{equation}
        \begin{split}
            \Gamma^{\mu_1\dots\mu_n\,Q} &= \\
            & \hspace{-1em} Q^\alpha\left( \Gamma^{\mu_1...\mu_n\alpha}_\perp + \sum_{m=1}^n\slashed{Q}\,Q^{\mu_m}\, \Gamma^{\mu_{1}...\widehat{\mu_m}...\mu_{n}\alpha}_\perp (-1)^{m+1} + \slashed{Q}Q^\alpha \Gamma^{\mu_1...\mu_n}_\perp(-1)^{n+2}\right) \\
                 & \hspace{20em}= \slashed{Q} \Gamma^{\mu_1...\mu_n}_\perp(-1)^{n},
        \end{split}
    \end{equation}
where we have used \cref{eq:gammaproof-gammatotrans} and then deleted any terms where $Q$ was contracted into $\Gamma_\perp$.

\bibliographystyle{JHEP}
\bibliography{refs}
 
\end{document}